\PassOptionsToPackage{pdfpagelabels=false}{hyperref}
\expandafter\def\csname ver@fixltx2e.sty\endcsname{}

\documentclass[fleqn,usenatbib]{mnras}

\usepackage[utf8]{inputenc}

\usepackage{longtable}
\usepackage{amssymb}

\usepackage{silence} 
\usepackage{times}
\usepackage{enumitem}
\usepackage{graphicx}
\usepackage{amsmath}
\usepackage{bm}
\usepackage{multicol}
\usepackage{graphics}
\usepackage{url}
\usepackage{epsfig}
\usepackage{wrapfig}
\usepackage{datetime}
\usepackage{ulem}
\usepackage{verbatim}
\usepackage{float}
\usepackage{lastpage}
\usepackage{enumitem}
\usepackage{deluxetable}
\usepackage{CJK}

\usepackage{natbib}

\providecommand{\mnras}{MNRAS}
\providecommand{\apj}{ApJ}
\providecommand{\apjl}{ApJ}
\providecommand{\aj}{AJ}
\providecommand{\apjs}{ApJS}
\providecommand{\aap}{A\&A}
\providecommand{\araa}{ARA\&A}
\providecommand{\nat}{Nature}
\providecommand{\pasp}{PASP}
\providecommand{\pasa}{PASA}

\providecommand\apss{ApSS}

\providecommand\pasj{PASJ}

\def\simlt{\mathrel{\hbox{\rlap{\hbox{\lower4pt\hbox{$\sim$}}}\hbox{$<$}}}}
\def\simgt{\mathrel{\hbox{\rlap{\hbox{\lower4pt\hbox{$\sim$}}}\hbox{$>$}}}}

\def\hst{{\it HST}}
\def\spitzer{{\it Spitzer}}

\def\I{\,\textsc{i}}
\def\II{\,\textsc{ii}}
\def\III{\,\textsc{iii}}

\def\arcsec{$^{\,\prime\prime}$}

\title[SN\,2020jfo Explosion and Progenitor System]{Type II-P Supernova Progenitor Star Initial Masses and SN\,2020jfo: Direct Detection, Light Curve Properties, Nebular Spectroscopy, and Local Environment}

\def\northwestern{1}
\def\dark{2}
\def\ucla{3}
\def\ifa{4}
\def\ucsc{5}
\def\toronto{6}
\def\auburn{7}
\def\NCU{8}
\def\stsci{9}
\def\jhu{10}
\def\thacher{11}
\def\psu{12}
\def\icds{13}
\def\igc{14}
\def\illinois{15}
\def\nhfp{\dagger}

\author[Kilpatrick~et~al.]{\href{https://orcid.org/0000-0002-5740-7747
}{Charles D. Kilpatrick}$^{\northwestern}$\thanks{Email: ckilpatrick@northwestern.edu}, 
\href{https://orcid.org/0000-0001-9695-8472}{Luca Izzo}$^{\dark}$,
\href{https://orcid.org/0000-0001-7017-8582}{Rory~O.~Bentley}$^{\ucla}$,
\href{https://orcid.org/0000-0001-6965-7789}{Kenneth~C.~Chambers}$^{\ifa}$, \newauthor
\href{https://orcid.org/0000-0003-4263-2228}{David~A.~Coulter}$^{\ucsc}$,
\href{https://orcid.org/0000-0001-7081-0082}{Maria~R.~Drout}$^{\toronto}$,
\href{https://orcid.org/0000-0001-5486-2747}{Thomas~de~Boer}$^{\ifa}$,
\href{https://orcid.org/0000-0002-2445-5275}{Ryan~J.~Foley}$^{\ucsc}$,
\href{https://orcid.org/0000-0002-8526-3963}{Christa~Gall}$^{\dark}$, \newauthor
\href{https://orcid.org/0000-0002-5440-2350}{Melissa~R.~Halford}$^{\auburn}$,
\href{https://orcid.org/0000-0002-6230-0151}{David~O.~Jones}$^{\ucsc\nhfp}$,
\href{https://orcid.org/0000-0001-5710-8395}{Danial~Langeroodi}$^{\dark}$,
\href{https://orcid.org/0000-0002-7272-5129}{Chien-Cheng~Lin}$^{\ifa}$, \newauthor
\href{https://orcid.org/0000-0002-7965-2815}{Eugene~A.~Magnier}$^{\ifa}$,
\href{https://orcid.org/0000-0002-1052-6749}{Peter~McGill}$^{\ucsc}$,
\href{https://orcid.org/0000-0002-7296-6547}{Anna~J.~G.~O`Grady}$^{\toronto}$,
\href{https://orcid.org/0000-0001-8415-6720}{Yen-Chen Pan}\begin{CJK*}{UTF8}{bsmi}(潘彥丞)\end{CJK*}$^{\NCU}$, \newauthor
\href{https://orcid.org/0000-0003-2558-3102}{Enrico~Ramirez-Ruiz}$^{\ucsc}$,
\href{https://orcid.org/0000-0002-4410-5387}{Armin~Rest}$^{\stsci,\jhu}$,
\href{https://orcid.org/0000-0002-9486-818X}{Jonathan~J.~Swift}$^{\thacher}$,
\href{https://orcid.org/0000-0002-1481-4676}{Samaporn~Tinyanont}$^{\ucsc}$, \newauthor
\href{https://orcid.org/0000-0002-5814-4061}{V.~Ashley~Villar}$^{\psu,\icds,\igc}$,
\href{https://orcid.org/0000-0002-1341-0952}{Richard~J.~Wainscoat}$^{\ifa}$,
\href{https://orcid.org/0000-0002-4186-6164}{Amanda~Rose~Wasserman}$^{\illinois}$, \newauthor
\href{https://orcid.org/0000-0002-0840-6940}{S.~Karthik~Yadavalli}$^{\psu}$,
\href{https://orcid.org/0000-0001-7823-2627}{Grace~Yang}$^{\thacher}$ \\
$^\northwestern$Center for Interdisciplinary Exploration and Research in Astrophysics (CIERA) and Department of Physics and Astronomy, \\
$^\dark$DARK, Niels Bohr Institute, University of Copenhagen, Jagtvej 128, 2200 Copenhagen, Denmark, \\
$^\ucla$Physics and Astronomy Department, University of California, Los Angeles, CA 90095-1547, USA, \\
$^\ifa$Institute for Astronomy, University of Hawaii, 2680 Woodlawn Drive, Honolulu, HI 96822, USA, \\
$^\ucsc$Department of Astronomy and Astrophysics, University of California, Santa Cruz, CA 95064, USA, \\
$^\toronto$David A. Dunlap Department of Astronomy and Astrophysics, University of Toronto, 50 St. George Street, Toronto, Ontario, M5S 3H4, Canada, \\
$^\auburn$Physics Department, Leach Science Center, Auburn University, Auburn, AL 36849, USA, \\
$^\NCU$Graduate Institute of Astronomy, National Central University, 300 Jhongda Road, 32001 Jhongli, Taiwan, \\
$^\stsci$Space Telescope Science Institute, Baltimore, MD 21218, USA, \\
$^{\jhu}$Department of Physics and Astronomy, The Johns Hopkins University, Baltimore, MD 21218, USA, \\
$^{\thacher}$The Thacher School, 5025 Thacher Rd., Ojai, CA 93023, USA, \\
$^{\psu}$Department of Astronomy \& Astrophysics, The Pennsylvania State University, University Park, PA 16802, USA, \\
$^{\icds}$Institute for Computational \& Data Sciences, The Pennsylvania State University, University Park, PA, USA, \\
$^{\igc}$Institute for Gravitation and the Cosmos, The Pennsylvania State University, University Park, PA 16802, USA, \\
$^{\illinois}$Department of Astronomy, University of Illinois at Urbana-Champaign, 1002 W. Green St., IL 61801, USA, \\
$^{\dagger}$NASA Hubble Fellow
}

\begin{document}
\date{Accepted 0000, Received 0000, in original form 0000}
\pagerange{\pageref*{firstpage}--\pageref*{LastPage}} \pubyear{2022}
\maketitle
\label{firstpage}

\begin{abstract}

We present optical, ultraviolet, and infrared data of the type\,II supernova (SN\,II) 2020jfo at 14.5~Mpc.  This wealth of multiwavelength data allows to compare different metrics commonly used to estimate progenitor masses of SN\,II for the same object.  Using its early light curve, we infer SN\,2020jfo had a progenitor radius of $\approx$700~$R_{\odot}$, consistent with red supergiants of initial mass $M_{\rm ZAMS}=$11--13~$M_{\odot}$.  The decline in its late-time light curve is best fit by a ${}^{56}$Ni mass of 0.018$\pm$0.007~$M_{\odot}$ consistent with that ejected from SN\,II-P with $\approx$13~$M_{\odot}$ initial mass stars.  Early spectra and photometry do not exhibit signs of interaction with circumstellar matter, implying that SN\,2020jfo experienced weak mass loss within the final years prior to explosion.  Our spectra at $>$250~days are best fit by models from 12~$M_{\odot}$ initial mass stars.  We analyzed integral field unit spectroscopy of the stellar population near SN\,2020jfo, finding its massive star population had a zero age main sequence mass of 9.7$\substack{+2.5\\-1.3}~M_{\odot}$.  We identify a single counterpart in pre-explosion imaging and find it has an initial mass of at most $7.2\substack{+1.2\\-0.6}~M_{\odot}$.  We conclude that the inconsistency between this mass and indirect mass indicators from SN\,2020jfo itself is most likely caused by extinction with $A_{V}=2$--3~mag due to matter around the progenitor star, which lowered its observed optical luminosity.  As SN\,2020jfo did not exhibit extinction at this level or evidence for interaction with circumstellar matter between 1.6--450~days from explosion, we conclude that this material was likely confined within $\approx3000~R_{\odot}$ from the progenitor star.

\end{abstract}


\section{INTRODUCTION}\label{sec:introduction}

Type II supernovae \citep[SNe\,II; with broad lines of hydrogen,][]{filippenko97} are the terminal explosions of stars more massive than $\approx$8~$M_{\odot}$ that retain massive hydrogen envelopes \citep{Anderson08,smartt+09a,Arcavi16}.  While some peculiar SNe\,II are observed to explode from blue and yellow supergiants \citep[e.g., the B3 I progenitor star of SN\,1987A;][]{arnett87,Hillebrandt87,Woosley88,Podsiadlowski1992}, the vast majority of these transient sources are thought to be the terminal explosions of red supergiants (RSGs) that undergo nuclear burning up to the iron peak then unbind their outer layers via core-collapse and neutrino-driven explosions \citep{Burrows95}.  

In particular, the type II-P SN sub-class, which is characterized by a $\approx$50--100~day plateau in optical light curves \citep{barbon+79,kirshner+90,Valenti16,Arcavi16}, requires a progenitor star that retains a massive, extended hydrogen envelope that is first ejected and ionized by the SN explosion and then slowly recombines as the ejecta expand \citep{falk+73,chevalier+76,falk+78}.  Although there is some diversity in peak luminosity, plateau duration, and rise times even among SNe\,II-P \citep[e.g.,][]{hillier+19}, the overall similarity in SN\,II-P light curve properties points to the explosion of RSGs with varying hydrogen-envelope and oxygen-core masses \citep{Goldberg+20,Dessart21}.

One of the most significant open questions from the population of directly detected SN\,II-P and II-L \citep[a SN\,II sub-class with linearly declining light curves;][]{barbon+79} progenitor stars is why they all appear consistent with having $\log(L/L_{\odot})<5.2$ \citep[or $M_{\rm ZAMS}\lesssim17~M_{\odot}$ assuming stellar evolution tracks in][]{choi+16} while the overall population of RSGs extends to $\log(L/L_{\odot})\approx5.6$ \citep[or $M_{\rm ZAMS}\approx25$--$30~M_{\odot}$;][the so-called ``red supergiant problem'']{Smartt09,Elias-Rosa11,Davies18,Kochanek20}.  Confirmed progenitor stars to other core-collapse SN types do not exceed this luminosity and implied initial mass either \citep[e.g., counterparts to hydrogen and helium poor type IIb and Ib/c SN in][]{maund+11,van-dyk+11,cao+13,van-dyk+14,Kilpatrick21}. 

One favored explanation for this inconsistency is that high-mass stars do not explode but collapse directly into black holes \citep[``failed SNe'';][]{Sukhbold16}.  However, it remains possible that higher mass stars appear to have lower luminosities due to extreme variability similar to that observed from Betelgeuse \citep{Levesque20} and M51-DS1 \citep{Jencson22}, but this may affect a small fraction of the overall population \citet{Soraisam18}.  For stars with strong pre-SN mass, this could change the envelope structure and potentially lead to explosion as another SN subtype \citep[e.g.,][]{Smith14,Davies17,Beasor18,Zapartas21} or extinction from dense shells of circumstellar matter (CSM).  Critically, the slow-moving, relatively cool CSM of RSGs is known to contain significant dust masses \citep{verhoelst+09}, which could shift the peak of the spectral energy distributions well into the infrared and lower their optical luminosities \citep[although, see][for counter-arguments to this hypothesis]{Kochanek12,Walmswell12,Dwarkadas14,Kochanek17}.  Analysis of all nearby SNe with pre-explosion \hst\ imaging, especially those with rare or unusual spectroscopic and photometric evolution, is needed to address potential biases in the population of SN\,II-P with known progenitor stars and find potential missing high-mass progenitor stars.  The binary fraction among RSGs is also high \citep[16--41\% in M31 and M33;][]{Neugent20}, implying that mergers or Roche lobe overflow may reduce the number of apparently luminous RSGs in pre-explosion imaging \citep[see, e.g.,][]{Zapartas21}.

Detailed observations of SN\,II-P at early times are essential to resolve this problem. If RSGs are enshrouded in dense CSM but are not discovered for several days or weeks after core collapse, that material will be swept up by the SN shock and any dust contributing to circumstellar extinction will be destroyed, which is visible in both early-time photometry \citep[e.g.,][]{Hosseinzadeh18,Jacobson-Galan22,terreran+2022} and ``flash spectroscopy'' \citep[e.g.,][]{gal-yam+14,khazov+15}.  Furthermore, thousands of SN\,II-P are currently discovered per year by wide-field optical surveys such as the Zwicky Transient Facility and the Young Supernova Experiment and soon hundreds of thousands with the Vera C. Rubin Observatory's Legacy Survey of Space and Time \citep{LSST2009,Bellm19,Jones2021}.  Higher cadence, multi-wavelength observations of a nearby ($<$40~Mpc) sample of SNe\,II-P with pre-explosion counterpart detections can provide a baseline for understanding the mapping between a large sample of SNe and their explosion properties and the mass, pre-explosion evolution, and circumstellar environments of these stars.

Here we discuss the SN\,II-P 2020jfo discovered in M61 by the Zwicky Transient Facility on  2020 May 2 \citep{2020TNSTR1248....1N}.  SN\,2020jfo was classified as a SN\,II-P based on a spectrum obtained by the Liverpool Telescope and Spectrograph for Rapid Acquisition of Transients \citep[SPRAT;][]{2020TNSCR1259....1P} and presented in \citet{Sollerman21}.  The site of SN\,2020jfo was observed with the {\it Hubble Space Telescope} (\hst) in multiple optical and ultraviolet bands and epochs starting 24~yr before discovery, as well as {\it Spitzer} imaging from 16~yr before discovery.  We present additional optical imaging and spectroscopy of SN\,2020jfo spanning 1--450~days from discovery, as well as an analysis of all pre-explosion optical and mid-infrared constraints on the progenitor system of SN\,2020jfo.  Throughout this manuscript, our specific focus is on using follow-up data to inform constraints on the progenitor system (e.g., initial mass). We will compare and contrast to analysis presented in \citet{Sollerman21} and \citet{Teja22} when relevant in the sections below.

Throughout this paper, we assume a line-of-sight extinction of $A_{V}=0.064$~mag for the Milky Way \citep{Schlafly11}.  We also adopt a redshift $z=0.005224$ for M61 from \citet{Haynes18}. For consistency with \citet{Sollerman21}, we adopt a distance modulus of $\mu=30.81\pm0.20$ (14.5~Mpc), originally derived from the expanding photosphere method in \citet{Bose14} for SN\,2008in.  This estimate is broadly consistent with both the Tully-Fisher distance modulus to M61 of 30.21$\pm$0.70~mag \citep{Schoeniger97} and the mean distance modulus estimate to SN\,2020jfo of 31.06$\pm$0.36~mag used in \citet{Teja22}.

\section{OBSERVATIONS OF SN~2020\lowercase{jfo}}\label{sec:observations}

SN\,2020jfo was discovered and reported to the Transient Name Server by the Zwicky Transient Facility (ZTF)\footnote{SN\,2020jfo is also called ZTF20aaynrrh} Alert Management, Photometry and Evaluation of Lightcurves broker \citep[AMPEL;][]{Nordin19} on 2020 May 6.  We initiated observations by the Young Supernova Experiment \citep[YSE;][]{Jones2021} with the Pan-STARRS1 (PS1) telescope \citep{Kaiser2002} starting on 10 May 2020.  Below we describe our imaging and spectroscopic follow-up as well as archival data used in our analysis.

\begin{figure}
    \centering
    \includegraphics[width=0.47\textwidth]{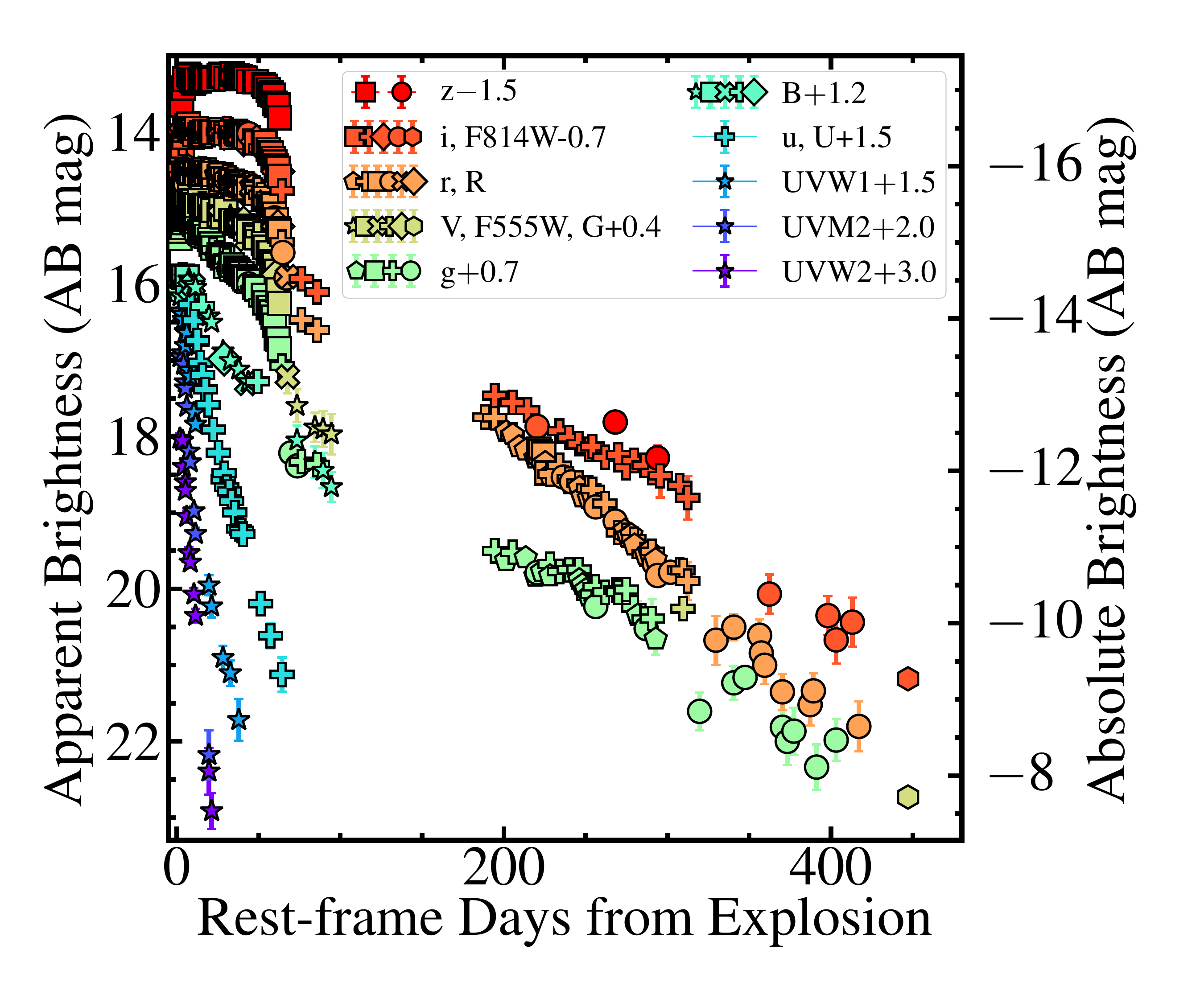}
    \includegraphics[width=0.47\textwidth]{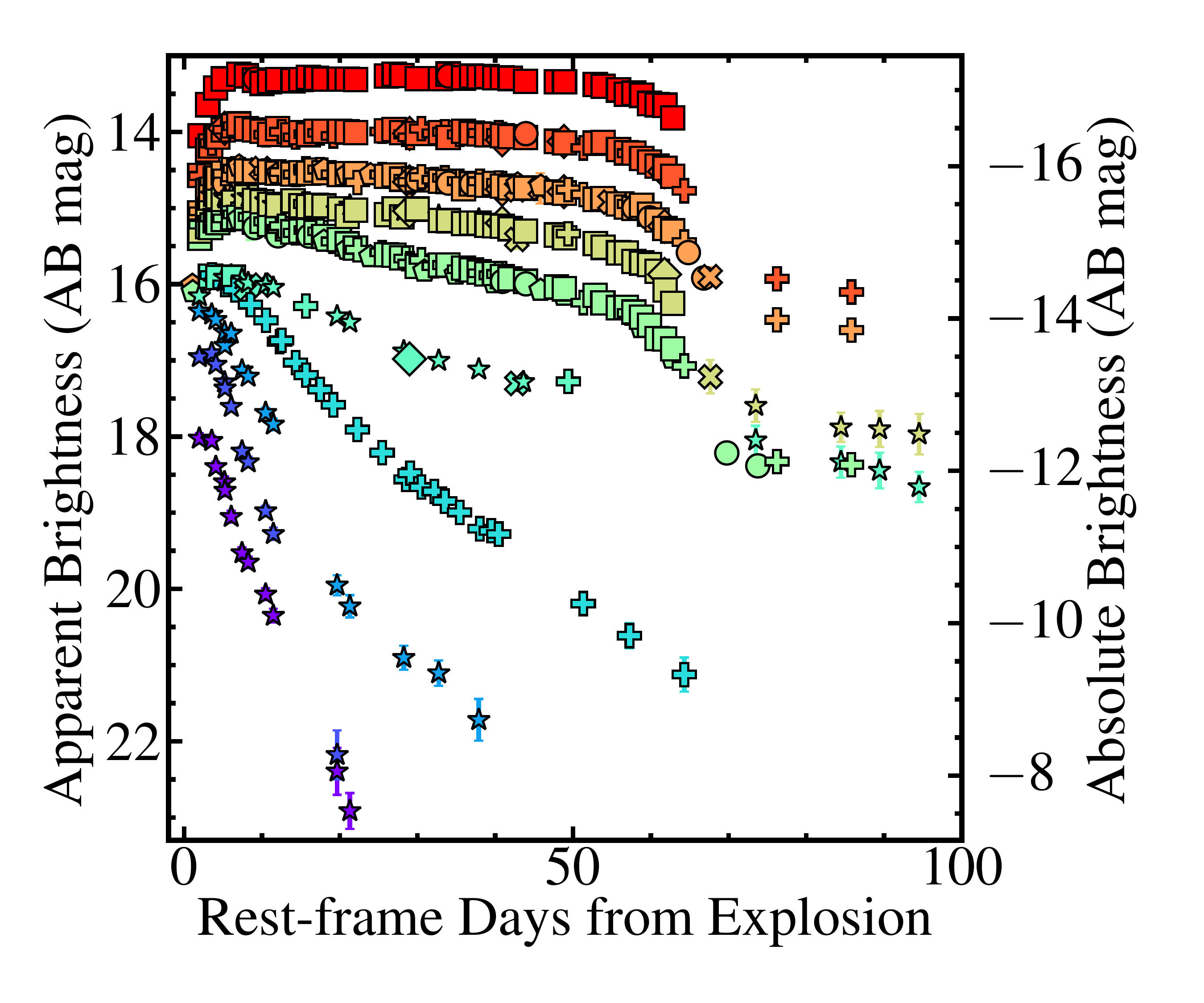}
    \caption{({\it Top panel}): Our ultraviolet and optical light curves of SN\,2020jfo.  We show all Auburn (solid X), {\it HST} (hexagons), Las Cumbres (pluses), Nickel (diamonds), Pan-STARRS (circles), {\it Swift} (stars), Thacher (squares), and ZTF (pentagons) detections of the SN on the same light curve.  All magnitudes are shown as observed in the AB system and before correcting for Milky Way extinction.  We show absolute magnitude on the right-hand axis assuming a distance modulus of 30.81~mag (14.5~Mpc).  ({\it Bottom panel}): Same as the upper panel but for the first 100 rest-frame days of data.}
    \label{fig:lightcurve}
\end{figure}

\subsection{Optical and Ultraviolet Imaging}\label{sec:imaging}

YSE imaging of SN\,2020jfo occurred from 2020 May 10 to 2021 July 10.  Initial reductions, including pixel-level corrections, were performed by the Image Processing Pipeline (IPP) as described in \citet{magnier13}.  We then processed all PS1 imaging of SN\,2020jfo with the {\tt photpipe} imaging and photometry package \citep{rest+05} as described in \citet{Jones2021}.  The {\tt photpipe} PS1 pipeline is a well-tested system used across several SN and transient studies \citep{Rest14,Foley18, Jones19}.  From the IPP images, each frame was aligned and resampled using {\tt SWarp} \citep{swarp} to skycells from the PS1 3$\pi$ survey \citep{Chambers2017}.  Initial photometry was performed with {\tt DoPhot} \citep{schechter+93} and calibrated using photometric standards from the PS1 DR2 catalog \citep{flewelling+16}.   Image subtraction was performed in {\tt hotpants} \citep{Becker15}, and final forced photometry was extracted using a custom version of {\tt DoPhot} at the location of SN\,2020jfo.  All SN\,2020jfo photometric data were analyzed using the open-source target and observation management system {\tt YSE-PZ} \citep{2022zndo...7278430C, Coulter23}.

We also observed SN\,2020jfo with the Sinistro imagers on the Las Cumbres Observatory (LCO) 1m telescope network \citep{Brown13} from 2020 May 7 to 2021 March 15 in $ugri$ bands.  The initial imaging was processed automatically by the LCO {\tt BANZAI} pipeline \citep{McCully18}, which performs pixel-level corrections and astrometric calibration.  Following methods described in \citet{Kilpatrick21}, we further processed these images using {\tt photpipe} and following procedures similar to those described above with PS1.  The final photometry of SN\,2020jfo was performed on the unsubtracted LCO images using the point-spread function methods in {\tt DoPhot} and calibrated using $gri$ Pan-STARRS1 and $u$-band SDSS photometric standards \citep{Alam15} observed in the same field as SN\,2020jfo.

We observed SN\,2020jfo with the 1\,m Nickel telescope at Lick Observatory on Mt.\ Hamilton, California from 2020 May 10 to July 6 in $BVri$ bands.  Following standard procedures in {\tt photpipe} as described for the LCO data, we calibrated and performed photometry on all Nickel images.  The magnitudes were calibrated using PS1 $gri$ photometric standards transformed into Johnson $BV$ bands.

The Lulin Compact Imager on the 1\,m telescope at Lulin Observatory observed SN\,2020jfo from 2020 May 8 to May 11 in $BVgri$ bands.  These data were calibrated following standard procedures, and we performed photometry in each image following the same methods described above for the LCO data.

SN\,2020jfo was observed with the Ultraviolet and Optical Telescope \citep[UVOT;][]{Roming05} on the {\it Neil Gehrels Swift Observatory} from 2020 May 7 to 2021 February 25.  We used {\tt uvotsource} in HEASoft v6.26 to perform aperture photometry within a 3\arcsec\ aperture centered on SN\,2020jfo on the UVOT data files as described in \citet{Brown14}.  In order to account for background emission in each UVOT band, we measured the total flux at the site of SN\,2020jfo in frames obtained from 2021 February 24--25 and subtracted this from all previous observations.  We detect significant emission in all early-time ($\Delta t<30$~days; \autoref{fig:lightcurve}) UVOT data, but in later imaging where we do not detect emission, we report the combined 3$\sigma$ magnitude limit from each epoch and the later template imaging in \autoref{tab:photometry}.

We also observed SN\,2020jfo in $griz$ bands with the 0.7\,m Thacher Observatory telescope \citep{Swift22} located in Ojai, CA from 7 May to 18 December 2020.  We reduced these images in {\tt photpipe} using bias, dark, and flat-field frames obtained in the same instrumental configuration and following procedures described in \citet{Jacobson-Galan20}.  We obtained photometry of SN\,2020jfo following the same procedure as the LCO imaging described above.

SN\,2020jfo was observed with the Auburn 10'' telescope located in Auburn, AL from 13 May to 17 July 2020 in $BGR$ bands\footnote{Filter transmission curves for the Auburn camera are available at \url{https://astronomy-imaging-camera.com/product/zwo-lrgb-31mm-filters-2}}.  Following standard procedures in {\tt astropy}, we aligned and stacked individual exposures from each date and bandpass.  We then performed photometry on the stacked frames with {\tt DoPhot} and calibrated the photometry using Pan-STARRS $griz$ standard star magnitudes transformed into Johnson $BVR$ magnitudes.  We note that the $G$ bandpass has an effective wavelength of 5290~\AA, close to that of Johnson $V$-band ($\approx$5480~\AA\ in this photometric system), and so we calibrate our $G$-band photometry with $V$-band magnitudes.

We also used the ZTF photometry of SN\,2020jfo presented in \citet{Sollerman21}.  

\subsection{{\it Swift}/XRT Observations}\label{sec:xrt}

SN\,2020jfo was observed over 29 epochs with {\it Swift}/XRT \citep{burrows05} from 2020 May 7 to 2021 February 25.  Analyzing these data with {\tt HEASOFT} v6.28 \citep{heasoft}, we do not detect emission at $>$3$\sigma$ in any of these epochs or in the total merged event file.  From these data, we infer upper limits on the total count rate of $1.6\times10^{-3}$ to $9\times10^{-2}$ counts~s$^{-1}$ in each epoch, consistent with analysis in \citet{Sollerman21}.

From our {\it Swift}/XRT limits, we derive an equivalent X-ray luminosity limit assuming a line-of-sight hydrogen column of $N_{H}=1.4\times10^{20}$~cm$^{2}$ \citep[using our Milky Way extinction and following][]{Guver09} and the distance given above.  Assuming a photon spectral index of $\Gamma=2$, these limits are equivalent to 1.0--6.9$\times$10$^{39}$~erg~s$^{-1}$ from 0.9--295.3~days from discovery.

\subsection{Spectroscopy}\label{sec:spectroscopy}

SN\,2020jfo was observed with the FLOYDS spectrograph on the Faulkes-North 2m telescope at Haleakal\={a}, Hawaii and the Faulkes-South 2m telescope at Siding Spring Observatory, Australia.  Both sets of spectra were processed using standard procedures for bias, flat-fielding, cosmic-ray rejection, aperture extraction, wavelength and flux calibration in {\tt IRAF}\footnote{{\tt IRAF} is distributed by the National Optical Astronomy Observatory, which is operated by the Association of Universities for Research in Astronomy (AURA) under a cooperative agreement with the National Science Foundation.}.  We performed telluric corrections and flux calibration using spectrophotometric standards observed at the same observatory and instrumental configuration within $\pm$3~days from our target spectra.

We also observed SN\,2020jfo with the Kast spectrograph on the Shane 3m telescope at Lick Observatory, California on 2020 May 23 and 2020 July 27.  We reduced these data using a {\tt pyraf}-based spectroscopic pipeline \citep{Siebert20}\footnote{\url{https://github.com/msiebert1/UCSC_spectral_pipeline}}, which performs corrections for bias, flat-fielding, amplifier crosstalk, background level, and telluric lines.  The flux calibration was performed using a standard star spectrum observed on the same night as both Shane/Kast spectra.

Finally, we observed SN\,2020jfo with the Low Resolution Imaging Spectrograph (LRIS) on the Keck-I 10m telescope on 2021 February 12 and 2021 May 11.  These data were reduced following the same procedure described above for the Shane/Kast spectroscopy.  All spectroscopic observations are described in \autoref{tab:spec}.

\begin{table*}
    \centering{Optical Spectroscopy of SN\,2020jfo}
    \begin{center}
    \begin{tabular}{cccccc}
 \hline
 Date & MJD & Phase & Telescope & Instrument & Wavelength Range \\
 (UT) &     & (days)&           &            & (\AA) \\ \hline\hline
2020-05-07 & 58976.545 & 2.9 & Faulkes-South & FLOYDS & 4800--10000 \\
2020-05-08 & 58977.547 & 3.9 & Faulkes-South & FLOYDS & 4800--10000 \\
2020-05-12 & 58981.360 & 7.7 & Faulkes-North & FLOYDS & 4800--10000 \\
2020-05-16 & 58985.491 & 11.8 & Faulkes-South & FLOYDS & 4800--10000 \\
2020-05-19 & 58988.503 & 14.8 & Faulkes-South & FLOYDS & 4800--10000 \\
2020-05-23 & 58992.191 & 18.4 & Shane         & Kast   & 3500--11000 \\
2020-05-24 & 58993.458 & 19.7 & Faulkes-South & FLOYDS & 4800--10000 \\
2020-05-28 & 58997.483 & 23.7 & Faulkes-South & FLOYDS & 4800--10000 \\
2020-06-05 & 59005.482 & 31.7 & Faulkes-South & FLOYDS & 4800--10000 \\
2020-06-13 & 59013.303 & 39.4 & Faulkes-North & FLOYDS & 4800--10000 \\
2020-06-26 & 59026.273 & 52.3 & Faulkes-North & FLOYDS & 4800--10000 \\
2020-07-06 & 59036.425 & 62.4 & Faulkes-South & FLOYDS & 4800--10000 \\
2020-07-14 & 59044.262 & 70.2 & Faulkes-North & FLOYDS & 4800--10000 \\
2020-07-23 & 59053.252 & 79.2 & Faulkes-North & FLOYDS & 4800--10000 \\
2020-07-27 & 59057.184 & 83.1 & Shane         & Kast   & 3500--11000 \\
2020-08-01 & 59062.352 & 88.2 & Faulkes-South & FLOYDS & 4800--10000 \\
2021-02-12 & 59257.504 & 282.3 & Keck-I       & LRIS   & 3200--10800 \\
2021-05-11 & 59345.335 & 369.6 & Keck-I       & LRIS   & 3200--10800 \\ \hline
\end{tabular}
\caption{Phase is given in rest-frame days relative to the derived epoch of explosion at MJD = 58973.65 (see \autoref{sec:shock}).\label{tab:spec}}
\end{center}
 \end{table*}

\subsection{Pre-Explosion Archival Data}\label{sec:archival}

\subsubsection{{\it HST}}

The site of SN\,2020jfo was observed with the {\it Hubble Space Telescope}/WFPC2, ACS, and WFC3 over five epochs from 15 March 1996 to 29 March 2020, or 24.1 to 0.1~years before discovery (\autoref{tab:hst}).  Following methods described in \citet{Kilpatrick21b} and \citet{Kilpatrick21}, we used our custom {\tt python}-based pipeline {\tt hst123}\footnote{\url{https://github.com/charliekilpatrick/hst123}} to download, register, drizzle \citep[for details, see][]{drizzlepac}, and perform photometry in {\tt dolphot} on all pre-explosion {\it HST} imaging \citep{dolphot}.  We used recommended {\tt dolphot} settings for each imager as described in the respective manual\footnote{\url{americano.dolphinsim.com/dolphot}}.

\begin{figure*}
    \centering
    \includegraphics[width=\textwidth]{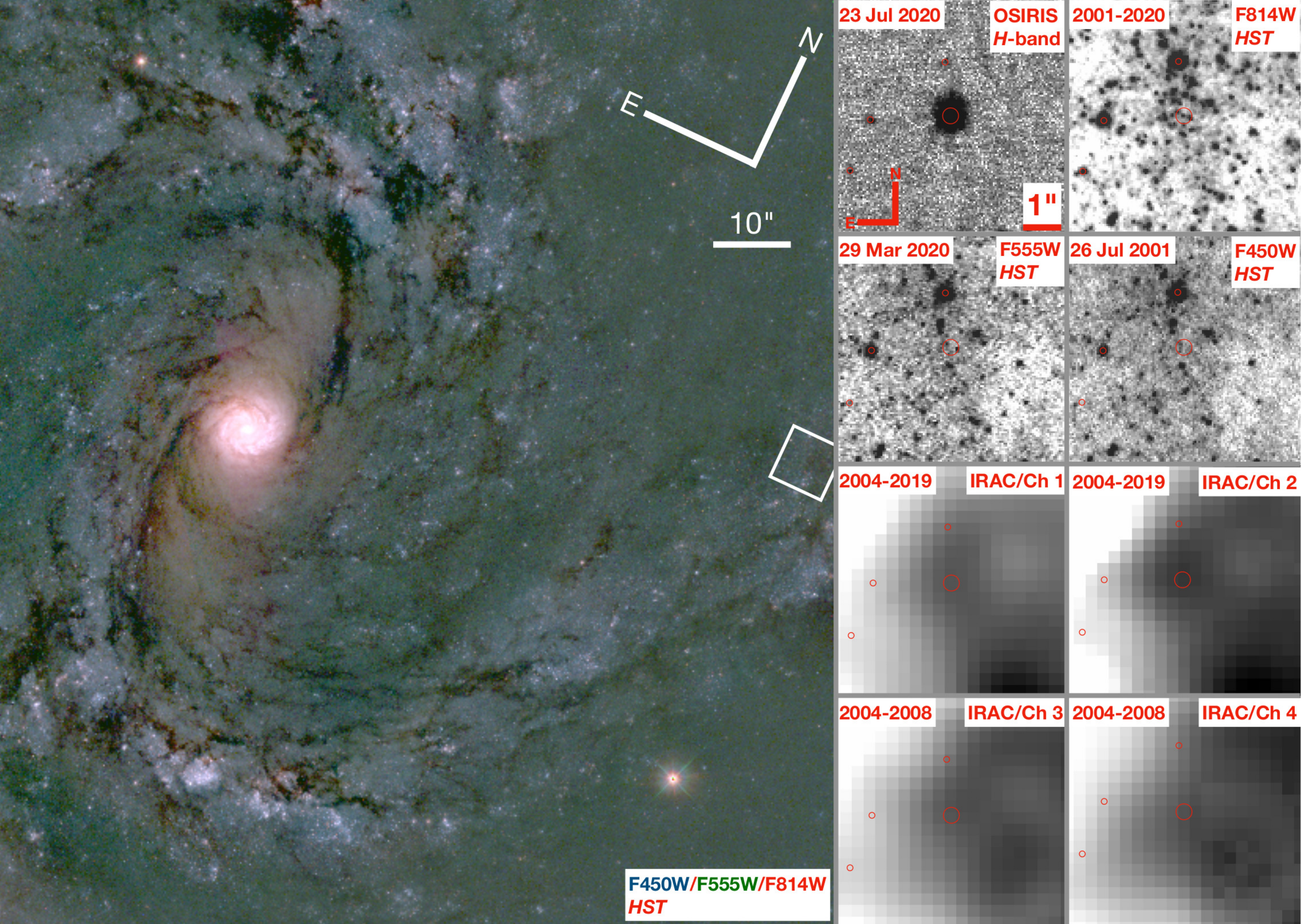}
    \caption{({\it Left}): {\it HST}/WFC3 imaging of M61 in F450W (blue), F555W (green), and F814W (red).  We show the approximate explosion site of SN\,2020jfo as a 10\arcsec$\times$10\arcsec\ white square in the outer western arm of M61.  ({\it Right panels}): Post-explosion OSIRIS $H$-band imaging (\autoref{sec:aoimaging}) and pre-explosion {\it HST} in F450W, F555W, and F814W and {\it Spitzer}/IRAC Ch 1--4 imaging in the 10\arcsec$\times$10\arcsec\ square centered on the explosion site of SN\,2020jfo.  In red, we show the locations of point sources in the {\it HST} and {\it Spitzer} frames used to align the OSIRIS imaging.}
    \label{fig:astrometry}
\end{figure*}

The final stacked imaging of M61 is shown in \autoref{fig:astrometry}.  We show a RGB colour image consisting of every F814W, F555W, and F450W frame, regardless of \hst\ instrument, that covers the site of SN\,2020jfo.

\begin{table*}
    \centering{\hst\ Photometry of the SN\,2020jfo Progenitor Candidate} \\
    \begin{tabular}{lccccc}
\hline MJD         & Phase & Instrument & Filter& Magnitude & Uncertainty \\
 & (days) & & & (AB mag) & (mag) \\\hline\hline
49509.8033 & -9416.58 & WFPC2 & F606W & $>$25.96 & -- \\
50157.9519 & -8771.66 & WFPC2 & F218W & $>$26.32 & -- \\
52116.0901 & -6823.31 & WFPC2 & F450W & $>$25.93 & -- \\
52116.1005 & -6823.30 & WFPC2 & F814W & 25.74 & 0.28 \\
56071.1848 & -2887.97 & ACS/WFC & F435W & $>$27.11 & -- \\
56071.2446 & -2887.91 & ACS/WFC & F814W & 25.62 & 0.08 \\
58580.8248 & -390.86 & ACS/WFC & F814W & 25.96 & 0.07 \\
58937.7315 & -35.74 & WFC3/UVIS & F814W & 25.60 & 0.09 \\
58937.7364 & -35.73 & WFC3/UVIS & F438W & $>$26.60 & -- \\
58937.7423 & -35.73 & WFC3/UVIS & F336W & $>$26.66 & -- \\
58937.7656 & -35.71 & WFC3/UVIS & F275W & $>$27.00 & -- \\
58937.7758 & -35.70 & WFC3/UVIS & F555W & $>$26.90 & -- \\
\hline
\end{tabular}
\centering
\caption{Phase is in rest-frame days relative to the explosion date MJD=58973.65 (\autoref{sec:shock}.  All magnitudes are in the AB system.}\label{tab:hst}
\end{table*}

\subsubsection{{\it Spitzer}/IRAC}

The {\it Spitzer Space Telescope} ({\it Spitzer}) observed M61 with the Infrared Array Camera (IRAC) over multiple epochs from 2004 June 10 to 2019 October 29 \citep[see, e.g.,][]{Crockett11,Fox15,Szalai19}.  These observations include both cold and warm {\it Spitzer}/IRAC data and cover all four IRAC channels (3.6, 4.5, 5.8, and 8~$\mu$m as in \autoref{tab:spitzer}).

Starting from the basic calibrated data ({\tt cbcd}) frames, we reduced these data using the {\tt photpipe} imaging and photometry package as described in \citet{Kilpatrick21}.  Each frame was calibrated using the zero points for the warm and cold {\it Spitzer} mission where appropriate and stacked together into a regridded image with pixel scale 0.6\arcsec~pixel$^{-1}$.  We performed photometry on each stacked frame with {\tt DoPhot} and recalculated the final zero point for each image using photometric standards from the Spitzer Enhanced Imaging Products source list \citep{SEIP}.  The maximum equivalent exposure time across each stacked {\it Spitzer}/IRAC image is 4.4, 5.1, 0.3, and 0.2~hr in Bands 1--4, respectively, with limiting magnitudes (5$\sigma$; AB mag) of 25.1, 25.1, 21.9, and 22.5~mag.  We show each stacked image centered on the site of SN\,2020jfo in \autoref{fig:astrometry}.

\begin{table}
    \centering{\spitzer/IRAC Photometry of the SN\,2020jfo Progenitor Candidate}
    \begin{tabular}{ccc}
\hline Filter & Magnitude & Uncertainty \\
& (AB mag) & (mag) \\\hline\hline
Ch 1 (3.6~$\mu$m) & 24.85 & 0.28 \\
Ch 2 (4.5~$\mu$m) & 25.01 & 0.25 \\
Ch 3 (5.8~$\mu$m) & $>$21.5 & -- \\
Ch 4 (8.0~$\mu$m) & $>$22.2 & -- \\
\hline
\end{tabular}
\caption{\spitzer/IRAC photometry and limits averaged across all epochs for the SN\,2020jfo counterpart.  Each limit is based on data obtained from 2004 June 10 to 2019 October 29, 15.9 to 0.5 years before explosion of SN\,2020jfo.}\label{tab:spitzer}
\end{table}

\subsection{VLT/MUSE}\label{sec:muse-obs}

M61 and the site of SN\,2020jfo were also observed with the Multi-Unit Spectroscopic Explorer \citep[MUSE;][]{Bacon2010} installed at the Yepun UT4 telescope of the European Southern Observatory (ESO) Very Large Telescope, under the program ID 1100.B-0651 (PI: Schinnerer) as a part of the PHANGS-MUSE survey \citep{Emsellem2021}. MUSE is an integral-field unit spectrograph covering the wavelength range from 4750--9350~\AA\ with a spectral resolution of 1.25~\AA. The field of view of MUSE, in the wide-field mode, covers a size of a 1'$\times$1'. The observations consist of multiple 3$\times$3 pointings that almost cover the entirety of M61 and observed on 2019 May 10, or 362~days before discovery of SN\,2020jfo. For the analysis presented in this paper, we obtained the reduced datacube publicly available on the ESO archive\footnote{\url{archive.eso.org/scienceportal/home}}. Details on the data reduction are described in \citet{Williams2021}.

\subsection{Adaptive Optics Imaging}\label{sec:aoimaging}

We observed SN\,2020jfo in $H$-band on 2020 July 23 with the Keck-I telescope from Mauna Kea, Hawaii and the OH-Suppressing Infra-Red Imaging Spectrograph \citep[OSIRIS;][]{OSIRIS} in conjunction with the Keck Adaptive Optics and laser guide star system \citep{LGSAO}.  We used SN\,2020jfo itself to perform tip-tilt corrections, which was $r=16.5$~mag at that epoch, resulting in $\approx$150~mas full-width at half-maximum point spread function (PSF).  Our SN\,2020jfo imaging consisted of a 9-point dither pattern consisting of 6$\times$10~s co-adds, or 540~s of effective exposure on the field surrounding SN\,2020jfo.  We processed the imaging using a {\tt python}-based reduction pipeline, including dark current subtraction and flat-fielding using calibration images obtained on the same night and instrumental configuration.  As the laser corrections resulted in a relatively broad PSF compared with optimal Keck-I/OSIRIS performance, we drizzled the final image to a pixel scale of 0.05\arcsec\ to match the pre-explosion {\it HST}/WFC3 image we used for alignment.  The final image is shown in \autoref{fig:astrometry}, in which we denote SN\,2020jfo and sources used to align to pre-explosion WFC3 imaging as described in \autoref{sec:alignment}.

\section{Physical Properties of SN~2020\lowercase{jfo}}\label{sec:physical}

\subsection{Line-of-sight Extinction}\label{sec:extinction}

Based on the Milky Way extinction maps of \citet{Schlafly11}, the Galactic reddening to SN\,2020jfo is only $E(B-V)=0.019$~mag, implying $A_{V}=0.06$~mag.  Thus any additional line-of-sight extinction due to a column of dust in M61 or in the circumstellar environment around SN\,2020jfo may dominate extinction indicators such as Na\I~D \citep{Poznanski12} or colour curves \citep{Galbany16}.  Assuming this material comes from interstellar dust in M61, it would contribute the same extinction to SN\,2020jfo and any pre-explosion counterpart, providing an independent constraint on the extinction and reddening we see from that source.

\begin{figure}
    \centering
    \includegraphics[width=0.49\textwidth]{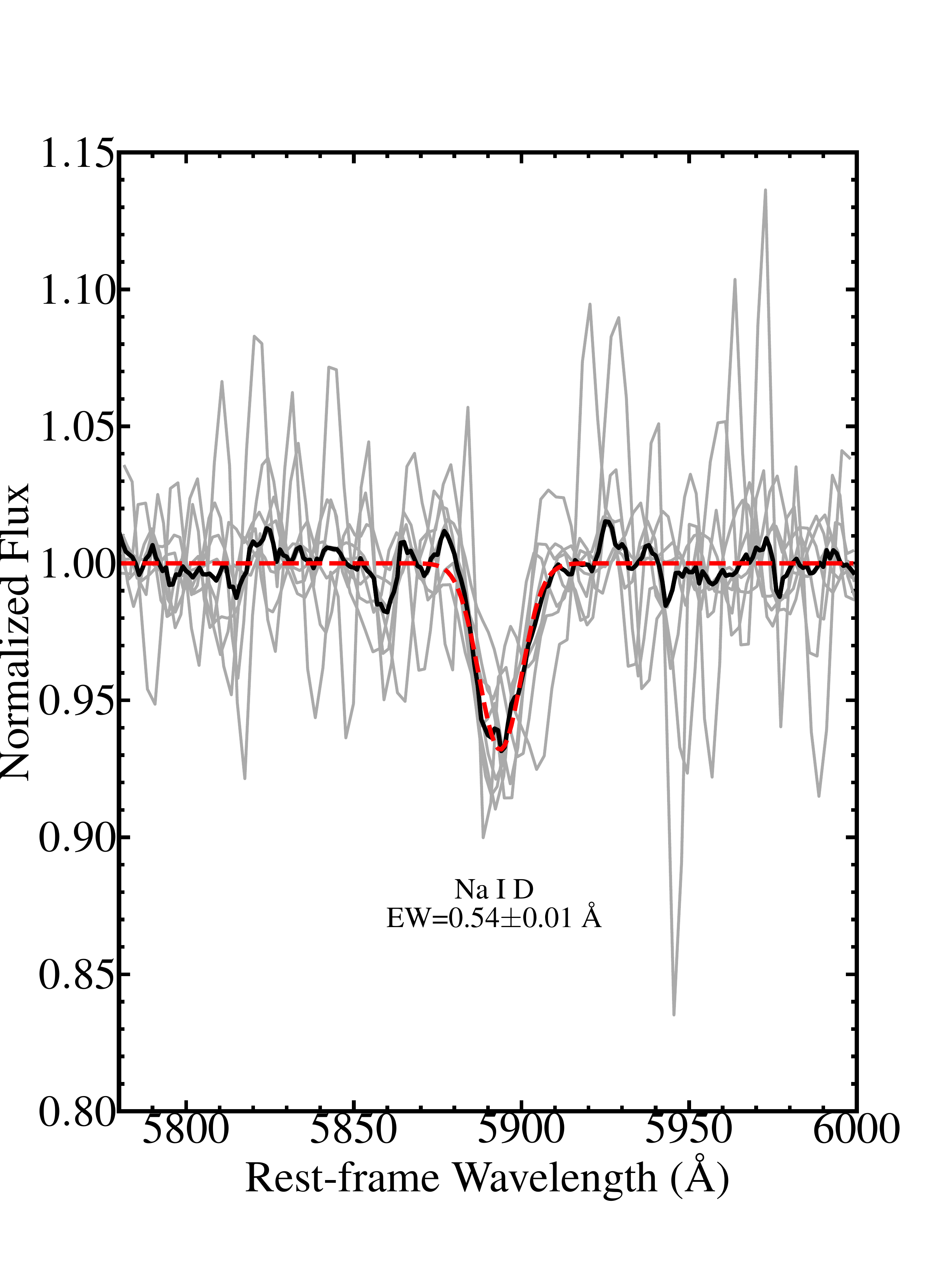}
    \caption{Our first eight epochs of spectra of SN\,2020jfo centered at the rest wavelength between Na\I~D1 and D2 (light grey).  We optimally stack all eight epochs (black) and fit a Gaussian profile to the Na\I~D feature (red) from which we derive an equivalent width of 0.54$\pm$0.01~\AA.}
    \label{fig:naid}
\end{figure}

We analyzed our first eight epochs of spectra of SN\,2020jfo from FLOYDS-N, FLOYDS-S, and Shane/KAST spectrographs around the rest wavelength of Na\I~D in M61 ($\approx$5922~\AA), which is dominated by continuum emission but with detectable absorption (\autoref{fig:naid}).  We normalized the continuum in our spectra using a third-order polynomial fit weighted by the signal-to-noise between 5700 and 6000~\AA.  We then combined these data accounting for the relative flux errors in each epoch and found that there is a clear signature of Na\I~D absorption with a Gaussian equivalent width of $0.54\pm0.01$~\AA\ for both the D1 and D2 components.  We note that this estimate is consistent with \citet{Sollerman21}, who find an equivalent width of $<$0.7~\AA\ but not of \citet{Teja22} where they estimate a Na\I~D equivalent width from the early-time spectra of 1.14$\pm$0.04~\AA.  In the latter, the authors use spectra from 11--15~days from explosion compared with $\approx$3--24~days in our analysis.  If the difference is physical and indicative of variable Na\I~D absorption in the spectra, this may be due to absorption from gas in CSM confined closely around the SN\,2020jfo progenitor star that was swept up and ionised by the SN itself.  However, in the absence of a higher resolution spectrum at early times, we rely on our Na\I~D absorption estimate and other extinction indicators to infer the total line-of-sight extinction.

Based on the empirical relation between Na\I~D column density and extinction presented in \citet{Phillips13}, this value corresponds to a line-of-sight extinction of $A_{V}=0.20\substack{+0.09\\-0.08}$~mag\footnote{The uncertainties here are dominated by the systematic uncertainty for the extinction relations in \citet{Phillips13}.}, which suggests that there is significant excess extinction and reddening beyond the Milky Way contribution.  We emphasize that this absorption feature is redshifted to the wavelength of Na\I~D in M61, and so our measurement constrains excess extinction in that galaxy beyond what we assume for the Milky Way.  

In the regime where the reddening parameter $R_{V}$ has a maximal value of 5.1 \citep[the maximum host value inferred for SNe in][]{Stritzinger18}, the corresponding extinction in {\it Swift}/UVOT UVW2 is $A_{\rm UVW2}=0.3$~mag compared with $A_{\rm UVW2}=0.7$~mag for $R_{V}=2.1$.  Thus if we assume a typical Milky Way value of $R_{V}=3.1$, the systematic uncertainty on host extinction is at most $\approx$0.2~mag and significantly less than this in most bands.

There is a wide diversity in SN\,II light curve properties even among SNe with similar peak magnitudes \citep{Valenti16,deJaeger18}, but comparing our SN\,2020jfo light curves to other SNe\,II with low extinction can provide a rough check for whether our extinction estimate is accurate.  Assuming $E(B-V)_{\rm host}=0.06$~mag and $R_{V}=3.1$, the peak $B$- and $V$-band absolute magnitudes of SN\,2020jfo are $M_{B}=-16.36$~mag and $M_{V}=-16.62$~mag, close to the average peak magnitudes of $\langle M_{B} \rangle =-16.39$~mag and $\langle M_{V} \rangle =-16.53$~mag for SNe\,II in \citet{deJaeger19}.  Our extinction-corrected light curves are roughly consistent with this peak magnitude and colour, validating our assumption of host extinction to SN\,2020jfo.

\subsection{Early-time Light Curve and Shock Cooling Models}\label{sec:shock}

We analyzed the SN\,2020jfo ultraviolet and optical light curves from within the first 16~days of observations using shock cooling models developed in \citet{Sapir17} and implemented in a {\tt python}-based Monte Carlo Markov Chain (MCMC) code and presented in \citet{Hosseinzadeh18}, \citet{Hosseinzadeh19} and \citet{Andrews20}\footnote{\url{github.com/griffin-h/lightcurve_fitting}}.  This analytical model assumes that emission within the first several days of explosion is dominated by a modified thermal spectrum that arises when the SN shock wave deposits energy into the progenitor star's envelope.  The model is parameterized by the time of explosion ($t_0$), the luminosity of the SN at 1~day post-explosion ($L_1$), the temperature at 1~day post-explosion ($T_1$), and the time from explosion at which the envelope becomes transparent ($t_{\rm tr}$).  We fit the observed photometry\footnote{Note that the {\tt lightcurve\_fitting} module assumes {\it Swift}/UVOT photometry are in Vega mag, and so before running the MCMC we took photometry from \autoref{tab:photometry} and applied the Vega-AB transformations from \url{swift.gsfc.nasa.gov/analysis/uvot_digest/zeropts.html}.} using an $n=1.5$ polytopic model for the SN\,2020jfo ejecta and assuming the Milky Way extinction, host galaxy extinction, and distance given above but with no additional sources of extinction.

\begin{figure}
    \centering
    \includegraphics[width=0.49\textwidth]{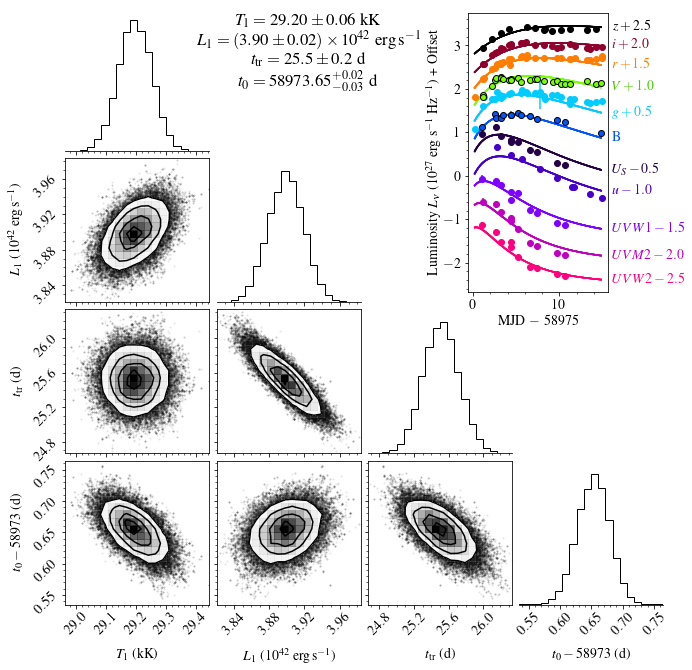}
    \caption{Corner plot and light curves showing the results of our analytical fit to the early-time (MJD=58975--58990) light curve of SN\,2020jfo in ultraviolet and optical bands as described in \autoref{sec:shock}.  We indicate the best-fitting parameters and 1-$\sigma$ uncertainties for each model parameter, $T_1$ (blackbody temperature at 1~day), $L_1$ (bolometric luminosity at 1~day), $t_{\rm tr}$ (time at which the outer envelope becomes transparent), and $t_0$ (the time of explosion) as described in \autoref{sec:shock}.  Note that the uncertainty on $L_1$ (luminosity at 1~day from explosion) does not include the uncertainty for distance modulus.}
    \label{fig:shock-cooling}
\end{figure}

The MCMC converged with best-fitting parameters $L_{1}=(3.90\pm0.02)\times10^{42}~\text{erg s}^{-1}$ and $T_{1}=29,200\pm60$~K as shown in \autoref{fig:shock-cooling}.  The transparency timescale is $t_{\rm tr}=25.5\pm0.2$~days from explosion, which is moderately longer but comparable to other SNe\,II to which this model has been applied such as SN\,2017gmr \citep[9.6~days;][]{Andrews20} and SN\,2016bkv \citep[17.2~days;][]{Hosseinzadeh18}.

The derived model parameters imply that SN\,2020jfo had a photospheric radius of $\approx1250~R_{\odot}$ at 1~day, and following scaling relations in \citet{Sapir17} and the formalism adopted in \citet{Hosseinzadeh18}, we can express the initial progenitor radius for an $n=1.5$ polytrope as

\begin{equation}
    R = 370~R_{\odot} \left(\frac{\kappa}{0.34~\text{cm}^{2}~\text{g}^{-1}}\right) \left(\frac{T_{1}}{25000~\mathrm{K}}\right)^{4.08}.
\end{equation}

\noindent We assume a standard opacity ($\kappa$) of 0.34~cm$^{2}$~g$^{-1}$ for a fully ionised hydrogen atmosphere, from which we derive an initial progenitor radius of $700\pm10~R_{\odot}$.  We emphasize that the uncertainties here are entirely due to the parameter uncertainties on $T_1$ from our MCMC and so do not include systematic uncertainties in the model or the opacity.  If the SN\,2020jfo progenitor star had this radius and a relatively cool photospheric temperature of $\approx$3300--3500~K \citep[comparable to the coolest RSGs in][]{Davies13,Smartt15}, this would imply the progenitor star had $\log(L/L_{\odot})=4.7$--4.8, while a warmer temperature would imply a significantly more luminous and massive star.  From single-star models derived in \citet{choi+16}, such a RSG would have $M_{\rm ZAMS}=$11--13~$M_{\odot}$.  Thus on the low-mass end, this analysis implies that the SN\,2020jfo progenitor star could not have been significantly less massive than 11~$M_{\odot}$.

Based on the \citet{Sapir17} prescription (see their equation 13), the SN envelope mass ($M_{\rm env}$) can be expressed in terms of the transparency timescale, opacity ($\kappa$), and shock velocity ($v_{s}$) as

\begin{multline} M_{\rm env} = 1.0~M_{\odot} \left(\frac{t_{\rm tr}}{19.5~\text{days}}\right)^{2}
 \left(\frac{\kappa}{0.34~\text{cm}^{2}~\text{g}^{-1}}\right)^{-1} \\
    \left(\frac{v_{s}}{3200~\text{km s}^{-1}}\right).
\end{multline}

\noindent This implies that the envelope mass of SN\,2020jfo is 1.7~$M_{\odot}$ assuming $\kappa=0.34$~cm$^{2}$~g$^{-1}$ and $v_{s}=3200$~km~s$^{-1}$, the latter of which is consistent with SNe\,II modeled in \citet{Matzner99}.  Although this value is dependent on the assumption of a shock velocity and opacity, it is quite low compared with the diversity of hydrogen-envelope masses derived for SN\,II in studies such as \citet{hillier+19} and \citet{Martinez22} and instead more typical of SN\,II-L \citep[e.g.,][]{Blinnikov93}.  We revisit this inconsistency in the context of the progenitor star in \autoref{sec:connecting}.

The implied explosion date for SN\,2020jfo (MJD $58973.65\substack{+0.03\\-0.02}$) is consistent with our latest Pan-STARRS1 non-detection at MJD 58970.31 with $m_{i}>21.92$~mag (3$\sigma$) as well as ZTF non-detections at MJD 58971.50 with $m_{r}>19.75$~mag and 58971.57 with $m_{g}>19.70$~mag (\autoref{tab:photometry}).  This suggests the SN was discovered only 38 rest-frame hours after explosion.  Thus, while the shock cooling model is a relatively good fit to our early time data until they no longer match the validity criterion \citep[photospheric temperature $kT<0.7$~eV as in][which occurs at $\approx$17.5 rest-frame days from explosion]{Sapir17}, the gap between explosion and discovery and the simplifying assumptions inherent in assuming a blackbody spectrum from UV to $z$-band suggests there may be some systematic uncertainties in the parameters derived above.  We compare our inference of the progenitor star and SN explosion properties from the early light curve to the rest of the light curve and our other observations below.

In \citet{Teja22}, the authors argue that the high peak luminosity and steep decline in the early-time light curve are indicative of CSM interaction with $\approx$0.2~$M_{\odot}$ of material confined within $40$~AU around the SN\,2020jfo progenitor star.  If the photosphere we see in our shock cooling analysis arises in the CSM rather than the shocked envelope of the progenitor star, this would imply that the progenitor star has a radius smaller than 700~$R_{\odot}$ and thus the star has a lower mass than 11--13~$R_{\odot}$.  We further discuss the implications of this point in Section~\ref{sec:connecting}.

\subsection{Bolometric Light Curve and Nickel Mass}\label{sec:bol}

Given the abundance of UV and optical photometry covering the majority of radiation from SN\,2020jfo, we can calculate the bolometric luminosity contemporaneous with our observations and over our observing bands.  We use the photometric analysis code {\tt superbol} \citep{Nicholl16,Nicholl18} to perform this analysis using our $g$-band light curve as a reference.  We interpolate the in-band light curves linearly in time and derive a bolometric correction by fitting a blackbody at each $g$-band epoch.

\begin{figure}
    \centering
    \includegraphics[width=0.49\textwidth]{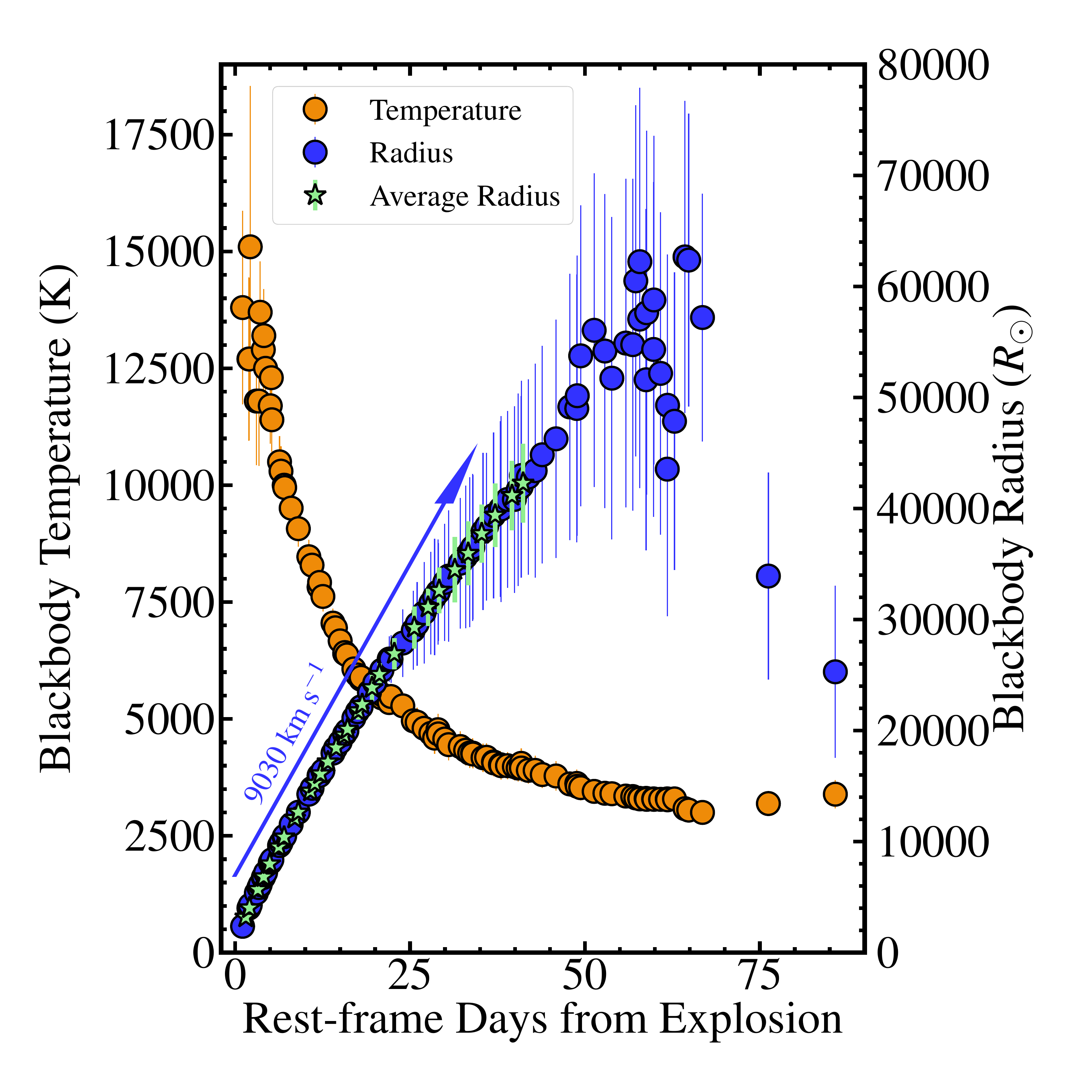}
    \includegraphics[width=0.40\textwidth]{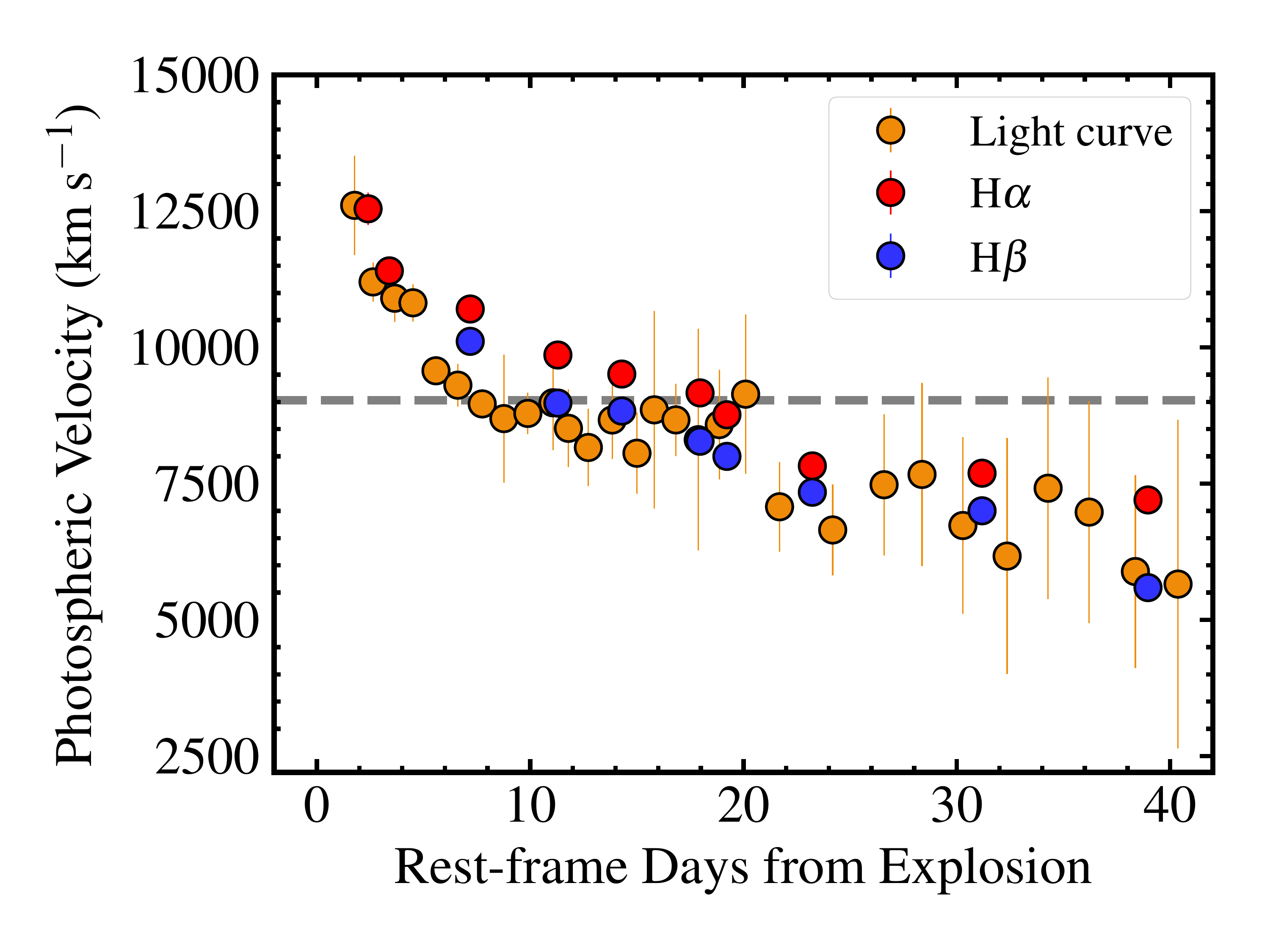}
    \caption{({\it Upper panel}): The derived blackbody temperature (orange circles; left axis) and radius (blue circles; right axis) based on our fits to the bolometric light curve of SN\,2020jfo at $<$90 rest-frame days from explosion as described in \autoref{sec:bol}.  We calculate the average radius (green stars) by averaging the blackbody radii with a 2~day top-hat window from 0--20 rest-frame days from explosion and a 4~day top-hat window from 20--40 rest-frame days from explosion.  The light curve exhibits a rapidly expanding blackbody radius, from which we derive an average photospheric velocity ($v_{\rm ph})$ of 9030~km~s$^{-1}$ at $<$40~days.  ({\it Lower panel}): The photospheric velocity of SN\,2020jfo estimated from the derivative of the radius in the upper panel compared with the velocity estimated from H$\alpha$ and H$\beta$ as described in \autoref{sec:spec-analysis}.  For comparison to the average, early-time velocity in the upper panel, we show a horizontal line at 9030~km~s$^{-1}$ (dashed gray line).}
    \label{fig:bolometric-parameters}
\end{figure}

We present the derived blackbody temperature and radius based on our bolometric fits and for rest-frame epochs at $<$90~days in \autoref{fig:bolometric-parameters}.  These data correspond to all epochs before SN\,2020jfo was close to the Sun and unobservable with ground-based optical telescopes starting in 2020 August and when the event was optically thick, as shown by the derived blackbody radius, which peaks around 60 rest-frame days from explosion.  Fitting to the derived blackbody radius at $<$40~days from explosion when the photospheric radius appears to increase linearly, we derive a photospheric velocity of 9030~km~s$^{-1}$ at early times as shown in \autoref{fig:bolometric-parameters} but with a declining velocity throughout this phase.  This is broadly consistent with the range of velocities derived for SN\,II in \citet{Bose14}.

We further estimate the photospheric velocity from our blackbody radii by averaging these radii within a 2~day top-hat window from 0--20 rest-frame days from explosion and a 4~day top-hat window from 20--40 rest-frame days from explosion.  These average radii are shown as green stars in \autoref{fig:bolometric-parameters}.  We convert these values into velocities by taking the first derivative along this sequence of average blackbody radii, which is shown in the bottom panel of \autoref{fig:bolometric-parameters} as the photospheric velocity inferred from our light curves.  Although these values have large uncertainties due to the distance and blackbody fitting, the velocity appears to decline as expected for SN\,II \citep[e.g., similar to those in][]{Bose14} from 12500--6000~km~s$^{-1}$ from 3--40 rest-frame days from explosion.

\begin{figure}
    \centering
    \includegraphics[width=0.47\textwidth]{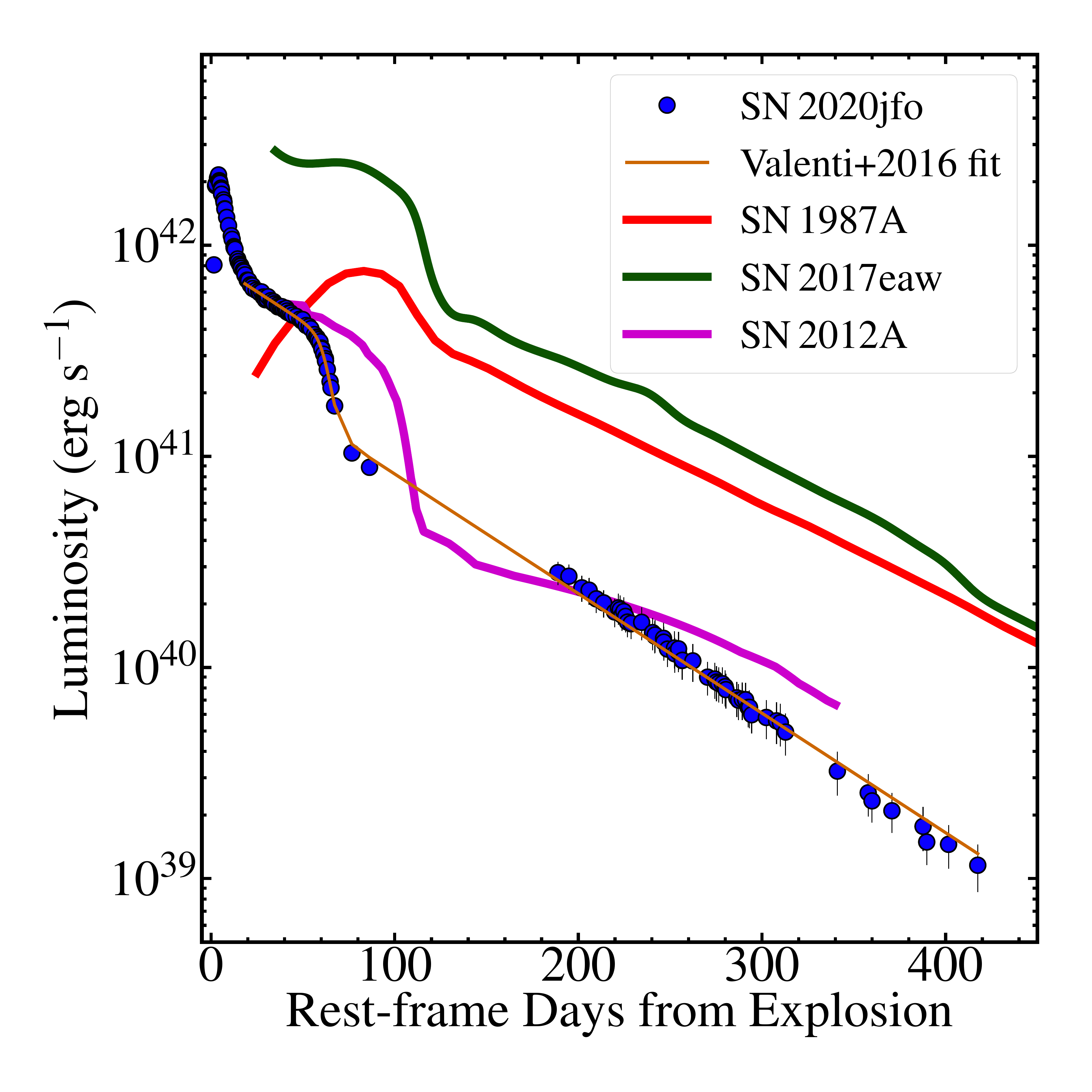}
    \caption{Bolometric luminosity from SN\,2020jfo (blue circles) as derived in \autoref{sec:bol}.  We compare to the parametric light curve model in \citet{Valenti16} (orange line) as well as bolometric light curves from SN\,1987A \citep[red;][]{Suntzeff90}, SN\,2012A \citep[magenta;][]{Tomasella13}, and SN\,2017eaw \citep[green;][]{Buta19}.}
    \label{fig:bolometric}
\end{figure}

The overall bolometric luminosity we derive for SN\,2020jfo is relatively low compared with other SNe\,II as shown in \autoref{fig:bolometric}, consistent with findings in \citet{Sollerman21}.  We also find an unusually short plateau time compared to the distribution of objects in \citet{Valenti16}, with $t_{\rm PT}=65.3\substack{+1.4\\-0.7}$~days following their formalism for characterising the SN\,II-P plateau time.  This plateau duration is within the lower 5th percentile for the sample of SN\,II studied by \citet{Valenti16} and points to a significant amount of stripping in the envelope of the SN\,2020jfo progenitor star.  However, the SN\,2020jfo progenitor star clearly retained a sufficiently massive hydrogen envelope to remain a SN\,II-P as opposed to a SN\,IIb.

We use this light curve at times $>$185 rest-frame days from explosion to estimate the total $^{56}$Ni mass produced in SN\,2020jfo following methods in \citet{hamuy+03} and \citet{Tinyanont21}.  Assuming that the emission at these times is entirely powered by the decay of $^{56}$Co$\rightarrow$$^{56}$Fe, we estimate the initial $^{56}$Ni mass ($M_{\rm Ni}$) as

\begin{equation}
    M_{\rm Ni} = \frac{L(t)}{\epsilon_{\rm Co}} \frac{\lambda_{\rm Co} - \lambda_{\rm Ni}}{\lambda_{\rm Ni}} \left( e^{-\lambda_{\rm Ni} t} - e^{-\lambda_{\rm Co} t} \right)^{-1}
\end{equation}\label{eqn:ni}

\noindent where $L(t)$ is the luminosity at a rest-frame time from explosion $t$, $\epsilon_{\rm Co}=6.8\times10^{9}$~erg~s$^{-1}$~g$^{-1}$ is the heating rate due to the decay of $^{56}$Co, and $\lambda_{\rm Co}=1/111.4$~days$^{-1}$ $\lambda_{\rm Ni}=1/8.8$~days$^{-1}$ are the radioactive decay timescales for $^{56}$Co and $^{56}$Ni, respectively.  Applying this equation to our light curve, we find that $M_{\rm Ni}=0.013\pm0.005~M_{\odot}$, which is low compared to previous estimates in \citet{Sollerman21} (0.024~$M_{\odot}$) and \citet{Teja22} (0.019~$M_{\odot}$). 

As noted by \citet{Sollerman21}, there may be incomplete trapping of $\gamma$-rays at later times based on the steeper slope of the SN\,2020jfo light curve ($\approx$1.3 mag / 100 days) compared with that theoretically expected for $^{56}$Co decay with complete trapping \citep[0.98 mag / 100 days given $\lambda_{\rm Co}=1/111.4$~days; see also][]{Woosley89}.  From our fit to the SN\,2020jfo light curve using the \citet{Valenti16} model, we find a similarly steep decline rate of 1.27$\pm$0.06~mag / (100~day).

We therefore consider the formalism in \citet{Sollerman98} to account for gamma-ray leakage using by modifying eqn.~\ref{eqn:ni} with an additional parameter $(1-0.965 \times \exp(-(t_1 / t)^{2})$, where $t_1$ is the timescale for the optical depth to gamma-rays to decrease to 1.  Similar to findings in \citet{Sollerman21}, we derive $t_1=170\pm22$~days and $M_{\rm Ni}=0.018\pm0.007~M_{\odot}$, in 1-$\sigma$ agreement with \citet{Sollerman21} and \citet{Teja22}.

Finally, we note that ${}^{56}$Ni has been used to empirically estimate the progenitor mass using a mapping derived from SNe\,II with direct progenitor star detections \citep[in][]{Eldridge19}.  This relation has significant scatter, as well as large systematic uncertainties on the initial masses of the progenitor, stars themselves, but in general they imply that SNe with a ${}^{56}$Ni mass $<$0.01~$M_{\odot}$ are likely from stars with initial masses $\lesssim$12~$M_{\odot}$.  Our estimate of $M_{\rm Ni}=0.018\pm0.007~M_{\odot}$ implies that the progenitor star was likely $>$12~$M_{\odot}$, although we acknowledge that progenitor star mass estimates from \citet{Davies18} would allow a lower mass star for this ${}^{56}$Ni mass.

\subsection{Spectroscopic Evolution of SN~2020\lowercase{jfo}}\label{sec:spec-analysis}

\begin{figure}
    \centering
    \includegraphics[width=0.49\textwidth]{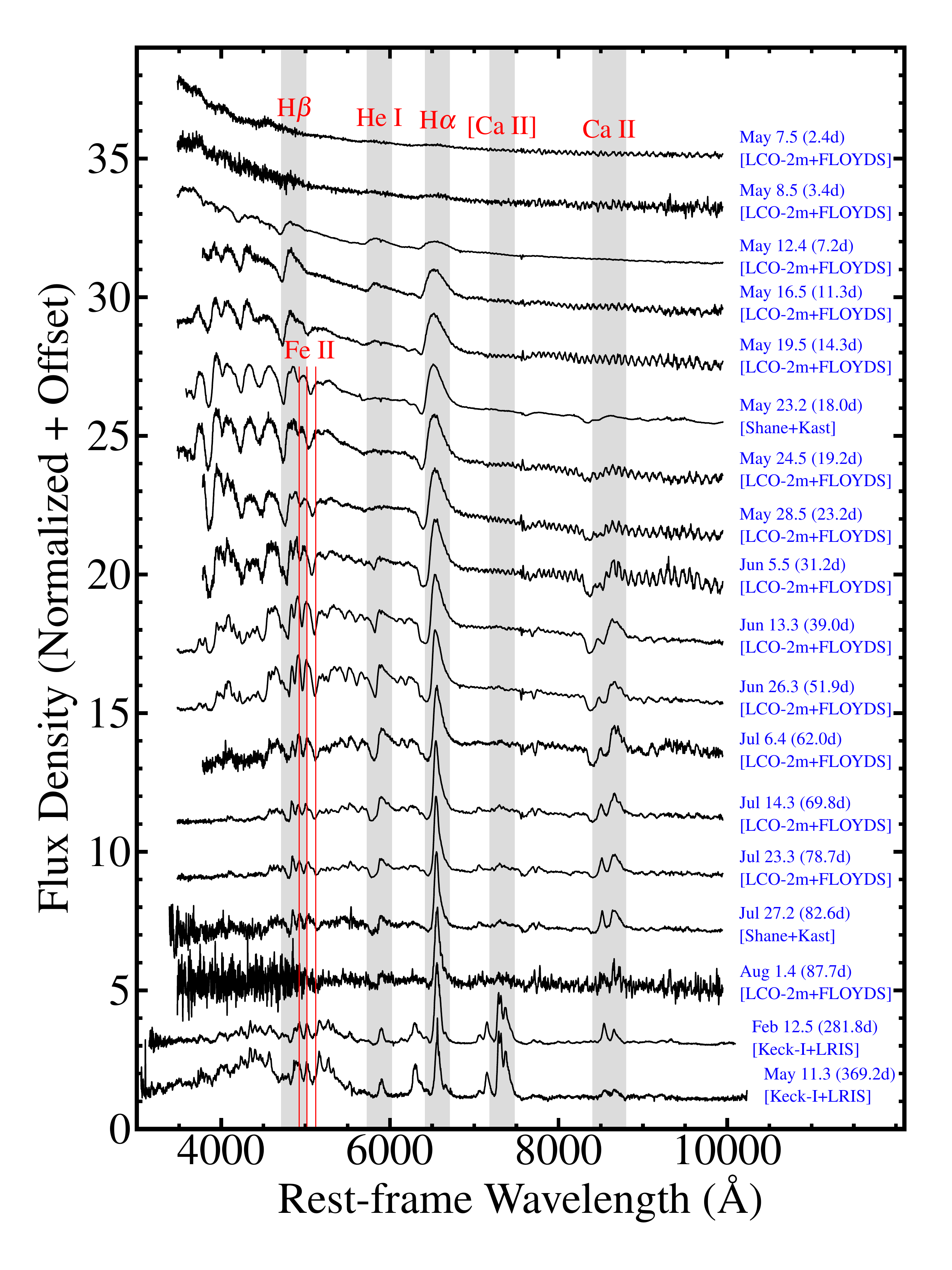}
    \caption{Our FLOYDS, Kast, and LRIS optical spectra of SN\,2020jfo from 2.4--369.2 rest-frame days after explosion.  We highlight the presence of broad Balmer, He\I, Fe\II, and Ca emission that begin to arise at $>$7.2 rest-frame days.  However, we note that in the earliest epochs, SN\,2020jfo does not exhibit signatures of narrow flash ionization features apparent in other SN\,II spectra at $<$3~days \citep[e.g., in][]{gal-yam+14,khazov+15,Tartaglia21}}
    \label{fig:spectra}
\end{figure}

The evolution of SN\,2020jfo from 2.4--370 days is typical for that of a SN\,II-P with relatively featureless blue continuum emission during the earliest phases followed by the emergence of broad, P-Cygni lines of hydrogen and helium (\autoref{fig:spectra}).  The overall continuum temperature rapidly cools from its initial temperature of $\approx$12,000~K, consistent with a shocked and expanding photosphere and our photometry around 2.4~days from the explosion (\autoref{fig:bolometric-parameters}).  The bluest wavelengths are dominated by metal lines in a pseudo-continuum, particularly those of Fe\II, which are notable in the spectra from 12.4--51.9~days after explosion.

We can independently measure the photospheric velocity of SN\,2020jfo using the absorption minima of high signal-to-noise, unblended lines in our spectra.  Following \citet{Szalai19} for SN\,2017eaw and \citet{Teja22} for SN\,2020jfo, we use H$\alpha$ and H$\beta$ for this purpose in order to compare with the photospheric velocity inferred from our light curve out to 40 rest-frame days from explosion.

At later times when the photosphere becomes optically thin, the Fe\II\ 5169~\AA\ line is typically used to track the photospheric velocity \citep[e.g., in][]{Nugent+06}.  However, this line is not detectable in our earliest spectroscopic epochs and the photosphere is more optically thick at these times.  We use the Balmer lines at these epochs to trace the photospheric velocity at $<$40 rest-frame days from explosion, although we note that the velocities may overestimate the velocity as traced by the optical light curve in the later epochs.

We measure H$\alpha$ velocities of 12000--5600~km~s$^{-1}$ for our spectra observed at these epochs (\autoref{fig:bolometric-parameters}), in good agreement with our estimate of the photospheric velocity from the light curve.  Similar to the expanding photosphere method for SN\,II \citep{Bose14}, this implies that our distance assumption above, which factors into the radius and thus velocity scale for our light curve, is self consistent for this event.

It is notable that, although our shock cooling model implies SN\,2020jfo was discovered within a few days of explosion, our earliest spectra do not show any signature of ``flash ionization'' or $<$100~km~s$^{-1}$, typically high ionization features observed in some SN\,II spectra \citep[e.g., SNe\,2013fs, 2016bkv, 2017ahn, 2020tlf, 2022pni, and others analyzed in][]{gal-yam+14,khazov+15,Hosseinzadeh18,Tartaglia21,Jacobson-Galan22,terreran+2022}.  These narrow features are not ubiquitous, even when it is clear that SNe\,II are observed within days of explosion \citep[other examples without strong flash ionisation signatures are SN\,2016X, 2017eaw, 2020fqv, and 2021yja in][]{Huang18,Szalai19,Tinyanont21,Hosseinzadeh22}.

Flash ionisation is characterized by narrow Balmer and He\II\ features as well as high-ionization lines such as those from C\III\ and N\III\ near 4640~\AA.  The lack of any such features in SN\,2020jfo points to $\lesssim$0.2~$M_{\odot}$ of CSM in its immediate environment.  Given that our light curves imply we initially detected SN\,2020jfo at 38~hours after explosion, it remains possible that there is a steep density gradient in the CSM and we missed an epoch where relatively weak flash ionisation line emission was produced in this material \citep[similar to findings for SN\,2020fqv in][]{Tinyanont21}.  In particular, the epoch of our first spectrum at 2.4~days can rule out dense CSM within $\approx$3000~$R_{\odot}$ (14~AU) assuming this observation probes the photosphere of ejecta sweeping outward at 10,000~km~s$^{-1}$.  This is likely only a few times the radius of the SN\,2020jfo progenitor star based on our shock cooling estimates above, and so we can rule out a relatively massive, compact shell of CSM \citep[similar to those observed in][]{Yaron17,Kilpatrick18:17eaw,Morozova18}.

\subsection{Nebular Spectroscopy of SN~2020jfo}\label{sec:nebular}

Evolved SNe\,II exhibit emission lines of oxygen and calcium ([O\I] at 5577, 6300, and 6364~\AA\ and [Ca\II] at 7292 and 7324~\AA) that correlate with the total ejecta mass from these atoms \citep[see][for a discussion of these features]{jerkstrand12}.  Using non-local thermodynamic equilibrium radiative transfer and nucleosynthetic yields for core-collapse SNe, synthetic spectra can be generated to estimate the total ejecta mass for SNe that explode from a massive star with a particular initial mass \citep[e.g.,][]{jerkstrand14,terreran2016,silverman17,bostroem19,Szalai19,hiramatsu21,Tinyanont21}.  Here we compare our spectra from $>$250~days post-explosion to nebular spectral models to estimate the initial mass of the SN\,2020jfo progenitor star.

Following methods in \citet{Tinyanont21}, we take high initial mass models from \citet{jerkstrand14} and low initial mass models from \citet{jerkstrand18} and scale them to the distance of SN\,2020jfo.  In addition, we scale the total flux by the ratio of the nickel mass inferred above for SN\,2020jfo to that assumed in each model.  Finally, we account for differences in the rest-frame epoch for each model and that of our SN\,2020jfo spectra by taking the closest model spectrum in time and scaling its overall flux by $\exp(-\delta t/111.3~\mathrm{days})$ where $\delta t$ is the difference between the phase of our spectrum and that of the model and 111.3~days is the decay constant for ${}^{56}$Co.

Because the fit between our spectra and the synthetic spectra from \citet{jerkstrand14} and \citet{jerkstrand18} is sensitive to the overall flux calibration of our spectra, we calibrate these data by performing synthetic photometry with {\tt pysynphot} in each spectrum and scaling to the flux implied by our light curves.  Both of our late-time LRIS spectra span a wavelength range covering $>$99\% of the flux in the Pan-STARRS $gri$ bands, and so we interpolate our light curves at $>$250~days to infer the SN flux in each band at 281.8 and 369.2 rest-frame days from explosion.  Scaling our spectra to match this in-band flux derived from {\tt pysynphot}, it is likely that both spectra have an absolute flux scale that is accurate to $\approx$10\%.

\begin{figure*}
    \centering
    \includegraphics[width=0.75\textwidth]{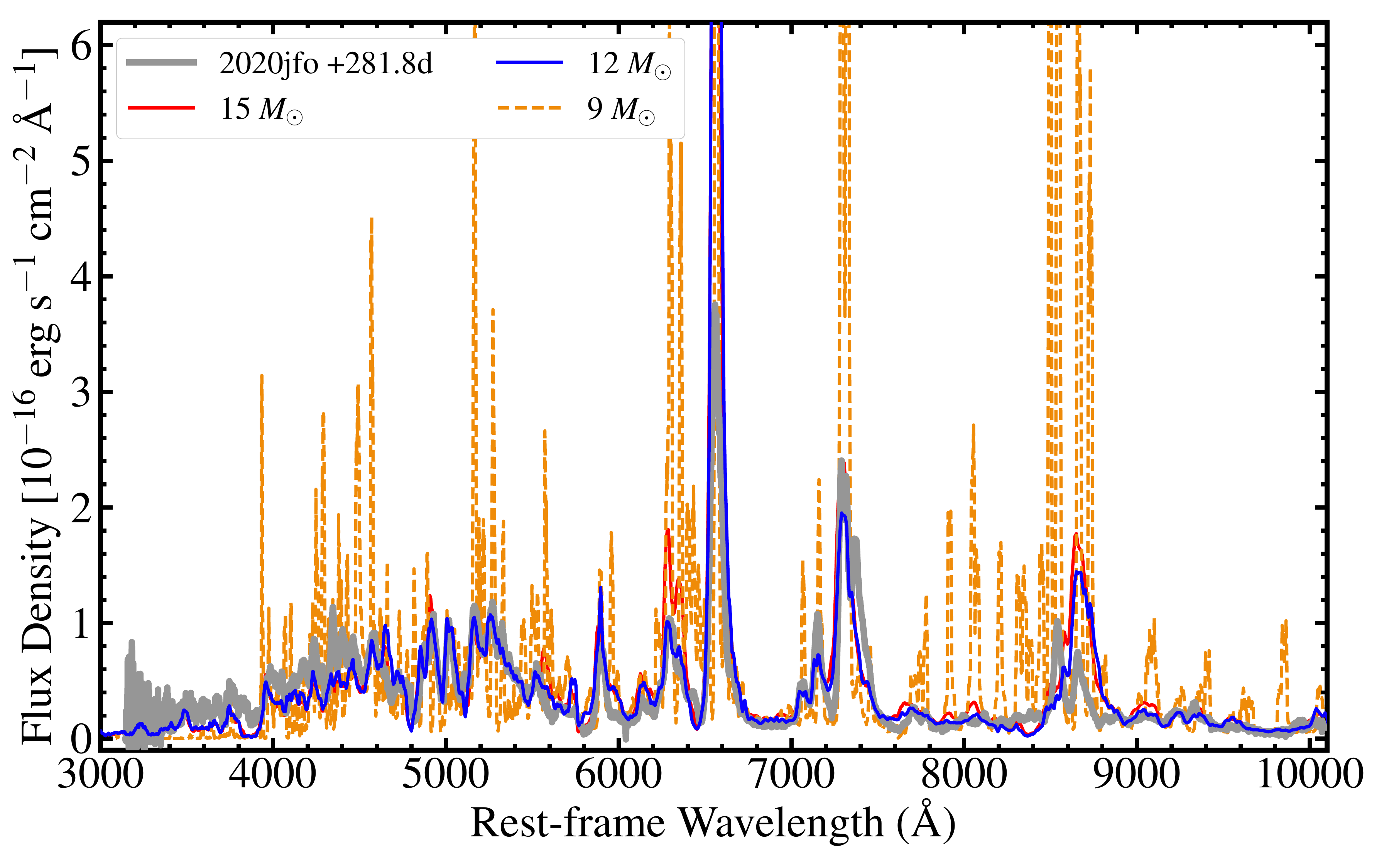}
    \includegraphics[width=0.75\textwidth]{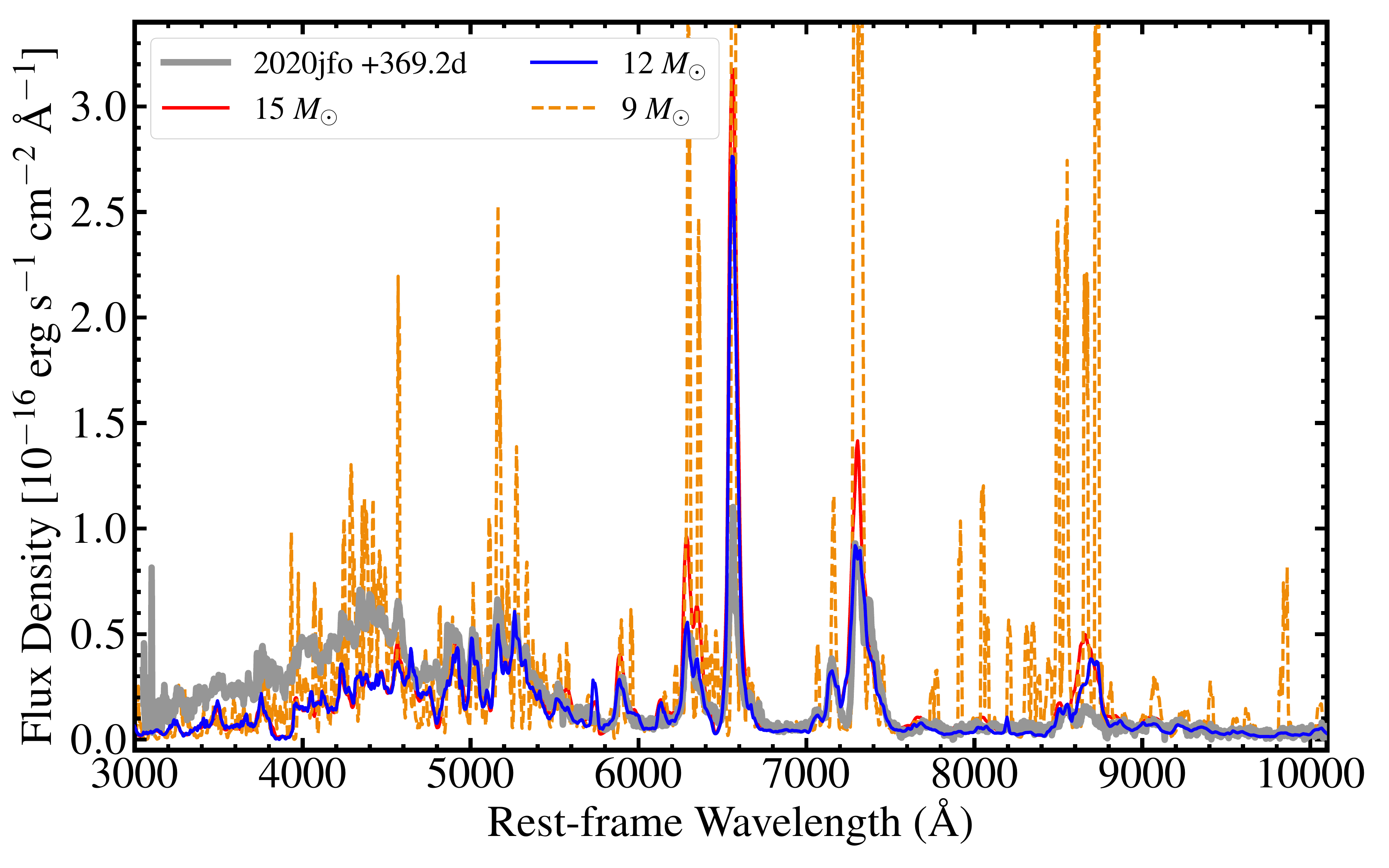}
    \caption{Our LRIS spectrum of SN 2020jfo (grey) from 2021 February 21 at +281.8 rest-frame days from explosion (upper panel) and 2021 May 11 at +369.2 rest-frame days from explosion (bottom panel). We compare to model nebular spectra for SN\,II from \citet{jerkstrand12} and \citet{jerkstrand18} for 9 (dashed orange), 12 (blue), and 15 (red) $M_{\odot}$ initial mass models scaled to the distance, nickel mass, and phase of our SN\,2020jfo spectrum.  Based on the fitting method described in \autoref{sec:nebular}, we find the best fit to the 12~$M_{\odot}$ spectrum.}
    \label{fig:nebular}
\end{figure*}

The flux-calibrated spectra are shown in \autoref{fig:nebular}.  The best-fitting scaled model in both cases corresponds to an initial mass of 12~$M_{\odot}$.  There is close agreement in the total flux from the [O\I] and  [Ca\II] features, albeit with significant divergence in the total flux from the Ca IR triplet near 8500~\AA, which is sensitive to the density and ionisation state of calcium and often is not reproduced in nebular spectral models \citep[see, e.g.,][]{jerkstrand12}.  Also notable is the fact that the model does not closely fit the Fe-dominated pseudo-continuum blueward of 5000~\AA, which is significantly brighter for SN\,2020jfo compared with the model spectrum.  This may indicate that there is a residual source of continuum emission in SN\,2020jfo \citep[such as late-time CSM interaction observed in otherwise normal SN\,II-P;][]{Weil20,Szalai21} and thus we are not seeing the entire ejecta profile.  However, in this scenario, the effect on the resulting line emission would be non-trivial.  Some ejecta may be obscured by a residual photosphere.  While our data imply an initial mass of $\approx$12~$M_{\odot}$, the true initial mass could be slightly higher (with a lower relative oxygen flux) or lower (with a higher relative oxygen flux).

It is notable that the 9~$M_{\odot}$ model implies a significantly higher [O\I] and [Ca\II] flux given our assumed distance and ${}^{56}$Ni mass.  There are several features in the model spectrum that are also be detectable at this epoch, such as [Ni\II] $\lambda$7378 noted by \citet{Sollerman21} and \citet{Teja22} and  [Fe\II] $\lambda$7155.  While the 12~$M_{\odot}$ spectrum is a relatively good match overall and a moderately higher oxygen flux could fit to the higher mass 15~$M_{\odot}$ spectrum, we emphasize that our spectra strongly disagree with the lower mass model.

\section{The Host Galaxy and Local Environment of SN~2020\lowercase{jfo}}\label{sec:environment}

M61 is a barred spiral galaxy that is seen almost face on, and this is an ideal configuration for the study of the physical properties of the stellar population and the gas distribution with MUSE. M61 is a prolific producer of SNe, given that SN\,2020jfo is the eighth SN observed in almost 100 years \citep[e.g.,][]{Humason62,Li07,Roy11,Foley14}, with the first event recorded being SN 1926I. Being in the Virgo cluster at 14.5~Mpc, M61 has been the subject of several observational campaigns, mainly at optical, infrared and radio frequencies \citep{Pastorini2007,Momose2010,Magrini2011,Yajima2019}. We focus on the immediate environment of SN\,2020jfo in the outer western arm of M61 as observed by MUSE. 

We analysed the entire datacube using a standard procedure that proceeds for several steps \citep[see, e.g. ][]{Gagliano2022}: after correcting the entire datacube for the Galactic extinction value ($E(B-V)=0.019$~mag), and using a \citet{cardelli+89} extinction law, we have implemented a Voronoi spatial binning over the entire datacube, using the prescriptions given in \citet{Cappellari2003}. This method provides a final datacube characterised by spatial bins with a similar signal-to-noise value. Given the large data volume of the MUSE datacube, for the analysis of the entire host galaxy, we have first considered a S/N value of 100, estimated in a spectral range free from emission or absorption line features. 

To study the stellar population properties, we have applied the stellar population synthesis code \texttt{STARLIGHT} \citep{CidFernandes2005} to fit the spectrum at each spatial bin. The fitting procedure consists in searching for the best combination of template stellar spectra that best fits the observed spectrum at each bin. The base spectral library consists of 25 stellar spectra that was generated using a \citet{Chabrier2003} initial mass function with ages varying between 10$^6$ yrs and 1.8 $\times$ 10$^{10}$ yrs, for six different metallicity values spanning from $Z = 10^{-4}$ to $Z = 0.05$ (with the Solar value assumed to be $Z_{\odot}$ = 0.02). The best-fitting combination also allows us to reconstruct the star-formation history at each spatial bin, and then in the entire host galaxy. We consider the mass-weighted age of all stars with initial mass $>$7~$M_{\odot}$ in this population in order to infer the most likely main sequence turnoff age for the SN\,2020jfo progenitor and find that it is 25$\substack{+14\\-6}$~Myr.  This corresponds to a star with an initial mass of 9.7$\substack{+2.5\\-1.3}$~$M_{\odot}$, implying a relatively low-mass progenitor star.

\begin{figure}
    \centering
    \includegraphics[width=0.48\textwidth]{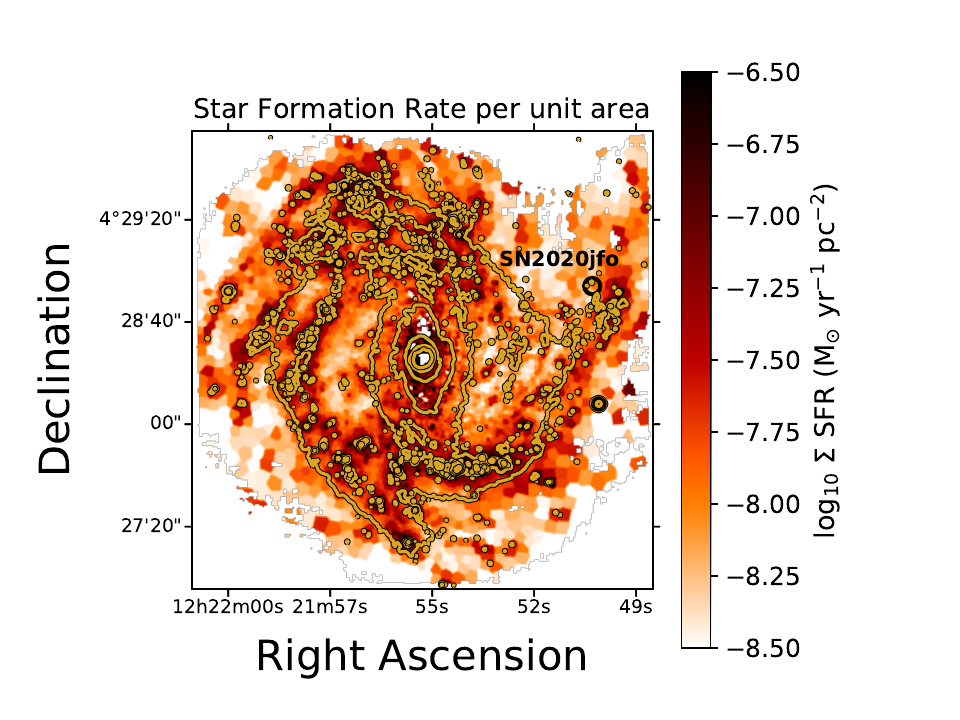}
    \includegraphics[width=0.48\textwidth]{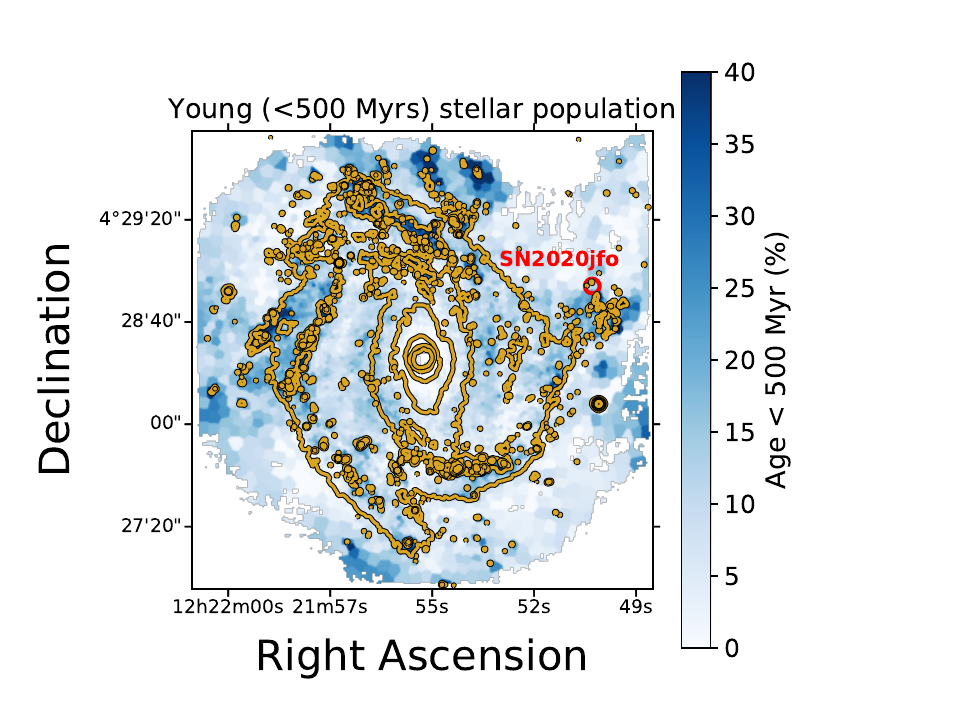}
    \includegraphics[width=0.48\textwidth]{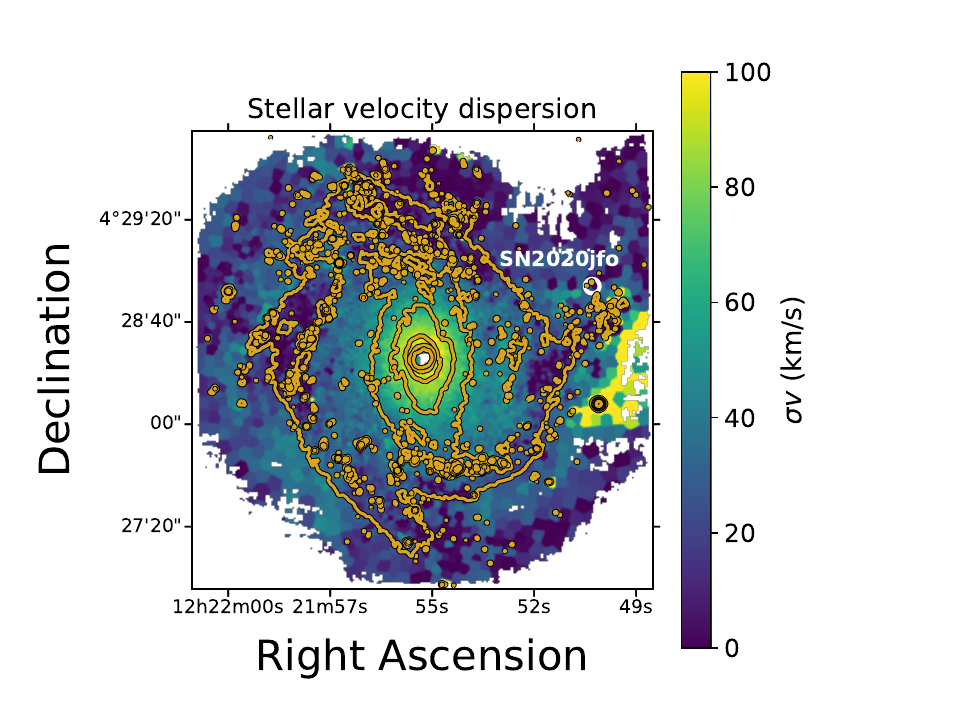}
    \caption{\textit{(Top)} The star-formation rate per unit area as estimated from the MUSE data analysis, using the \citet{Kennicutt} method based on the luminosity of the H$\alpha$ line. \textit{(Middle)} The distribution of the young ($<$ 500 Myr) stellar population fraction in M61. Regions with a large percentage of young stars are observed in the close vicinity to regions with high SFR values. \textit{(Bottom)} The stellar velocity dispersion as inferred from the analysis with \texttt{STARLIGHT}. The location of SN\,2020jfo is reported with a colored circle.}
    \label{fig:MUSE}
\end{figure}

We have studied the properties of the gas using emission lines in the residual spectrum resulting from the subtraction of the \texttt{STARLIGHT} best-fit from the observed spectrum.  The analysis of MUSE data demonstrates that the star-formation rate (SFR) is more intense along the bright \ion{H}{II} regions distributed along the spiral arms: we have measured the SFR using the de-reddened H$\alpha$ flux and the \citet{Kennicutt} method. These are also the regions where the younger stars in the galaxy are more concentrated, as it is clearly shown in \autoref{fig:MUSE} where the distribution of stars with age less than 5$\times$10$^8$ years, resulting from \texttt{STARLIGHT} is reported. SN\,2020jfo is located 5.0\arcsec\ in projection ($\approx$300~pc) from the most active star-forming region in this arm of M61.

The radial velocity distribution from our stellar population analysis shows the typical ``spider'' diagram observed for spiral galaxies. The velocity dispersion $\sigma_{v*}$ map shows an excess value at the center of the galaxy, due to the presence of a massive black hole \cite{Pastorini2007}, but interestingly, we note an excess of $\sigma_{v*}$ at the border of the galaxy western arm \citep{Yajima2019}, which is the arm where SN 2020jfo has been observed in, see \autoref{fig:MUSE}. A possible origin of this excess can be found in a current inflow of gas from a nearby companion; M61 has been proposed to be in interaction with two nearby smaller galaxies, namely NGC\,4292 and NGC\,4303A \citep{Binggeli1985}. Moreover, the detection of high velocity clouds in the spectra of the background QSO Q1219+047 \citep{Bowen1996}, located in the same direction of the dispersion velocity excess, led to the possible conclusion that an on-going interaction between these galaxies is currently happening.

\begin{figure}
    \centering
    \includegraphics[width=0.49\textwidth]{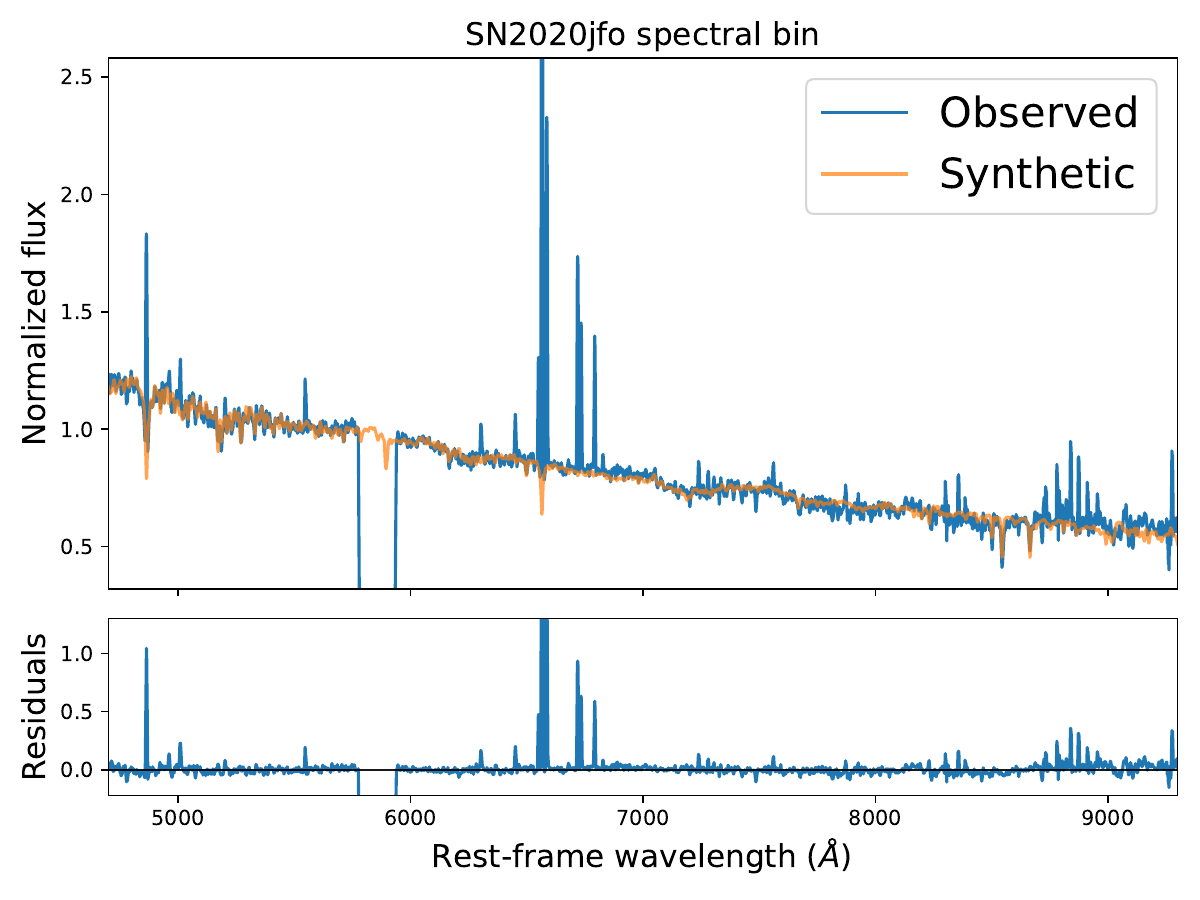}
    \includegraphics[width=0.49\textwidth]{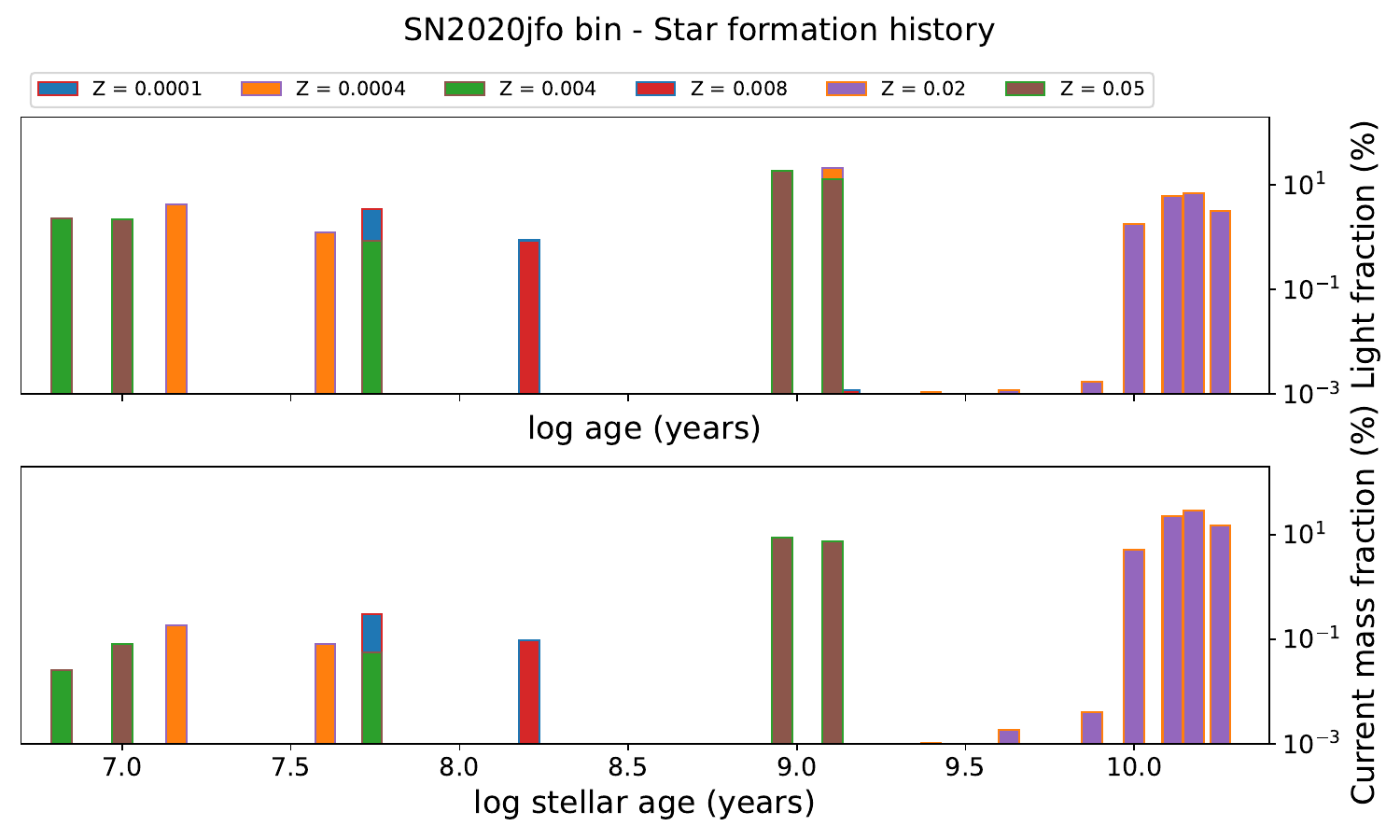}
    \caption{\textit{(Top panel)} The MUSE spectrum (blue curve) at the position of SN 2020jfo, obtained using the prescriptions described in the main text. The orange curve is the resulting best-fit spectrum obtained with \texttt{STARLIGHT} with the difference between the two shown below. \textit{(Bottom panel)} The star-formation history at the location of SN 2020jfo in terms of light fraction and mass-fraction, and for different stellar ages, of the stellar templates used during the \texttt{STARLIGHT} analysis.  We divide each quantity into the six metallicity bins ($Z=0.0001-0.05$) used in our analysis.}
    \label{fig:SL}
\end{figure}

We then proceed to the extraction of the spectrum at the position corresponding to SN\,2020jfo's location.  We emphasize that this data cube was obtained before the explosion of SN\,2020jfo and so does not contain any SN flux, only flux from the stellar population and interstellar gas around those stars in M61.  The size of the spatial bin used for the spectrum extraction must take into account possible effects of the PSF. Using the code described in \citet{Fusco2020}, we reconstruct the atmospheric conditions at the epochs observed, deriving a PSF FWHM of 0.53 arcsec, estimated at 7000\AA. Then, we extracted a spectrum from the MUSE datacube at the location of SN\,2020jfo and with an extraction radius of 0.5 arcsec (2.5 spaxel). We fit the spectrum with \texttt{STARLIGHT} and then we have reconstructed the star-formation history from the  best-fit result, assuming the spectral library described above. The results are shown in \autoref{fig:SL}. The results show a heterogeneous distribution in ages and metallicity at the position of SN 2020jfo. This points to a continuous star-formation activity in the last few Gyrs (we measure a light-weighted age of 9.52 $\times$ 10$^{8}$~yrs), but this cannot exclude that multiple consecutive episodes of SF bursts happened surrounding the location of SN\,2020jfo.

We note in Section~\ref{sec:alignment} that there is significant crowding of stars around the location of SN\,2020jfo in the {\it HST} imaging, which are unresolved by the MUSE data.  The ages and metallicities of these stars may bias our estimate of the properties of the SN\,2020jfo progenitor star if they arise from other stellar populations unassociated with that star.  While we consider all possible stellar ages to infer the main sequence turnoff mass for the SN\,2020jfo progenitor star, we also note that the gradients of stellar ages in spiral galaxies occur over much larger size scales than the MUSE PSF size \citep[0.53 arcsec or $\approx$40~pc at the distance of M61 compared with hundreds of pc in, e.g.,][]{Bell00,Tissera16}.  Thus even if emission in the MUSE data are not dominated by the SN\,2020jfo progenitor star at that spaxel, it is likely that the other stars are representative of the stellar population that gave rise to that system.

Finally, we have used emission line indicators to measure the gas metallicity at the position of SN\,2020jfo. We used the empirical equations derived in \citet{Marino2013} based on H$\alpha$, H$\beta$, [N II] and [O III] line flux to derive the N2 and O3N2 indicators, widely used in literature to study the metallicity conditions of SN environments \citep{Galbany2016}. We find 12 + log(O/H) = 8.55 $\pm$ 0.01 and 12 + log(O/H) = 8.56 $\pm$ 0.03, for the N2 and O3N2 respectively (we assume a Solar value of 12 + log(O/H) = 8.69 \citep{Asplund2009}). These values are also consistent with the template mass-weighted metallicity value ($\log(Z/Z_{\odot}) = -0.146$) inferred from the \texttt{STARLIGHT} results. The SFR value at the position of SN\,2020jfo, inferred from H$\alpha$ line, is SFR=(4.18 $\pm$ 0.11) $\times$ 10$^{-7}$ M$_{\odot}$~yr$^{-1}$~pc$^{-2}$, comparable to values inferred for SN\,II in \citet{galbany+16}.

\section{The Progenitor Star of SN~2020\lowercase{jfo}}\label{sec:progenitor}

\subsection{Aligning Pre- and Post-Explosion Imaging}\label{sec:alignment}

We aligned our Keck/OSIRIS image to the \hst\ frames using common alignment stars to pre- and post-explosion frames.  Using 22--30 common sources in both sets of imaging, we calculated an alignment solution from OSIRIS$\rightarrow$\hst, achieving 0.02--0.03\arcsec\ root-mean-square alignment precision.

Taking the location of SN\,2020jfo in the OSIRIS image, we found that this position corresponds to that of a single point source in each of our F814W frames, which is shown in \autoref{fig:astrometry}.  SN\,2020jfo and the pre-explosion counterpart are nominally separated by 0.005\arcsec, or 24\% of the alignment uncertainty between the two images.  We therefore consider this to be a credible pre-explosion counterpart to SN\,2020jfo.  This source is characterized as a point source by {\tt dolphot} with ${\rm sharpness} = -0.007$ and ${\rm crowding} = 0.689$.  Thus while the source is in a relatively crowded field, it appears consistent with a bright point source.  We give the photometry across each F814W detection in \autoref{tab:hst}, but we do not detect the source in any other band.  In these latter cases, we instead derive upper limits on the presence of a point source using the {\tt dolphot} artificial star tool as described in \citet{Kilpatrick18:17ejb,Kilpatrick21b}.  Finally, we further note that this counterpart and our F814W detections agree with the findings of \citet{Sollerman21}.

We perform a similar analysis on the stacked {\it Spitzer}/IRAC data to constrain the location of SN\,2020jfo in our pre-explosion mid-infrared imaging.  Identifying 17--24 common sources in the OSIRIS and IRAC imaging, we align OSIRIS$\rightarrow$IRAC with 0.04--0.06\arcsec\ root-mean-square precision.  This location corresponds to a single, marginally detected (3--5$\sigma$) source in IRAC Channels 1 and 2 based on our {\tt photpipe} photometry, but we do not detect a counterpart above the background level in Channels 3 and 4.  We note that in Channels 1 and 2, the nominal separation between the position of SN\,2020jfo is 0.014\arcsec\ and 0.016\arcsec, respectively, which corresponds to 31\% and 40\% the combined astrometric uncertainty for alignment and the centroid of the IRAC counterparts, which has significant astrometric uncertainty given its low detection threshold.  This source appears to be point-like based on the {\tt DoPhot} detection, but it is not detected with high enough significance to be classified as a bright point source \citep[i.e., {\tt DoPhot} object type 1 in][]{schechter+93}.  We provide our photometry for the Channels 1 and 2 detections in \autoref{tab:spitzer}, and our upper limits for Channels 3 and 4 correspond to the approximate 3$\sigma$ limiting magnitude for detections within 30\arcsec\ of SN\,2020jfo.

Based on the crowding in the {\it HST} imaging and the faint nature of the {\it Spitzer} counterpart, we consider that this source may not be dominated by flux from the progenitor star of SN\,2020jfo.  In this case, the {\it Spitzer} Channel 1 and 2 fluxes would be upper limits on the nature of any pre-explosion counterpart.  Below we take the {\it Spitzer} photometry to be the true flux the F814W counterpart in the IRAC bands.  However, if these detections are upper limits instead, then for a RSG-like counterpart with an effective temperature $T_{\rm eff}<4500$~K, the bolometric luminosity of the F814W counterpart would be even lower than we model below.

The field around SN\,2020jfo and in the outer spiral arm of M61 (\autoref{fig:astrometry}) appears relatively crowded, and the {\tt dolphot} crowding parameter is also high.  Thus the likelihood of a chance coincidence may be non-negligible.  We estimate this likelihood by considering the number of point sources found by {\tt dolphot} in the F814W image and within 10\arcsec\ of the nominal counterpart at $>$3$\sigma$ significance is 2157.  Thus, at most 2.5\% of the solid angle within 10\arcsec\ of the counterpart is within 3$\sigma$ of a point source, a conservative upper limit on the probability of chance coincidence.  A similar analysis for {\it Spitzer}/IRAC Channel 1 yields 104 sources detected at $>$3$\sigma$ within 30\arcsec\ of SN\,2020jfo.  Using the correspondingly larger astrometric precision, we estimate an upper limit of 0.6\% that SN\,2020jfo aligns with this counterpart by chance in the IRAC imaging. It therefore remains possible that the counterpart aligns with SN\,2020jfo by chance.  Validating that it is the counterpart by its disappearance \citep[as in, e.g.,][]{Maund09} will solidify our association between SN\,2020jfo and this counterpart, but in the analysis below we assume this source represents the true progenitor system.  We further note that our analysis of the progenitor star identification and its photometry are in agreement with \citet{Sollerman21}, who used an independently obtained high-resolution AO image to associate SN\,2020jfo with its candidate progenitor star.

\subsection{The Spectral Energy Distribution of the SN~2020jfo Progenitor System}\label{sec:sed}

We consider the photometry of the SN\,2020jfo progenitor candidate in the context of model spectral energy distributions (SEDs) following methods in \citet{Kilpatrick21}.  The {\it HST} detection was previously analyzed by \citet{Sollerman21}, who derived $M_{\rm F814W}=-5.4$~AB~mag with non-detections at bluer wavelengths.  The authors acknowledge that this is $>$1~mag fainter than expected for stars in the 10--15~$M_{\odot}$ range that are consistent with their nebular spectra.  Here we explore a detailed SED model that can account for this inconsistency while agreeing with the {\it HST} and {\it Spitzer} detections.

These models include a blackbody, stellar SEDs from \citet{pickles+98}, and a dust-obscured RSG model originally presented in \citet{Kilpatrick18:17eaw}.  We fit all models to the joint set of photometry in \autoref{tab:hst} and \autoref{tab:spitzer}, including all individual detections in F814W and limits.  We use a Monte Carlo Markov Chain (MCMC) approach using the {\tt python}-based package {\tt emcee} to forward model each SED using the appropriate {\it HST} and {\it Spitzer} filter functions as given in {\tt pysynphot}, the distance and Milky Way extinction to M61 as given above, and the host extinction we derive from Na\I~D in \autoref{sec:extinction} and assuming $R_{V}=3.1$.  We then calculate a log-likelihood for {\tt emcee} by minimizing the value of $\chi^{2} = \sum_{i} (m_{{\rm obs}, i} - m_{{\rm model}, i})^{2}/\sigma_{i}^{2}$ where $m_{{\rm obs}, i}$ and $m_{{\rm model}, i}$ are the observed and forward-modeled magnitudes for each progenitor candidate observation $i$ and $\sigma_{i}$ is the corresponding observed magnitude uncertainty.  We handle our limiting magnitudes by setting the value of $\chi^{2}\rightarrow\infty$ when the forward-modeled magnitude for any observation is less than the observed magnitude.

\begin{figure}
    \centering
    \includegraphics[width=0.49\textwidth]{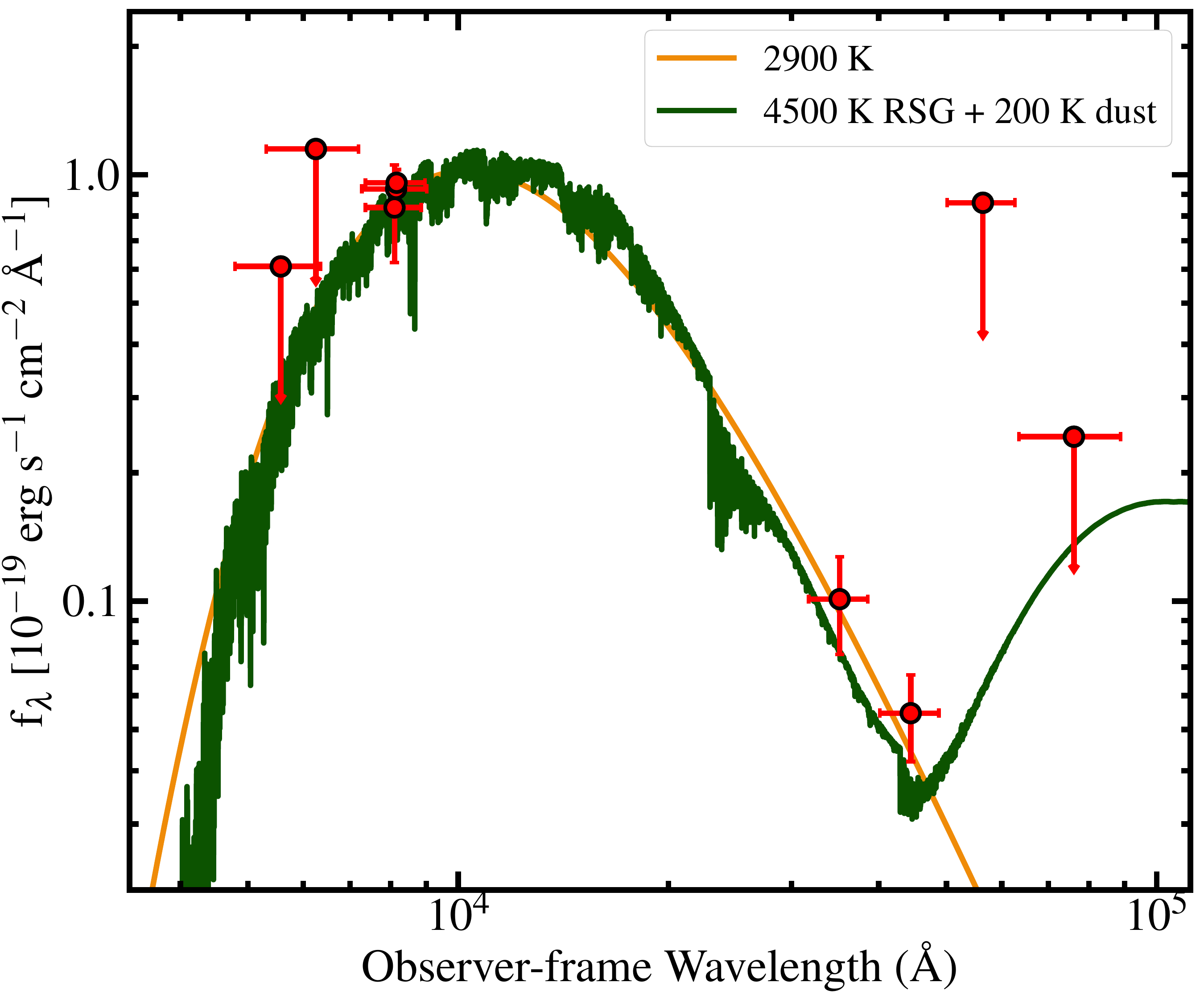}
    \caption{Spectral energy distribution of the pre-explosion counterpart to SN\,2020jfo (red circles; described in \autoref{sec:sed}).  We fit the {\it HST} and {\it Spitzer} photometry with a 2900~K blackbody (orange line), which describes the data but is much cooler than typical effective temperatures for the RSG progenitor stars of SN\,II \citep[e.g., in][]{Smartt15}.  We also show a RSG spectral energy distribution with a mid-infrared excess due to an 200~K extended dust shell \citep[green; from][]{Kilpatrick18:17eaw}.}
    \label{fig:sed}
\end{figure}

Following these methods, we find that the SN\,2020jfo progenitor photometry is consistent with a blackbody with $\log(L/L_{\odot})=3.9\pm0.3$ and $T_{\rm eff}=2890\pm60$~K as shown in \autoref{fig:sed}.  Assuming a star with $\log(L/L_{\odot})$ exploded as a RSG and comparing to MESA Isochrone \& Stellar Tracks (MIST) models from \citet{choi+16} with Solar metallicity and rotating at $v/v_{\rm crit}=0.4$, we find that such a star would have an initial mass of $\approx$5~$M_{\odot}$ and thus well below the mass range where stars are thought to explode as SNe\,II.  This tension can be partly relieved by adopting a much farther distance \citep[e.g., 20.7~Mpc to M61 as in]{Rodriguez14}.  However, even this farther distance places SN\,2020jfo in the range of terminal RSGs with initial masses 7--8~$M_{\odot}$ and a terminal luminosity of $\log(L/L_{\odot})=4.4$.  We also note that the distance-independent blackbody temperature is significantly cooler the values derived for all other SN\,II progenitor stars \citep[e.g., in][]{smartt+15}, even accounting for the fact that RSGs likely evolve to later spectral types as they approach core collapse \citep[e.g., M6--M8 with $T_{\rm eff}\approx$3100--3200~K;][]{Davies13,Negueruela13,Davies18}.

An alternative explanation for this unusually low luminosity and cool temperature is that the star is obscured by a significant amount of extinction, such as from a compact shell of dust or interstellar material in M61.  Based on MIST stellar evolution tracks for a single star with initial mass 12~$M_{\odot}$, its terminal RSG would have $M_{\rm F814W}=-6.8$~AB~mag, whereas the average F814W magnitude of the progenitor counterpart is 25.7~AB~mag or -5.2~AB~mag correcting for distance and the extinction assumed above, implying an excess of $A_{\rm F814W}=1.6$~mag extinction ($A_{V}=2.8$~mag for $R_{V}=3.1$) assuming the progenitor star has this much brighter intrinsic luminosity.

If the excess extinction is due to circumstellar matter, then emission arising from interaction between this shell and the SN\,2020jfo shock would not necessarily be visible in early-time photometry and spectra, especially if the shell were confined to within $\approx10$~AU (2000~$R_{\odot}$) of the progenitor star where the bulk of the ejecta were at the time of discovery.  On the other hand, if this material were at a much larger separation from SN\,2020jfo, then the SN would appear intrinsically reddened long after explosion, which appears inconsistent with the relatively weak Na\I~D features in its spectra \autoref{sec:extinction}, its extremely blue colours in early light curves and near peak light \autoref{sec:shock}, and the strong blue continuum in its early spectra \autoref{sec:spec-analysis}.  Thus any material providing this extinction would need to be overrun at early times so that it does not provide significant extinction that would be observable in the later data.

Motivated by this hypothesis, we apply a dust-obscured RSG model described in \citet{Kilpatrick18:17eaw} to our photometry.  This model uses realistic MARCS RSG photospheres surrounded by a circumstellar dust shell to infer the intrinsic stellar properties from photometry.  The model is parameterized by the intrinsic stellar luminosity, intrinsic RSG photospheric temperature, $V$-band optical depth due to extinction in the dust shell, and effective dust blackbody temperature.  The MARCS models correspond to RSGs and yellow supergiants with $T_{\rm eff}=2600$--$8000$~K and surface gravities $\log g = -0.5$--1.0 in steps of 0.5.  In addition, the dust types and shell geometries are parameterized by silicate and carbonaceous dust grains with inner radii ($R_{\rm inner}$) to outer radii ($R_{\rm outer}$) ratios of $R_{\rm outer}/R_{\rm inner}=2$ and 10 as in \citet{Kochanek12}.  We do not directly fit for surface gravity, dust type, or geometry, but instead we consider all combinations of each parameter and choose the best-fitting model \citep[as in][]{Kilpatrick18:17eaw}.

We find the best fits with the surface gravity $\log g=0.5$, $R_{\rm outer}/R_{\rm inner}=2$, and silicate dust model.  The best-fitting parameters imply a progenitor star with $\log(L/L_{\odot})=4.1\pm0.4$ and $T_{\rm eff}=4700\substack{+1800\\-900}$~K.  The resulting dust shell has $\tau_{V}=1.9\substack{+1.9\\-1.1}$ (corresponding to $A_{V}=1.5\substack{+1.5\\-0.9}$) and $T_{d}=1200\substack{+500\\-800}$~K, implying a shell with a photospheric radius of $R_{d}=1200\substack{+1800\\-100}~R_{\odot}$, a mass of $2.8\substack{+2.8\\-1.2}\times10^{-6}~M_{\odot}$ and a pre-explosion mass-loss rate of $\dot{M}=5.1\substack{+4.9\\-2.4}\times10^{-7}~M_{\odot}~\text{yr}^{-1}$, where we have assumed that the progenitor star's wind speed is 10~km~s$^{-1}$ to calculate the latter quantity \citep[following][]{Kochanek12,Kilpatrick18:17eaw}.  

\begin{figure}
    \centering
    \includegraphics[width=0.49\textwidth]{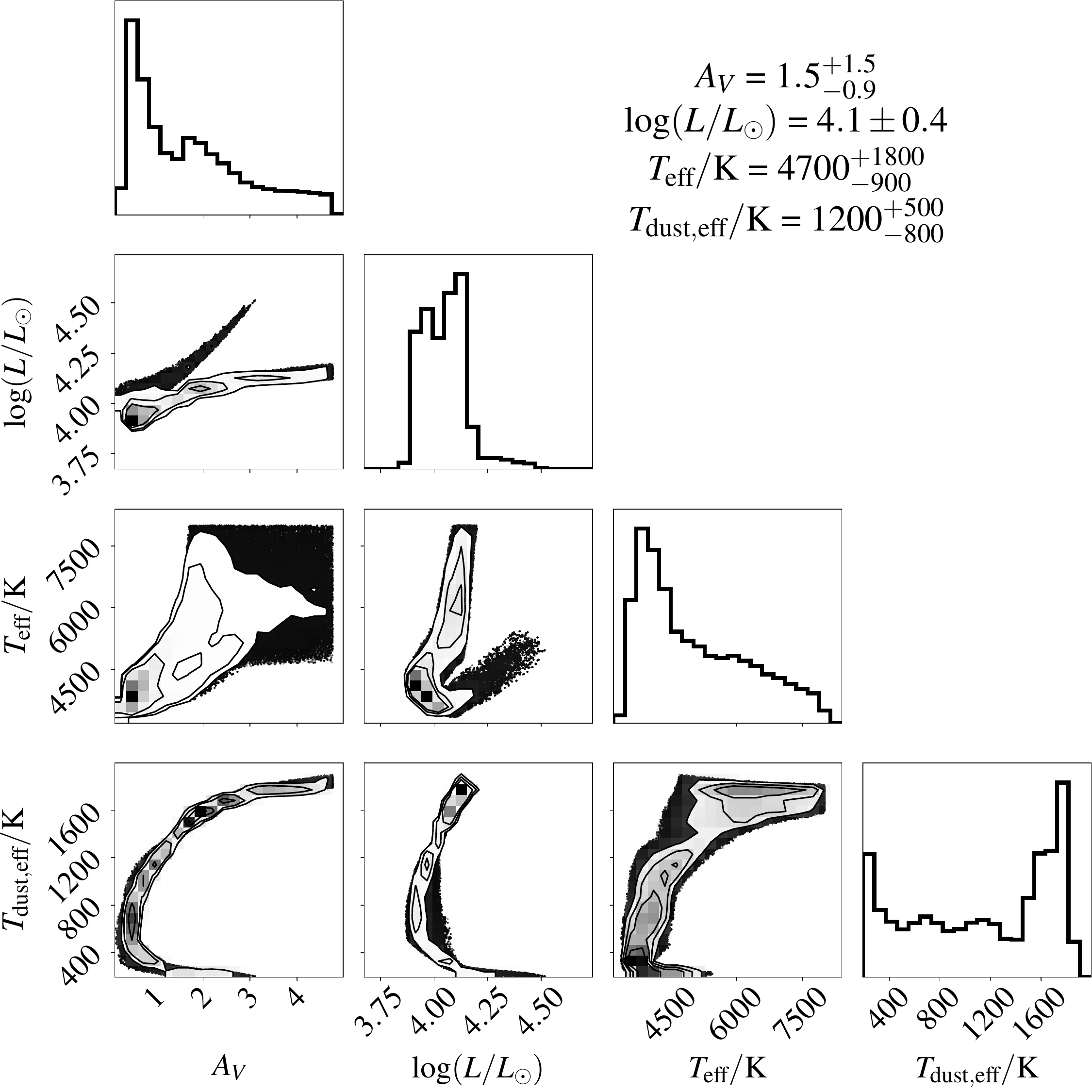}
    \caption{Results from the Monte Carlo Markov Chain fit for RSG spectral energy distributions with a shell of circumstellar dust to our SN\,2020jfo progenitor candidate photometry \citep[described in \autoref{sec:progenitor} and originally presented in][]{Kilpatrick18:17eaw}.  We parameterize the spectral energy distribution by the overall luminosity of the progenitor star ($\log(L/L_{\odot})$), its effective temperature ($T_{\rm eff}$), the total $V$-band extinction due to CSM ($A_{V}$), and the effective temperature of the circumstellar dust ($T_{\rm dust, eff}$).}
    \label{fig:rsg-corner}
\end{figure}

The highest luminosity progenitor stars level correspond to dust shells with a cool temperature ($\approx$200~K) and moderately high extinction ($A_{V}\approx3$) that pushes a large fraction of the spectral energy distribution into the mid-infrared, which we show in \autoref{fig:sed}. We show these correlations in the corner plot for each parameter derived from our MCMC in \autoref{fig:rsg-corner}.  The full luminosity range for our MCMC corresponds to a RSG mass of $7.2\substack{+1.2\\-0.6}~M_{\odot}$ based on MIST evolutionary tracks we used for comparison \citep[][]{choi+16}.  This estimate does not change significantly if we instead compare to Geneva rotating star models \citep[we find 7.1~$M_{\odot}$ for][]{Hirschi04} or KEPLER models \citep[we find 7.4~$M_{\odot}$;][]{woosley+07}.  In summary, our counterpart photometry and modeling predict an extremely low luminosity and thus initial mass.

Thus while circumstellar dust extinction can partly account for the unusually low luminosity and photospheric temperature of the SN\,2020jfo progenitor system, these values do not extend to the luminosities of $\approx12~M_{\odot}$ RSGs suggested by our analysis of SN\,2020jfo itself.  This can be observed from our corner plot \autoref{fig:rsg-corner}, where even in the most extreme case with the coolest dust ($T_{\rm eff}\approx200$~K) that provides a high extinction $A_{V}\approx3$~mag, the luminosity extends only to $\log(L/L_{\odot})=4.4$.  If we incorporate the uncertainty on distance placing M61 closer to the high end of the uncertainty on $\mu=30.81\pm0.20$~mag, then the counterpart could be as luminous as $\log(L/L_{\odot})=4.5$.  Accepting the most extreme distances to M61 \citep[e.g., 20.7~Mpc based on one estimate from the expanding photosphere method for SN\,2008in;][]{Bose14}, this luminosity would then be $\log(L/L_{\odot})=4.7$, consistent with $M_{\rm ZAMS}\approx11~M_{\odot}$ terminal RSGs from the MIST models \citep{choi+16}.  We emphasize that this assumes the {\it Spitzer}/IRAC photometry is dominated by this counterpart.  If this is not the case, these luminosities would simply be upper limits.

Moreover, the temperatures of the most extreme RSG models ($T_{\rm eff}=6500$~K) are hotter than those inferred for other SNe\,II-P \citep[even for the relatively warm RSG SEDs modeled in][]{Davies13}, with a correspondingly small envelope that is inconsistent with inferences from observations of the early-time photometry and shock cooling models (\autoref{sec:shock}) and expectations for the envelope sizes of SN\,II-P progenitor stars \citep[e.g.,][]{Dessart13,Gonzalez15}.

Using our preferred distance and assuming the {\it Spitzer}/IRAC photometry corresponds to the F814W counterpart brightness, we consider our best-fitting luminosity estimate to be $\log(L/L_{\odot}=4.1\pm0.4$) and mass estimate to be $7.2\substack{+1.2\\-0.6}~M_{\odot}$.  We compare these to our indirect mass estimates and consider the implication for SN\,II-P and II-L progenitor systems below. 

\section{Connecting the Progenitor Stars of Type~II Supernovae to their Terminal Explosions}\label{sec:connecting}

Including SN\,2020jfo, there are 17 SN\,II-P and II-L with progenitor star detections in the literature \citep[counting the 14 objects used in previous samples as well as SN\,2017eaw, 2018aoq, and 2020jfo;][]{Smartt15,Davies18,Kilpatrick18:17eaw,ONeill19,Sollerman21}.  This uniform sample enables direct comparisons between massive stars and their resulting explosions in a statistical sense.  One of the most significant examples of these comparisons is the ``red supergiant problem'', that is the inconsistency between the observed population of type II supernova progenitor stars \citep[which are mostly red supergiants with luminosities from $\log(L/L_{\odot})=4.4$--$5.2$;][]{Smartt09,Smartt15} and the population of red supergiants observed in the Milky Way, Magellanic Clouds, and other nearby galaxies \citep{Humphreys79,Massey03,Levesque06}, whose luminosities extend to $\log(L/L_{\odot})=5.6$.  

However, the physical properties of SNe\,II as inferred from large samples of directly-identified progenitor stars can be used to identify the incidence and rates of statistical outliers in peak luminosity, nickel mass, local environment, and nebular spectra at distances inaccessible to direct progenitor star detection studies.  Current SN surveys detect and classify hundreds of SN\,II per year \citep[e.g., ATLAS, PSST, YSE, ZTF;][]{Tonry11,Chambers16,Bellm19,Jones2021}, and the Vera C. Rubin Legacy Survey of Space and Time is expected to increase that rate to $>$10000~yr$^{-1}$ with photometrically-classified transients \citep{LSST2009}.

Efforts to establish a connection between the small, well-observed sample of nearby SN\,II with direct progenitor star detections to physical properties in their explosions \citep[e.g.,][]{Eldridge19} indicate that there is a significant amount of scatter in relationships between explosion energy, ejecta mass, nickel mass, and inferred progenitor star mass.  This may be an indication that there remain systematic uncertainties that bias these parameters, such as mass loss, binary interactions, or uncertainties in spectral energy distribution modeling that significantly affect the implied initial mass for progenitor star detections \citep{Smartt15}.  

For SN\,2020jfo, we summarize our direct and indirect mass estimates in \autoref{tab:progenitor-mass}.  In particular, we highlight the fact that although both {\it HST} and {\it Spitzer} photometry imply a relatively low-mass progenitor star with an initial mass of 7--8~$M_{\odot}$, our analysis of the shock breakout light curve and nebular spectroscopy support a higher initial mass around 10--12~$M_{\odot}$.  The only exception is our analysis of the local environment, which implies a main sequence turnoff age that is slightly lower in mass but nominally inconsistent with our pre-explosion photometry at the 1$\sigma$ level.  Even with the highest mass estimates from our pre-explosion photometry of $\log(L/L_{\odot})=4.5$ and thus an initial mass of $8.4~M_{\odot}$ and the lowest mass estimates from our indirect methods (8.7--11~$M_{\odot}$) from all methods), these two sets of methods are inconsistent.

This matches analysis in both \citet{Sollerman21} and \citet{Teja22} despite the use of independent data sets and slightly different assumptions about distance and line-of-sight extinction, implying that our results are relatively insensitive to these details in our analysis.  Moreover, the nickel mass we infer from the light curves of SN\,2020jfo is $0.018\pm0.007~M_{\odot}$, while slightly lower than derived in these previous studies, also suggests that SN\,2020jfo exploded from a higher initial mass star around 13~$M_{\odot}$ when compared to the empirical relationship derived by \citet{Eldridge19}.  This tension would only be exacerbated by adopting a larger nickel mass as in \citet{Sollerman21} and \citet{Teja22}, implying a larger initial mass progenitor star.

\begin{table}
\renewcommand{\arraystretch}{1.3}
    \begin{tabular}{ccc}
\hline Mass estimate & Method & Section \\
($M_{\odot}$) & & \\\hline\hline
11--13 & Shock breakout radius & \autoref{sec:shock} \\
$>$12 & ${}^{56}$Ni mass & \autoref{sec:bol} \\
12\tablenotemark{a} & Nebular spectroscopy & \autoref{sec:nebular} \\
9.7$\substack{+2.5\\-1.3}$ & Turnoff age & \autoref{sec:environment} \\
7.2$\substack{+1.2\\-0.6}$ & Pre-explosion counterpart & \autoref{sec:progenitor} \\
\hline
\end{tabular}
\caption{Direct and Indirect Progenitor Mass Estimates for SN\,2020jfo}\label{tab:progenitor-mass}
\vspace{-20pt}
\tablenotetext{a}{See discussion in \autoref{sec:nebular} for consistency with the closest 9~$M_{\odot}$ and 15~$M_{\odot}$ nebular spectroscopic models.  While a moderate increase in flux for our spectra could be consistent with models at higher mass, the flux level and absence of O and Fe lines in our spectra suggest that SN\,2020jfo is strongly inconsistent with the lower mass 9~$M_{\odot}$ model.}
\end{table}

Furthermore, it is difficult to reconcile the tension between these properties with line-of-sight extinction due to a massive circumstellar dust shell in part because our {\it Spitzer} photometry has already ruled out that scenario and because our light curve (\autoref{sec:shock}) and spectral (\autoref{sec:spec-analysis}) modeling does not support the presence of any such material.

However, we note another unusual feature from our analysis in the relatively depleted envelope of SN\,2020jfo compared with other SN\,II, which is reflected in the short plateau time ($t_{\rm PT}=65.3\substack{+1.4\\-0.7}$~days) and a low envelope mass derived from its early time light curve (1.7~$M_{\odot}$).  

We consider the possibility that the SN\,2020jfo progenitor star began as a 12~$M_{\odot}$ initial mass star but was stripped via radiative, wave-driven, or explosive mass loss or Roche lobe overflow onto a companion star.  This interpretation is complex as our direct detection suggest that it terminated with a much lower luminosity. Given that luminosity on the RSG branch is primarily tied to He-core mass this implies that the previous evolution of the progenitor system must have led to a He-core mass much smaller than expected for stars with $M_{\rm ZAMS}$=12~$M_{\odot}$. However, a small He-core mass would also imply a smaller O-core mass, leading to a lower $^{56}$Ni ejecta mass and lower [O\I] luminosity in nebular spectra than expected for a 12~$M_{\odot}$ star. This is in conflict with our analysis above.  Overall, we conclude that there is no evolutionary path through which a single 12~$M_{\odot}$ could evolve and terminate with an {\it intrinsically} lower luminosity consistent with the {\it HST} and {\it Spitzer} detection but still produce a high $^{56}$Ni mass and [O\I] luminosity.

The majority of massive stars are found in binary systems \citep{Sana12}, so it is natural to consider the effect of binary evolution on the final core and hydrogen-envelope mass of a 12~$M_{\odot}$ star.  \citet{Zapartas21} consider the effect of close binary interactions on the final effective core mass of Type II SN progenitor stars.  Their findings demonstrate that the He-core mass is significantly broadened for primary stars with binary interactions compared with single-star models.  This effect results from mass transfer early in the evolution of the primary star, which significantly changes the He core structure and leads to an intrinsically lower luminosity star than would be predicted from the main sequence turnoff age of the surrounding stellar population.  This effect could explain the discrepancy between the stellar population around SN\,2020jfo and its low luminosity, but the core structure would be inconsistent with the [O\I] luminosity and moderately large nickel mass.

As a quantitative examination of the hypothesis that SN\,2020jfo originated from a binary star, we examined all BPASS v2.2 models \citep{eldridge+17} for binary star systems at Solar metallicity.  BPASS enables a direct comparison to our photometry, although the predicted SEDs do not extend to {\it Spitzer} bandpasses, and so we instead fit to our F814W detection ($M_{\rm F814W}=-5.2\pm0.4$~mag accounting for distance uncertainty) and required only that the models have a terminal effective temperature for the primary star $T_{\rm eff}<4800$~K, consistent with the 1$\sigma$ uncertainties from our MCMC fit to the progenitor candidate of SN\,2020jfo (\autoref{sec:progenitor}).  We found models with primary star masses from 3.2--7.0~$M_{\odot}$ consistent with these criteria, with the higher mass models spanning 7+1.4 to 7+6.3~$M_{\odot}$ binaries and relatively wide separations ($\log(P/1~{\rm day})>3$).  No models in the 10--12~$M_{\odot}$ terminated with such low luminosities, although the 12,663 models we considered span a limited range and density of initial mass, mass ratio, and orbital period compared with the 150$\times$150$\times$150 models in \citet{Zapartas21}.  Extending this analysis to larger ($Z=0.02--0.04$) or smaller ($Z=0.001-0.010$) does not reveal any other models with $M_{\rm ZAMS}>10~M_{\odot}$ consistent with the observed progenitor photometry and implied effective temperature.

Constraints on the pre-explosion mass loss rate from late-time analysis of the SN\,2020jfo light curve or a search for a surviving companion star to this system would help to confirm the exact pre-explosion evolutionary scenario.  To date, surviving companion star candidates have been identified for SN\,IIb and SN\,Ib explosions \citep{Fox14,Ryder18,Fox22}, but there is not yet any evidence for such companion stars at SN\,II-P explosion sites, possibly pointing to the merger scenario for some SN\,II-P progenitor systems in cases where the progenitor star appears more massive than the surrounding stellar population would suggest \citep[e.g.,][]{Zapartas19}.  The increasing tension between normal SN\,II explosion properties and pre-explosion constraints on the nature of their progenitor stars \citep[also observed with SN\,2020fqv and 2021yja;][]{Tinyanont21,Hosseinzadeh22} may point to the influence of companion stars via binary interactions, which is expected for a significant fraction of massive stars overall \citep{Sana12}.

The remaining explanations we consider for the low luminosity of SN\,2020jfo is a compact circumstellar shell or variability that can explain the $A_{V}=2$--3~mag difference between our observed photometry and the expected luminosity based on our indirect mass estimates in \autoref{tab:progenitor-mass}.  Variability at this level is extremely rare for RSGs observed in situ in other galaxies \citep[e.g., $<$2\% of RSGs have $V$-band variations at $>$2~mag in][]{Conroy18}.  However, as we calculate in \autoref{sec:sed}, the SN\,2020jfo progenitor star would require a circumstellar shell of only $3$--$6\times10^{-6}~M_{\odot}$ to provide the necessary extinction to lower a terminal $M_{\rm ZAMS}=$12~$M_{\odot}$ star's F814W brightness to the value we measure in pre-explosion {\it HST} imaging.  The timescale of our detections of the pre-explosion counterpart in F814W, from 18.7 to 0.1 rest-frame years before explosion, suggests that this material is released in a steady wind or another continuous process such as wave-driven mass loss \citep{fuller+17}.  Light curves of the progenitor star, such as those available from multi-epoch {\it HST} and {\it Spitzer} imaging of SN\,2017eaw \citep{Kilpatrick18:17eaw} or future LSST imaging of extragalactic SN progenitor stars, is needed to constrain the timing and mode of mass loss leading to this material.

Finally, we consider the effect of similar circumstellar extinction on all RSGs observed with pre-explosion imaging.  The vast majority of Type II SN progenitor stars, including SN\,2020jfo, are identified in F814W imaging \citep[][and references therein]{Smartt15}.  An additional $A_{V}=2$--3~mag line-of-sight extinction would correspond to 1--1.5~mag extinction in F814W assuming $R_{V}=3.1$, or a 0.4--0.6~dex increase in luminosity after making a bolometric correction \citep[see, e.g., analysis in][]{Davies18}.  Increasing the luminosities of all RSG progenitor stars to SN\,II by such an extreme amount \citep[compared to the more typical RSG circumstellar dust masses assumed in][]{Walmswell12} would essentially ``solve'' the red supergiant problem, but this would require extreme mass loss and dust production timed almost immediately before the SN.  \citet{Teja22} find SN\,2020jfo appears to require a large mass ($\approx$0.2~$M_{\odot}$) within 40~AU of the progenitor star, but other SN\,II with well-constrained explosion times and early rises lack evidence for such CSM \citep{Hosseinzadeh22}.  In future, evidence for the persistence of the red supergiant problem may be better tested using a more uniform sample of SN\,II with well-sampled early light curves or pre-explosion mid-IR data from {\it JWST} in which the total circumstellar dust mass can be directly probed.

\section{Conclusions}\label{sec:conclusions}

We have presented a comparison between the explosion properties of the type II-P SN\,2020jfo as inferred from UV and optical observations, its host environment, and photometry of its pre-explosion counterpart observed in archival {\it HST} and {\it Spitzer} imaging.  From this data set, we infer that:

\begin{enumerate}
    \item UV and optical light curves within 16~days of discovery support a progenitor star radius of 700~$R_{\odot}$ for SN\,2020jfo, implying that this star had a luminosity of $\log(L/L_{\odot})=4.7$--4.8 if its photosphere was $T_{\rm eff}=$3300--3500~K.  This would place it in the range of $M_{\rm ZAMS}=11$--13~$M_{\odot}$ for terminal RSGs.
    \item The light curve of SN\,2020jfo had a short plateau phase with $t_{\rm PT}=65.3\substack{+1.4\\-0.7}$~days following \citet{Valenti16}, in the lower 5th percentile of all events studied in that analysis, pointing to a relatively low-mass hydrogen envelope.  In addition, its late-time luminosity can be explained with an initial ${}^{56}$Ni mass of $0.018\pm0.007~M_{\odot}$.  The latter quantity is consistent with a $\approx$13~$M_{\odot}$ progenitor star following the empirical relations in \citet{Eldridge19}.
    \item Modeling of our nebular spectra at $>$250~days with models from \citet{jerkstrand14} and \citet{jerkstrand18} points to a 12~$M_{\odot}$ initial mass progenitor star for SN\,2020jfo.  This is consistent with previous findings for nebular spectral analysis of SN\,2020jfo in \citet{Sollerman21} and \citet{Teja22}.
    \item Modeling of the SN\,2020jfo spiral host galaxy M61 via VLT/MUSE IFU spectroscopy points to a near Solar metallicity close to the explosion site and a SFR of (4.18$\pm$0.11)$\times$10$^{-7}$~$M_{\odot}$~yr$^{-1}$~pc$^{-1}$, similar to those observed for other SN\,II in \citet{galbany+16}.  In addition, we model the stellar population near SN\,2020jfo with {\tt STARLIGHT} and find that the most likely main sequence turnoff age of 35$\pm$10~Myr corresponds to a main sequence turnoff age of 8.7$\substack{+1.0\\-0.9}~M_{\odot}$, implying a relatively low mass progenitor star.
    \item We detect a counterpart to SN\,2020jfo in pre-explosion {\it HST} and {\it Spitzer} imaging by aligning adaptive optics imaging of SN\,2020jfo itself.  Modeling of the spectral energy distribution to this counterpart supports a low-mass progenitor star with the most likely luminosity being $\log(L/L_{\odot})=4.1$.  Modeling with a mid-infrared excess due to a compact shell of circumstellar gas and dust implies the pre-explosion counterpart was at most $\log(L/L_{\odot})=4.5$, and thus had an initial mass of 7--8~$M_{\odot}$, making it among the lowest luminosity pre-explosion counterparts to a SN\,II.
    \item Our constraints on the progenitor mass from direct detection ($M_{\rm ZAMS}<9~M_{\odot}$) is in tension with all other initial mass estimates from indirect methods, which favor a significantly more massive progenitor star around $M_{\rm ZAMS}=9$--13~$M_{\odot}$.
    \item We find that the tension between the explosion properties of SN\,2020jfo, its local environment, and its pre-explosion counterpart for the implied initial mass of the progenitor star cannot be explained through an intrinsic change in the luminosity of that star.  The most likely explanation for SN\,2020jfo is the explosion of a more luminous star exploding inside of a compact circumstellar shell that would have been swept up before its discovery and first spectroscopic observations.
\end{enumerate}

\bigskip\bigskip\bigskip
\noindent {\bf ACKNOWLEDGMENTS}
\smallskip

We thank K. Clever, C. Smith, and E. Strasburger for help with obtaining our Nickel observations.
C.D.K. acknowledges support from HST program AR-16136 and from a CIERA postdoctoral fellowship.
C.G. acknowledges support from a VILLUM FONDEN Young Investor Grant (project number 25501).
D.O.J. acknowledges support provided by NASA Hubble Fellowship grant HST-HF2-51462.001, which is awarded by the Space Telescope Science Institute, operated by the Association of Universities for Research in Astronomy, Inc., for NASA, under contract NAS5-26555.
A.J.G.O. acknowledges support from the Lachlan Gilchrist Fellowship Fund.
M.R.D. acknowledges support from the NSERC through grant RGPIN-2019-06186, the Canada Research Chairs Program, the Canadian Institute for Advanced Research (CIFAR), and the Dunlap Institute at the University of Toronto.
Pan-STARRS is a project of the Institute for Astronomy of the University of Hawaii, and is supported by the NASA SSO Near Earth Observation Program under grants 80NSSC18K0971, NNX14AM74G, NNX12AR65G, NNX13AQ47G, NNX08AR22G, 80NSSC21K1572 and by the State of Hawaii.
Some of the data presented herein were obtained at the W. M. Keck Observatory, which is operated as a scientific partnership among the California Institute of Technology, the University of California and the National Aeronautics and Space Administration. The Observatory was made possible by the generous financial support of the W. M. Keck Foundation.
The authors wish to recognize and acknowledge the very significant cultural role and reverence that the summit of Maunakea has always had within the indigenous Hawaiian community.  We are most fortunate to have the opportunity to conduct observations from this mountain.
This work makes use of observations from the LCOGT network through programs NOAO2020A-008, NOAO2020B-009 (PI Kilpatrick), NOAO2020A-012, and NOAO2020B-011 (PI Foley).
Based on observations made with the NASA/ESA Hubble Space Telescope, obtained from the data archive at the Space Telescope Science Institute. STScI is operated by the Association of Universities for Research in Astronomy, Inc. under NASA contract NAS 5-26555.
This work is based in part on observations made with the Spitzer Space Telescope, which was operated by the Jet Propulsion Laboratory, California Institute of Technology under a contract with NASA.
This publication has made use of data collected at Lulin Observatory, partly supported by MoST grant 108-2112-M-008-001.

\textit{Facilities}: {\it HST} (WFPC2, ACS, WFC3), Keck (LRIS, OSIRIS), LCOGT (Sinistro, FLOYDS), Nickel (Direct-2K), PS1 (GPC1), Shane (Kast), {\it Spitzer} (IRAC), {\it Swift} (XRT, UVOT), Thacher (ACP), VLT (MUSE)

\section*{Data Availability}

All imaging, spectroscopy, and relevant data products presented in this article are available upon request.  The {\it Hubble Space Telescope} and {\it Spitzer Space Telescope} data are publicly available and can be accessed from the Mikulski Archive for Space Telescopes (\url{https://archive.stsci.edu/hst/}) and Spitzer Heritage Archive (\url{http://sha.ipac.caltech.edu/applications/Spitzer/SHA/}), respectively.  

\bibliography{2020jfo}

\begin{thebibliography}{}
\makeatletter
\relax
\def\mn@urlcharsother{\let\do\@makeother \do\$\do\&\do\#\do\^\do\_\do\%\do\~}
\def\mn@doi{\begingroup\mn@urlcharsother \@ifnextchar [ {\mn@doi@}
  {\mn@doi@[]}}
\def\mn@doi@[#1]#2{\def\@tempa{#1}\ifx\@tempa\@empty \href
  {http://dx.doi.org/#2} {doi:#2}\else \href {http://dx.doi.org/#2} {#1}\fi
  \endgroup}
\def\mn@eprint#1#2{\mn@eprint@#1:#2::\@nil}
\def\mn@eprint@arXiv#1{\href {http://arxiv.org/abs/#1} {{\tt arXiv:#1}}}
\def\mn@eprint@dblp#1{\href {http://dblp.uni-trier.de/rec/bibtex/#1.xml}
  {dblp:#1}}
\def\mn@eprint@#1:#2:#3:#4\@nil{\def\@tempa {#1}\def\@tempb {#2}\def\@tempc
  {#3}\ifx \@tempc \@empty \let \@tempc \@tempb \let \@tempb \@tempa \fi \ifx
  \@tempb \@empty \def\@tempb {arXiv}\fi \@ifundefined
  {mn@eprint@\@tempb}{\@tempb:\@tempc}{\expandafter \expandafter \csname
  mn@eprint@\@tempb\endcsname \expandafter{\@tempc}}}

\bibitem[\protect\citeauthoryear{{Alam} et~al.,}{{Alam} et~al.}{2015}]{Alam15}
{Alam} S.,  et~al., 2015, \mn@doi [\apjs] {10.1088/0067-0049/219/1/12}, \href
  {https://ui.adsabs.harvard.edu/abs/2015ApJS..219...12A} {219, 12}

\bibitem[\protect\citeauthoryear{{Anderson} \& {James}}{{Anderson} \&
  {James}}{2008}]{Anderson08}
{Anderson} J.~P.,  {James} P.~A.,  2008, \mn@doi [\mnras]
  {10.1111/j.1365-2966.2008.13843.x}, \href
  {https://ui.adsabs.harvard.edu/abs/2008MNRAS.390.1527A} {390, 1527}

\bibitem[\protect\citeauthoryear{{Andrews} et~al.,}{{Andrews}
  et~al.}{2019}]{Andrews20}
{Andrews} J.~E.,  et~al., 2019, \mn@doi [\apj] {10.3847/1538-4357/ab43e3},
  \href {https://ui.adsabs.harvard.edu/abs/2019ApJ...885...43A} {885, 43}

\bibitem[\protect\citeauthoryear{{Arcavi}}{{Arcavi}}{2017}]{Arcavi16}
{Arcavi} I.,  2017, {Hydrogen-Rich Core-Collapse Supernovae}.
Cham, Switzerland : Springer, p.~239, \mn@doi{10.1007/978-3-319-21846-5\_39}

\bibitem[\protect\citeauthoryear{{Arnett}}{{Arnett}}{1987}]{arnett87}
{Arnett} W.~D.,  1987, \mn@doi [\apj] {10.1086/165439}, \href
  {http://adsabs.harvard.edu/abs/1987ApJ...319..136A} {319, 136}

\bibitem[\protect\citeauthoryear{{Asplund}, {Grevesse}, {Sauval}  \&
  {Scott}}{{Asplund} et~al.}{2009}]{Asplund2009}
{Asplund} M.,  {Grevesse} N.,  {Sauval} A.~J.,   {Scott} P.,  2009, \mn@doi
  [\araa] {10.1146/annurev.astro.46.060407.145222}, \href
  {https://ui.adsabs.harvard.edu/abs/2009ARA&A..47..481A} {47, 481}

\bibitem[\protect\citeauthoryear{{Bacon} et~al.,}{{Bacon}
  et~al.}{2010}]{Bacon2010}
{Bacon} R.,  et~al., 2010, in {McLean} I.~S.,  {Ramsay} S.~K.,   {Takami} H.,
  eds,  Society of Photo-Optical Instrumentation Engineers (SPIE) Conference
  Series Vol. 7735, Ground-based and Airborne Instrumentation for Astronomy
  III. p. 773508, \mn@doi{10.1117/12.856027}

\bibitem[\protect\citeauthoryear{{Barbon}, {Ciatti}  \& {Rosino}}{{Barbon}
  et~al.}{1979}]{barbon+79}
{Barbon} R.,  {Ciatti} F.,   {Rosino} L.,  1979, \aap, \href
  {http://adsabs.harvard.edu/abs/1979A%26A....72..287B} {72, 287}

\bibitem[\protect\citeauthoryear{{Beasor} \& {Davies}}{{Beasor} \&
  {Davies}}{2018}]{Beasor18}
{Beasor} E.~R.,  {Davies} B.,  2018, \mn@doi [\mnras] {10.1093/mnras/stx3174},
  \href {https://ui.adsabs.harvard.edu/abs/2018MNRAS.475...55B} {475, 55}

\bibitem[\protect\citeauthoryear{{Becker}}{{Becker}}{2015}]{Becker15}
{Becker} A.,  2015, {HOTPANTS: High Order Transform of PSF ANd Template
  Subtraction}, Astrophysics Source Code Library, record ascl:1504.004
  (\mn@eprint {ascl} {1504.004})

\bibitem[\protect\citeauthoryear{{Bell} \& {de Jong}}{{Bell} \& {de
  Jong}}{2000}]{Bell00}
{Bell} E.~F.,  {de Jong} R.~S.,  2000, \mn@doi [\mnras]
  {10.1046/j.1365-8711.2000.03138.x}, \href
  {https://ui.adsabs.harvard.edu/abs/2000MNRAS.312..497B} {312, 497}

\bibitem[\protect\citeauthoryear{{Bellm} et~al.,}{{Bellm}
  et~al.}{2019}]{Bellm19}
{Bellm} E.~C.,  et~al., 2019, \mn@doi [\pasp] {10.1088/1538-3873/aaecbe}, \href
  {https://ui.adsabs.harvard.edu/abs/2019PASP..131a8002B} {131, 018002}

\bibitem[\protect\citeauthoryear{{Bertin}}{{Bertin}}{2010}]{swarp}
{Bertin} E.,  2010, {SWarp: Resampling and Co-adding FITS Images Together}
  (\mn@eprint {ascl} {1010.068})

\bibitem[\protect\citeauthoryear{{Binggeli}, {Sandage}  \&
  {Tammann}}{{Binggeli} et~al.}{1985}]{Binggeli1985}
{Binggeli} B.,  {Sandage} A.,   {Tammann} G.~A.,  1985, \mn@doi [\aj]
  {10.1086/113874}, \href
  {https://ui.adsabs.harvard.edu/abs/1985AJ.....90.1681B} {90, 1681}

\bibitem[\protect\citeauthoryear{{Blinnikov} \& {Bartunov}}{{Blinnikov} \&
  {Bartunov}}{1993}]{Blinnikov93}
{Blinnikov} S.~I.,  {Bartunov} O.~S.,  1993, \aap, \href
  {https://ui.adsabs.harvard.edu/abs/1993A&A...273..106B} {273, 106}

\bibitem[\protect\citeauthoryear{{Bose} \& {Kumar}}{{Bose} \&
  {Kumar}}{2014}]{Bose14}
{Bose} S.,  {Kumar} B.,  2014, \mn@doi [\apj] {10.1088/0004-637X/782/2/98},
  \href {https://ui.adsabs.harvard.edu/abs/2014ApJ...782...98B} {782, 98}

\bibitem[\protect\citeauthoryear{{Bostroem} et~al.,}{{Bostroem}
  et~al.}{2019}]{bostroem19}
{Bostroem} K.~A.,  et~al., 2019, \mn@doi [\mnras] {10.1093/mnras/stz570}, \href
  {https://ui.adsabs.harvard.edu/abs/2019MNRAS.485.5120B} {485, 5120}

\bibitem[\protect\citeauthoryear{{Bowen}, {Blades}  \& {Pettini}}{{Bowen}
  et~al.}{1996}]{Bowen1996}
{Bowen} D.~V.,  {Blades} J.~C.,   {Pettini} M.,  1996, \mn@doi [\apjl]
  {10.1086/310370}, \href
  {https://ui.adsabs.harvard.edu/abs/1996ApJ...472L..77B} {472, L77}

\bibitem[\protect\citeauthoryear{{Brown} et~al.,}{{Brown}
  et~al.}{2013}]{Brown13}
{Brown} T.~M.,  et~al., 2013, \mn@doi [\pasp] {10.1086/673168}, \href
  {https://ui.adsabs.harvard.edu/abs/2013PASP..125.1031B} {125, 1031}

\bibitem[\protect\citeauthoryear{{Brown}, {Breeveld}, {Holland}, {Kuin}  \&
  {Pritchard}}{{Brown} et~al.}{2014}]{Brown14}
{Brown} P.~J.,  {Breeveld} A.~A.,  {Holland} S.,  {Kuin} P.,   {Pritchard} T.,
  2014, \mn@doi [\apss] {10.1007/s10509-014-2059-8}, \href
  {https://ui.adsabs.harvard.edu/abs/2014Ap&SS.354...89B} {354, 89}

\bibitem[\protect\citeauthoryear{{Burrows}, {Hayes}  \& {Fryxell}}{{Burrows}
  et~al.}{1995}]{Burrows95}
{Burrows} A.,  {Hayes} J.,   {Fryxell} B.~A.,  1995, \mn@doi [\apj]
  {10.1086/176188}, \href
  {https://ui.adsabs.harvard.edu/abs/1995ApJ...450..830B} {450, 830}

\bibitem[\protect\citeauthoryear{{Burrows} et~al.,}{{Burrows}
  et~al.}{2005}]{burrows05}
{Burrows} D.~N.,  et~al., 2005, \mn@doi [\ssr] {10.1007/s11214-005-5097-2},
  \href {https://ui.adsabs.harvard.edu/abs/2005SSRv..120..165B} {120, 165}

\bibitem[\protect\citeauthoryear{{Buta} \& {Keel}}{{Buta} \&
  {Keel}}{2019}]{Buta19}
{Buta} R.~J.,  {Keel} W.~C.,  2019, \mn@doi [\mnras] {10.1093/mnras/stz1291},
  \href {https://ui.adsabs.harvard.edu/abs/2019MNRAS.487..832B} {487, 832}

\bibitem[\protect\citeauthoryear{{Cao} et~al.,}{{Cao} et~al.}{2013}]{cao+13}
{Cao} Y.,  et~al., 2013, \mn@doi [\apjl] {10.1088/2041-8205/775/1/L7}, \href
  {https://ui.adsabs.harvard.edu/abs/2013ApJ...775L...7C} {775, L7}

\bibitem[\protect\citeauthoryear{{Cappellari} \& {Copin}}{{Cappellari} \&
  {Copin}}{2003}]{Cappellari2003}
{Cappellari} M.,  {Copin} Y.,  2003, \mn@doi [\mnras]
  {10.1046/j.1365-8711.2003.06541.x}, \href
  {https://ui.adsabs.harvard.edu/abs/2003MNRAS.342..345C} {342, 345}

\bibitem[\protect\citeauthoryear{{Cardelli}, {Clayton}  \& {Mathis}}{{Cardelli}
  et~al.}{1989}]{cardelli+89}
{Cardelli} J.~A.,  {Clayton} G.~C.,   {Mathis} J.~S.,  1989, \mn@doi [\apj]
  {10.1086/167900}, \href {http://adsabs.harvard.edu/abs/1989ApJ...345..245C}
  {345, 245}

\bibitem[\protect\citeauthoryear{{Chabrier}}{{Chabrier}}{2003}]{Chabrier2003}
{Chabrier} G.,  2003, \mn@doi [\pasp] {10.1086/376392}, \href
  {https://ui.adsabs.harvard.edu/abs/2003PASP..115..763C} {115, 763}

\bibitem[\protect\citeauthoryear{{Chambers} et~al.,}{{Chambers}
  et~al.}{2016}]{Chambers16}
{Chambers} K.~C.,  et~al., 2016, arXiv e-prints, \href
  {https://ui.adsabs.harvard.edu/abs/2016arXiv161205560C} {p. arXiv:1612.05560}

\bibitem[\protect\citeauthoryear{{Chambers} et~al.,}{{Chambers}
  et~al.}{2017}]{Chambers2017}
{Chambers} K.~C.,  et~al., 2017, Transient Name Server Discovery Report, \href
  {https://ui.adsabs.harvard.edu/abs/2017TNSTR.324....1C} {2017-324, 1}

\bibitem[\protect\citeauthoryear{{Chevalier}}{{Chevalier}}{1976}]{chevalier+76}
{Chevalier} R.~A.,  1976, \mn@doi [\apj] {10.1086/154557}, \href
  {http://adsabs.harvard.edu/abs/1976ApJ...207..872C} {207, 872}

\bibitem[\protect\citeauthoryear{{Choi}, {Dotter}, {Conroy}, {Cantiello},
  {Paxton}  \& {Johnson}}{{Choi} et~al.}{2016}]{choi+16}
{Choi} J.,  {Dotter} A.,  {Conroy} C.,  {Cantiello} M.,  {Paxton} B.,
  {Johnson} B.~D.,  2016, \mn@doi [\apj] {10.3847/0004-637X/823/2/102}, \href
  {http://adsabs.harvard.edu/abs/2016ApJ...823..102C} {823, 102}

\bibitem[\protect\citeauthoryear{{Cid Fernandes}, {Mateus}, {Sodr{\'e}},
  {Stasi{\'n}ska}  \& {Gomes}}{{Cid Fernandes} et~al.}{2005}]{CidFernandes2005}
{Cid Fernandes} R.,  {Mateus} A.,  {Sodr{\'e}} L.,  {Stasi{\'n}ska} G.,
  {Gomes} J.~M.,  2005, \mn@doi [\mnras] {10.1111/j.1365-2966.2005.08752.x},
  \href {https://ui.adsabs.harvard.edu/abs/2005MNRAS.358..363C} {358, 363}

\bibitem[\protect\citeauthoryear{{Conroy} et~al.,}{{Conroy}
  et~al.}{2018}]{Conroy18}
{Conroy} C.,  et~al., 2018, \mn@doi [\apj] {10.3847/1538-4357/aad460}, \href
  {https://ui.adsabs.harvard.edu/abs/2018ApJ...864..111C} {864, 111}

\bibitem[\protect\citeauthoryear{{Coulter} et~al.,}{{Coulter}
  et~al.}{2022}]{2022zndo...7278430C}
{Coulter} D.~A.,  et~al., 2022, {YSE-PZ: An Open-source Target and Observation
  Management System}, Zenodo, \mn@doi{10.5281/zenodo.7278430}

\bibitem[\protect\citeauthoryear{{Coulter} et~al.,}{{Coulter}
  et~al.}{2023}]{Coulter23}
{Coulter} D.~A.,  et~al., 2023, \mn@doi [arXiv e-prints]
  {10.48550/arXiv.2303.02154}, \href
  {https://ui.adsabs.harvard.edu/abs/2023arXiv230302154C} {p. arXiv:2303.02154}

\bibitem[\protect\citeauthoryear{{Crockett}, {Smartt}, {Pastorello},
  {Eldridge}, {Stephens}, {Maund}  \& {Mattila}}{{Crockett}
  et~al.}{2011}]{Crockett11}
{Crockett} R.~M.,  {Smartt} S.~J.,  {Pastorello} A.,  {Eldridge} J.~J.,
  {Stephens} A.~W.,  {Maund} J.~R.,   {Mattila} S.,  2011, \mn@doi [\mnras]
  {10.1111/j.1365-2966.2010.17652.x}, \href
  {https://ui.adsabs.harvard.edu/abs/2011MNRAS.410.2767C} {410, 2767}

\bibitem[\protect\citeauthoryear{{Davies} \& {Beasor}}{{Davies} \&
  {Beasor}}{2018}]{Davies18}
{Davies} B.,  {Beasor} E.~R.,  2018, \mn@doi [\mnras] {10.1093/mnras/stx2734},
  \href {https://ui.adsabs.harvard.edu/abs/2018MNRAS.474.2116D} {474, 2116}

\bibitem[\protect\citeauthoryear{{Davies} et~al.,}{{Davies}
  et~al.}{2013}]{Davies13}
{Davies} B.,  et~al., 2013, \mn@doi [\apj] {10.1088/0004-637X/767/1/3}, \href
  {https://ui.adsabs.harvard.edu/abs/2013ApJ...767....3D} {767, 3}

\bibitem[\protect\citeauthoryear{{Davies} et~al.,}{{Davies}
  et~al.}{2017}]{Davies17}
{Davies} B.,  et~al., 2017, \mn@doi [\apj] {10.3847/1538-4357/aa89ed}, \href
  {https://ui.adsabs.harvard.edu/abs/2017ApJ...847..112D} {847, 112}

\bibitem[\protect\citeauthoryear{{Dessart}, {Hillier}, {Waldman}  \&
  {Livne}}{{Dessart} et~al.}{2013}]{Dessart13}
{Dessart} L.,  {Hillier} D.~J.,  {Waldman} R.,   {Livne} E.,  2013, \mn@doi
  [\mnras] {10.1093/mnras/stt861}, \href
  {https://ui.adsabs.harvard.edu/abs/2013MNRAS.433.1745D} {433, 1745}

\bibitem[\protect\citeauthoryear{{Dessart}, {John Hillier}, {Sukhbold},
  {Woosley}  \& {Janka}}{{Dessart} et~al.}{2021}]{Dessart21}
{Dessart} L.,  {John Hillier} D.,  {Sukhbold} T.,  {Woosley} S.~E.,   {Janka}
  H.~T.,  2021, \mn@doi [\aap] {10.1051/0004-6361/202140839}, \href
  {https://ui.adsabs.harvard.edu/abs/2021A&A...652A..64D} {652, A64}

\bibitem[\protect\citeauthoryear{{Dolphin}}{{Dolphin}}{2016}]{dolphot}
{Dolphin} A.,  2016, {DOLPHOT: Stellar photometry} (\mn@eprint {ascl}
  {1608.013})

\bibitem[\protect\citeauthoryear{{Dwarkadas}}{{Dwarkadas}}{2014}]{Dwarkadas14}
{Dwarkadas} V.~V.,  2014, \mn@doi [\mnras] {10.1093/mnras/stu347}, \href
  {https://ui.adsabs.harvard.edu/abs/2014MNRAS.440.1917D} {440, 1917}

\bibitem[\protect\citeauthoryear{{Eldridge}, {Stanway}, {Xiao}, {McClelland},
  {Taylor}, {Ng}, {Greis}  \& {Bray}}{{Eldridge} et~al.}{2017}]{eldridge+17}
{Eldridge} J.~J.,  {Stanway} E.~R.,  {Xiao} L.,  {McClelland} L.~A.~S.,
  {Taylor} G.,  {Ng} M.,  {Greis} S.~M.~L.,   {Bray} J.~C.,  2017, \mn@doi
  [\pasa] {10.1017/pasa.2017.51}, \href
  {http://adsabs.harvard.edu/abs/2017PASA...34...58E} {34, e058}

\bibitem[\protect\citeauthoryear{{Eldridge}, {Guo}, {Rodrigues}, {Stanway}  \&
  {Xiao}}{{Eldridge} et~al.}{2019}]{Eldridge19}
{Eldridge} J.~J.,  {Guo} N.~Y.,  {Rodrigues} N.,  {Stanway} E.~R.,   {Xiao} L.,
   2019, \mn@doi [\pasa] {10.1017/pasa.2019.31}, \href
  {https://ui.adsabs.harvard.edu/abs/2019PASA...36...41E} {36, e041}

\bibitem[\protect\citeauthoryear{{Elias-Rosa} et~al.,}{{Elias-Rosa}
  et~al.}{2010}]{Elias-Rosa11}
{Elias-Rosa} N.,  et~al., 2010, \mn@doi [\apjl] {10.1088/2041-8205/714/2/L254},
  \href {https://ui.adsabs.harvard.edu/abs/2010ApJ...714L.254E} {714, L254}

\bibitem[\protect\citeauthoryear{{Emsellem} et~al.,}{{Emsellem}
  et~al.}{2021}]{Emsellem2021}
{Emsellem} E.,  et~al., 2021, arXiv e-prints, \href
  {https://ui.adsabs.harvard.edu/abs/2021arXiv211003708E} {p. arXiv:2110.03708}

\bibitem[\protect\citeauthoryear{{Falk}}{{Falk}}{1978}]{falk+78}
{Falk} S.~W.,  1978, in Bulletin of the American Astronomical Society. p.~425

\bibitem[\protect\citeauthoryear{{Falk} \& {Arnett}}{{Falk} \&
  {Arnett}}{1973}]{falk+73}
{Falk} S.~W.,  {Arnett} W.~D.,  1973, \mn@doi [\apjl] {10.1086/181154}, \href
  {http://adsabs.harvard.edu/abs/1973ApJ...180L..65F} {180, L65}

\bibitem[\protect\citeauthoryear{{Filippenko}}{{Filippenko}}{1997}]{filippenko97}
{Filippenko} A.~V.,  1997, \mn@doi [\araa] {10.1146/annurev.astro.35.1.309},
  \href {http://adsabs.harvard.edu/abs/1997ARA%26A..35..309F} {35, 309}

\bibitem[\protect\citeauthoryear{{Flewelling} et~al.,}{{Flewelling}
  et~al.}{2020}]{flewelling+16}
{Flewelling} H.~A.,  et~al., 2020, \mn@doi [\apjs] {10.3847/1538-4365/abb82d},
  \href {https://ui.adsabs.harvard.edu/abs/2020ApJS..251....7F} {251, 7}

\bibitem[\protect\citeauthoryear{{Foley}, {Van Dyk}, {Jha}, {Clubb},
  {Filippenko}, {Mauerhan}, {Miller}  \& {Smith}}{{Foley}
  et~al.}{2015}]{Foley14}
{Foley} R.~J.,  {Van Dyk} S.~D.,  {Jha} S.~W.,  {Clubb} K.~I.,  {Filippenko}
  A.~V.,  {Mauerhan} J.~C.,  {Miller} A.~A.,   {Smith} N.,  2015, \mn@doi
  [\apjl] {10.1088/2041-8205/798/2/L37}, \href
  {https://ui.adsabs.harvard.edu/abs/2015ApJ...798L..37F} {798, L37}

\bibitem[\protect\citeauthoryear{{Foley} et~al.,}{{Foley}
  et~al.}{2018}]{Foley18}
{Foley} R.~J.,  et~al., 2018, \mn@doi [\mnras] {10.1093/mnras/stx3136}, \href
  {https://ui.adsabs.harvard.edu/abs/2018MNRAS.475..193F} {475, 193}

\bibitem[\protect\citeauthoryear{{Fox} et~al.,}{{Fox} et~al.}{2014}]{Fox14}
{Fox} O.~D.,  et~al., 2014, \mn@doi [\apj] {10.1088/0004-637X/790/1/17}, \href
  {https://ui.adsabs.harvard.edu/abs/2014ApJ...790...17F} {790, 17}

\bibitem[\protect\citeauthoryear{{Fox} et~al.,}{{Fox} et~al.}{2016}]{Fox15}
{Fox} O.~D.,  et~al., 2016, \mn@doi [\apjl] {10.3847/2041-8205/816/1/L13},
  \href {https://ui.adsabs.harvard.edu/abs/2016ApJ...816L..13F} {816, L13}

\bibitem[\protect\citeauthoryear{{Fox} et~al.,}{{Fox} et~al.}{2022}]{Fox22}
{Fox} O.~D.,  et~al., 2022, \mn@doi [\apjl] {10.3847/2041-8213/ac5890}, \href
  {https://ui.adsabs.harvard.edu/abs/2022ApJ...929L..15F} {929, L15}

\bibitem[\protect\citeauthoryear{{Fuller}}{{Fuller}}{2017}]{fuller+17}
{Fuller} J.,  2017, preprint, \href
  {http://adsabs.harvard.edu/abs/2017arXiv170408696F} {} (\mn@eprint {arXiv}
  {1704.08696})

\bibitem[\protect\citeauthoryear{{Fusco} et~al.,}{{Fusco}
  et~al.}{2020}]{Fusco2020}
{Fusco} T.,  et~al., 2020, \mn@doi [\aap] {10.1051/0004-6361/202037595}, \href
  {https://ui.adsabs.harvard.edu/abs/2020A&A...635A.208F} {635, A208}

\bibitem[\protect\citeauthoryear{{Gagliano} et~al.,}{{Gagliano}
  et~al.}{2022}]{Gagliano2022}
{Gagliano} A.,  et~al., 2022, \mn@doi [\apj] {10.3847/1538-4357/ac35ec}, \href
  {https://ui.adsabs.harvard.edu/abs/2022ApJ...924...55G} {924, 55}

\bibitem[\protect\citeauthoryear{{Gal-Yam} et~al.,}{{Gal-Yam}
  et~al.}{2014}]{gal-yam+14}
{Gal-Yam} A.,  et~al., 2014, \mn@doi [\nat] {10.1038/nature13304}, \href
  {http://adsabs.harvard.edu/abs/2014Natur.509..471G} {509, 471}

\bibitem[\protect\citeauthoryear{{Galbany} et~al.,}{{Galbany}
  et~al.}{2016a}]{Galbany16}
{Galbany} L.,  et~al., 2016a, \mn@doi [\aj] {10.3847/0004-6256/151/2/33}, \href
  {https://ui.adsabs.harvard.edu/abs/2016AJ....151...33G} {151, 33}

\bibitem[\protect\citeauthoryear{{Galbany} et~al.,}{{Galbany}
  et~al.}{2016b}]{galbany+16}
{Galbany} L.,  et~al., 2016b, \mn@doi [\mnras] {10.1093/mnras/stv2620}, \href
  {http://adsabs.harvard.edu/abs/2016MNRAS.455.4087G} {455, 4087}

\bibitem[\protect\citeauthoryear{{Galbany} et~al.,}{{Galbany}
  et~al.}{2016c}]{Galbany2016}
{Galbany} L.,  et~al., 2016c, \mn@doi [\aap] {10.1051/0004-6361/201528045},
  \href {https://ui.adsabs.harvard.edu/abs/2016A&A...591A..48G} {591, A48}

\bibitem[\protect\citeauthoryear{{Goldberg} \& {Bildsten}}{{Goldberg} \&
  {Bildsten}}{2020}]{Goldberg+20}
{Goldberg} J.~A.,  {Bildsten} L.,  2020, \mn@doi [\apjl]
  {10.3847/2041-8213/ab9300}, \href
  {https://ui.adsabs.harvard.edu/abs/2020ApJ...895L..45G} {895, L45}

\bibitem[\protect\citeauthoryear{{Gonz{\'a}lez-Gait{\'a}n}
  et~al.,}{{Gonz{\'a}lez-Gait{\'a}n} et~al.}{2015}]{Gonzalez15}
{Gonz{\'a}lez-Gait{\'a}n} S.,  et~al., 2015, \mn@doi [\mnras]
  {10.1093/mnras/stv1097}, \href
  {https://ui.adsabs.harvard.edu/abs/2015MNRAS.451.2212G} {451, 2212}

\bibitem[\protect\citeauthoryear{{G{\"u}ver} \& {{\"O}zel}}{{G{\"u}ver} \&
  {{\"O}zel}}{2009}]{Guver09}
{G{\"u}ver} T.,  {{\"O}zel} F.,  2009, \mn@doi [\mnras]
  {10.1111/j.1365-2966.2009.15598.x}, \href
  {https://ui.adsabs.harvard.edu/abs/2009MNRAS.400.2050G} {400, 2050}

\bibitem[\protect\citeauthoryear{{Hack} et~al.,}{{Hack}
  et~al.}{2021}]{drizzlepac}
{Hack} W.~J.,  et~al., 2021, {spacetelescope/drizzlepac: Drizzlepac v3.3.0},
  Zenodo, \mn@doi{10.5281/zenodo.5534751}

\bibitem[\protect\citeauthoryear{{Hamuy}}{{Hamuy}}{2003}]{hamuy+03}
{Hamuy} M.,  2003, \mn@doi [\apj] {10.1086/344689}, \href
  {http://adsabs.harvard.edu/abs/2003ApJ...582..905H} {582, 905}

\bibitem[\protect\citeauthoryear{{Haynes} et~al.,}{{Haynes}
  et~al.}{2018}]{Haynes18}
{Haynes} M.~P.,  et~al., 2018, \mn@doi [\apj] {10.3847/1538-4357/aac956}, \href
  {https://ui.adsabs.harvard.edu/abs/2018ApJ...861...49H} {861, 49}

\bibitem[\protect\citeauthoryear{{Hillebrandt}, {Hoeflich}, {Weiss}  \&
  {Truran}}{{Hillebrandt} et~al.}{1987}]{Hillebrandt87}
{Hillebrandt} W.,  {Hoeflich} P.,  {Weiss} A.,   {Truran} J.~W.,  1987, \mn@doi
  [\nat] {10.1038/327597a0}, \href
  {https://ui.adsabs.harvard.edu/abs/1987Natur.327..597H} {327, 597}

\bibitem[\protect\citeauthoryear{{Hillier} \& {Dessart}}{{Hillier} \&
  {Dessart}}{2019}]{hillier+19}
{Hillier} D.~J.,  {Dessart} L.,  2019, \mn@doi [\aap]
  {10.1051/0004-6361/201935100}, \href
  {https://ui.adsabs.harvard.edu/abs/2019A&A...631A...8H} {631, A8}

\bibitem[\protect\citeauthoryear{{Hiramatsu} et~al.,}{{Hiramatsu}
  et~al.}{2021}]{hiramatsu21}
{Hiramatsu} D.,  et~al., 2021, \mn@doi [Nature Astronomy]
  {10.1038/s41550-021-01384-2}, \href
  {https://ui.adsabs.harvard.edu/abs/2021NatAs...5..903H} {5, 903}

\bibitem[\protect\citeauthoryear{{Hirschi}, {Meynet}  \& {Maeder}}{{Hirschi}
  et~al.}{2004}]{Hirschi04}
{Hirschi} R.,  {Meynet} G.,   {Maeder} A.,  2004, \mn@doi [\aap]
  {10.1051/0004-6361:20041095}, \href
  {https://ui.adsabs.harvard.edu/abs/2004A&A...425..649H} {425, 649}

\bibitem[\protect\citeauthoryear{{Hosseinzadeh}}{{Hosseinzadeh}}{2019}]{Hosseinzadeh19}
{Hosseinzadeh} G.,  2019, {Light Curve Fitting},
  \mn@doi{10.5281/zenodo.2639464}

\bibitem[\protect\citeauthoryear{{Hosseinzadeh} et~al.,}{{Hosseinzadeh}
  et~al.}{2018}]{Hosseinzadeh18}
{Hosseinzadeh} G.,  et~al., 2018, \mn@doi [\apj] {10.3847/1538-4357/aac5f6},
  \href {https://ui.adsabs.harvard.edu/abs/2018ApJ...861...63H} {861, 63}

\bibitem[\protect\citeauthoryear{{Hosseinzadeh} et~al.,}{{Hosseinzadeh}
  et~al.}{2022}]{Hosseinzadeh22}
{Hosseinzadeh} G.,  et~al., 2022, arXiv e-prints, \href
  {https://ui.adsabs.harvard.edu/abs/2022arXiv220308155H} {p. arXiv:2203.08155}

\bibitem[\protect\citeauthoryear{{Huang} et~al.,}{{Huang}
  et~al.}{2018}]{Huang18}
{Huang} F.,  et~al., 2018, \mn@doi [\mnras] {10.1093/mnras/sty066}, \href
  {https://ui.adsabs.harvard.edu/abs/2018MNRAS.475.3959H} {475, 3959}

\bibitem[\protect\citeauthoryear{{Humason}, {Kearns}  \& {Gomes}}{{Humason}
  et~al.}{1962}]{Humason62}
{Humason} M.~L.,  {Kearns} C.~E.,   {Gomes} A.~M.,  1962, \mn@doi [\pasp]
  {10.1086/127790}, \href
  {https://ui.adsabs.harvard.edu/abs/1962PASP...74..215H} {74, 215}

\bibitem[\protect\citeauthoryear{{Humphreys}}{{Humphreys}}{1978}]{Humphreys79}
{Humphreys} R.~M.,  1978, \mn@doi [\apjs] {10.1086/190559}, \href
  {https://ui.adsabs.harvard.edu/abs/1978ApJS...38..309H} {38, 309}

\bibitem[\protect\citeauthoryear{{Jacobson-Gal{\'a}n}
  et~al.,}{{Jacobson-Gal{\'a}n} et~al.}{2020}]{Jacobson-Galan20}
{Jacobson-Gal{\'a}n} W.~V.,  et~al., 2020, \mn@doi [\apj]
  {10.3847/1538-4357/ab9e66}, \href
  {https://ui.adsabs.harvard.edu/abs/2020ApJ...898..166J} {898, 166}

\bibitem[\protect\citeauthoryear{{Jacobson-Gal{\'a}n}
  et~al.,}{{Jacobson-Gal{\'a}n} et~al.}{2022}]{Jacobson-Galan22}
{Jacobson-Gal{\'a}n} W.~V.,  et~al., 2022, \mn@doi [\apj]
  {10.3847/1538-4357/ac3f3a}, \href
  {https://ui.adsabs.harvard.edu/abs/2022ApJ...924...15J} {924, 15}

\bibitem[\protect\citeauthoryear{{Jencson} et~al.,}{{Jencson}
  et~al.}{2022}]{Jencson22}
{Jencson} J.~E.,  et~al., 2022, \mn@doi [\apj] {10.3847/1538-4357/ac626c},
  \href {https://ui.adsabs.harvard.edu/abs/2022ApJ...930...81J} {930, 81}

\bibitem[\protect\citeauthoryear{{Jerkstrand}, {Fransson}, {Maguire}, {Smartt},
  {Ergon}  \& {Spyromilio}}{{Jerkstrand} et~al.}{2012}]{jerkstrand12}
{Jerkstrand} A.,  {Fransson} C.,  {Maguire} K.,  {Smartt} S.,  {Ergon} M.,
  {Spyromilio} J.,  2012, \mn@doi [\aap] {10.1051/0004-6361/201219528}, \href
  {https://ui.adsabs.harvard.edu/abs/2012A&A...546A..28J} {546, A28}

\bibitem[\protect\citeauthoryear{{Jerkstrand}, {Smartt}, {Fraser}, {Fransson},
  {Sollerman}, {Taddia}  \& {Kotak}}{{Jerkstrand} et~al.}{2014}]{jerkstrand14}
{Jerkstrand} A.,  {Smartt} S.~J.,  {Fraser} M.,  {Fransson} C.,  {Sollerman}
  J.,  {Taddia} F.,   {Kotak} R.,  2014, \mn@doi [\mnras]
  {10.1093/mnras/stu221}, \href
  {https://ui.adsabs.harvard.edu/abs/2014MNRAS.439.3694J} {439, 3694}

\bibitem[\protect\citeauthoryear{{Jerkstrand}, {Ertl}, {Janka}, {M{\"u}ller},
  {Sukhbold}  \& {Woosley}}{{Jerkstrand} et~al.}{2018}]{jerkstrand18}
{Jerkstrand} A.,  {Ertl} T.,  {Janka} H.~T.,  {M{\"u}ller} E.,  {Sukhbold} T.,
   {Woosley} S.~E.,  2018, \mn@doi [\mnras] {10.1093/mnras/stx2877}, \href
  {https://ui.adsabs.harvard.edu/abs/2018MNRAS.475..277J} {475, 277}

\bibitem[\protect\citeauthoryear{{Jones} et~al.,}{{Jones}
  et~al.}{2019}]{Jones19}
{Jones} D.~O.,  et~al., 2019, Transient Name Server AstroNote, \href
  {https://ui.adsabs.harvard.edu/abs/2019TNSAN.148....1J} {148, 1}

\bibitem[\protect\citeauthoryear{{Jones} et~al.,}{{Jones}
  et~al.}{2021}]{Jones2021}
{Jones} D.~O.,  et~al., 2021, \mn@doi [\apj] {10.3847/1538-4357/abd7f5}, \href
  {https://ui.adsabs.harvard.edu/abs/2021ApJ...908..143J} {908, 143}

\bibitem[\protect\citeauthoryear{{Kaiser} et~al.,}{{Kaiser}
  et~al.}{2002}]{Kaiser2002}
{Kaiser} N.,  et~al., 2002, in {Tyson} J.~A.,  {Wolff} S.,  eds,  Society of
  Photo-Optical Instrumentation Engineers (SPIE) Conference Series Vol. 4836,
  Survey and Other Telescope Technologies and Discoveries. pp 154--164,
  \mn@doi{10.1117/12.457365}

\bibitem[\protect\citeauthoryear{{Kennicutt}, {Tamblyn}  \&
  {Congdon}}{{Kennicutt} et~al.}{1994}]{Kennicutt}
{Kennicutt} Robert~C. J.,  {Tamblyn} P.,   {Congdon} C.~E.,  1994, \mn@doi
  [\apj] {10.1086/174790}, \href
  {https://ui.adsabs.harvard.edu/abs/1994ApJ...435...22K} {435, 22}

\bibitem[\protect\citeauthoryear{{Khazov} et~al.,}{{Khazov}
  et~al.}{2016}]{khazov+15}
{Khazov} D.,  et~al., 2016, \mn@doi [\apj] {10.3847/0004-637X/818/1/3}, \href
  {http://adsabs.harvard.edu/abs/2016ApJ...818....3K} {818, 3}

\bibitem[\protect\citeauthoryear{{Kilpatrick} \& {Foley}}{{Kilpatrick} \&
  {Foley}}{2018}]{Kilpatrick18:17eaw}
{Kilpatrick} C.~D.,  {Foley} R.~J.,  2018, \mn@doi [\mnras]
  {10.1093/mnras/sty2435}, \href
  {https://ui.adsabs.harvard.edu/abs/2018MNRAS.481.2536K} {481, 2536}

\bibitem[\protect\citeauthoryear{{Kilpatrick} et~al.,}{{Kilpatrick}
  et~al.}{2018}]{Kilpatrick18:17ejb}
{Kilpatrick} C.~D.,  et~al., 2018, \mn@doi [\mnras] {10.1093/mnras/sty2503},
  \href {https://ui.adsabs.harvard.edu/abs/2018MNRAS.481.4123K} {481, 4123}

\bibitem[\protect\citeauthoryear{{Kilpatrick} et~al.,}{{Kilpatrick}
  et~al.}{2021a}]{Kilpatrick21b}
{Kilpatrick} C.~D.,  et~al., 2021a, arXiv e-prints, \href
  {https://ui.adsabs.harvard.edu/abs/2021arXiv210906211K} {p. arXiv:2109.06211}

\bibitem[\protect\citeauthoryear{{Kilpatrick} et~al.,}{{Kilpatrick}
  et~al.}{2021b}]{Kilpatrick21}
{Kilpatrick} C.~D.,  et~al., 2021b, \mn@doi [\mnras] {10.1093/mnras/stab838},
  \href {https://ui.adsabs.harvard.edu/abs/2021MNRAS.504.2073K} {504, 2073}

\bibitem[\protect\citeauthoryear{{Kirshner}}{{Kirshner}}{1990}]{kirshner+90}
{Kirshner} R.~P.,  1990, in Supernovae. pp 59--75

\bibitem[\protect\citeauthoryear{{Kochanek}}{{Kochanek}}{2020}]{Kochanek20}
{Kochanek} C.~S.,  2020, \mn@doi [\mnras] {10.1093/mnras/staa605}, \href
  {https://ui.adsabs.harvard.edu/abs/2020MNRAS.493.4945K} {493, 4945}

\bibitem[\protect\citeauthoryear{{Kochanek}, {Khan}  \& {Dai}}{{Kochanek}
  et~al.}{2012}]{Kochanek12}
{Kochanek} C.~S.,  {Khan} R.,   {Dai} X.,  2012, \mn@doi [\apj]
  {10.1088/0004-637X/759/1/20}, \href
  {http://adsabs.harvard.edu/abs/2012ApJ...759...20K} {759, 20}

\bibitem[\protect\citeauthoryear{{Kochanek} et~al.,}{{Kochanek}
  et~al.}{2017}]{Kochanek17}
{Kochanek} C.~S.,  et~al., 2017, \mn@doi [\pasp] {10.1088/1538-3873/aa80d9},
  \href {https://ui.adsabs.harvard.edu/abs/2017PASP..129j4502K} {129, 104502}

\bibitem[\protect\citeauthoryear{{LSST Science Collaboration} et~al.,}{{LSST
  Science Collaboration} et~al.}{2009}]{LSST2009}
{LSST Science Collaboration} et~al., 2009, arXiv e-prints, \href
  {https://ui.adsabs.harvard.edu/abs/2009arXiv0912.0201L} {p. arXiv:0912.0201}

\bibitem[\protect\citeauthoryear{{Larkin} et~al.,}{{Larkin}
  et~al.}{2006}]{OSIRIS}
{Larkin} J.,  et~al., 2006, in {McLean} I.~S.,  {Iye} M.,  eds,  Society of
  Photo-Optical Instrumentation Engineers (SPIE) Conference Series Vol. 6269,
  Society of Photo-Optical Instrumentation Engineers (SPIE) Conference Series.
  p. 62691A, \mn@doi{10.1117/12.672061}

\bibitem[\protect\citeauthoryear{{Levesque} \& {Massey}}{{Levesque} \&
  {Massey}}{2020}]{Levesque20}
{Levesque} E.~M.,  {Massey} P.,  2020, \mn@doi [\apjl]
  {10.3847/2041-8213/ab7935}, \href
  {https://ui.adsabs.harvard.edu/abs/2020ApJ...891L..37L} {891, L37}

\bibitem[\protect\citeauthoryear{{Levesque}, {Massey}, {Olsen}, {Plez},
  {Meynet}  \& {Maeder}}{{Levesque} et~al.}{2006}]{Levesque06}
{Levesque} E.~M.,  {Massey} P.,  {Olsen} K.~A.~G.,  {Plez} B.,  {Meynet} G.,
  {Maeder} A.,  2006, \mn@doi [\apj] {10.1086/504417}, \href
  {https://ui.adsabs.harvard.edu/abs/2006ApJ...645.1102L} {645, 1102}

\bibitem[\protect\citeauthoryear{{Li}, {Wang}, {Van Dyk}, {Cuillandre}, {Foley}
   \& {Filippenko}}{{Li} et~al.}{2007}]{Li07}
{Li} W.,  {Wang} X.,  {Van Dyk} S.~D.,  {Cuillandre} J.-C.,  {Foley} R.~J.,
  {Filippenko} A.~V.,  2007, \mn@doi [\apj] {10.1086/516747}, \href
  {https://ui.adsabs.harvard.edu/abs/2007ApJ...661.1013L} {661, 1013}

\bibitem[\protect\citeauthoryear{{Magnier} et~al.,}{{Magnier}
  et~al.}{2013}]{magnier13}
{Magnier} E.~A.,  et~al., 2013, \mn@doi [\apjs] {10.1088/0067-0049/205/2/20},
  \href {https://ui.adsabs.harvard.edu/abs/2013ApJS..205...20M} {205, 20}

\bibitem[\protect\citeauthoryear{{Magrini} et~al.,}{{Magrini}
  et~al.}{2011}]{Magrini2011}
{Magrini} L.,  et~al., 2011, \mn@doi [\aap] {10.1051/0004-6361/201116872},
  \href {https://ui.adsabs.harvard.edu/abs/2011A&A...535A..13M} {535, A13}

\bibitem[\protect\citeauthoryear{{Marino} et~al.,}{{Marino}
  et~al.}{2013}]{Marino2013}
{Marino} R.~A.,  et~al., 2013, \mn@doi [\aap] {10.1051/0004-6361/201321956},
  \href {https://ui.adsabs.harvard.edu/abs/2013A&A...559A.114M} {559, A114}

\bibitem[\protect\citeauthoryear{{Martinez} et~al.,}{{Martinez}
  et~al.}{2022}]{Martinez22}
{Martinez} L.,  et~al., 2022, \mn@doi [\aap] {10.1051/0004-6361/202142555},
  \href {https://ui.adsabs.harvard.edu/abs/2022A&A...660A..42M} {660, A42}

\bibitem[\protect\citeauthoryear{{Massey} \& {Olsen}}{{Massey} \&
  {Olsen}}{2003}]{Massey03}
{Massey} P.,  {Olsen} K.~A.~G.,  2003, \mn@doi [\aj] {10.1086/379558}, \href
  {https://ui.adsabs.harvard.edu/abs/2003AJ....126.2867M} {126, 2867}

\bibitem[\protect\citeauthoryear{{Matzner} \& {McKee}}{{Matzner} \&
  {McKee}}{1999}]{Matzner99}
{Matzner} C.~D.,  {McKee} C.~F.,  1999, \mn@doi [\apj] {10.1086/306571}, \href
  {https://ui.adsabs.harvard.edu/abs/1999ApJ...510..379M} {510, 379}

\bibitem[\protect\citeauthoryear{{Maund} \& {Smartt}}{{Maund} \&
  {Smartt}}{2009}]{Maund09}
{Maund} J.~R.,  {Smartt} S.~J.,  2009, \mn@doi [Science]
  {10.1126/science.1170198}, \href
  {https://ui.adsabs.harvard.edu/abs/2009Sci...324..486M} {324, 486}

\bibitem[\protect\citeauthoryear{{Maund} et~al.,}{{Maund}
  et~al.}{2011}]{maund+11}
{Maund} J.~R.,  et~al., 2011, \mn@doi [\apjl] {10.1088/2041-8205/739/2/L37},
  \href {http://adsabs.harvard.edu/abs/2011ApJ...739L..37M} {739, L37}

\bibitem[\protect\citeauthoryear{{McCully}, {Volgenau}, {Harbeck}, {Lister},
  {Saunders}, {Turner}, {Siiverd}  \& {Bowman}}{{McCully}
  et~al.}{2018}]{McCully18}
{McCully} C.,  {Volgenau} N.~H.,  {Harbeck} D.-R.,  {Lister} T.~A.,  {Saunders}
  E.~S.,  {Turner} M.~L.,  {Siiverd} R.~J.,   {Bowman} M.,  2018, in {Guzman}
  J.~C.,  {Ibsen} J.,  eds,  Society of Photo-Optical Instrumentation Engineers
  (SPIE) Conference Series Vol. 10707, Software and Cyberinfrastructure for
  Astronomy V. p. 107070K (\mn@eprint {arXiv} {1811.04163}),
  \mn@doi{10.1117/12.2314340}

\bibitem[\protect\citeauthoryear{{Momose}, {Okumura}, {Koda}  \&
  {Sawada}}{{Momose} et~al.}{2010}]{Momose2010}
{Momose} R.,  {Okumura} S.~K.,  {Koda} J.,   {Sawada} T.,  2010, \mn@doi [\apj]
  {10.1088/0004-637X/721/1/383}, \href
  {https://ui.adsabs.harvard.edu/abs/2010ApJ...721..383M} {721, 383}

\bibitem[\protect\citeauthoryear{{Morozova}, {Piro}  \& {Valenti}}{{Morozova}
  et~al.}{2018}]{Morozova18}
{Morozova} V.,  {Piro} A.~L.,   {Valenti} S.,  2018, \mn@doi [\apj]
  {10.3847/1538-4357/aab9a6}, \href
  {https://ui.adsabs.harvard.edu/abs/2018ApJ...858...15M} {858, 15}

\bibitem[\protect\citeauthoryear{{Nasa High Energy Astrophysics Science Archive
  Research Center (Heasarc)}}{{Nasa High Energy Astrophysics Science Archive
  Research Center (Heasarc)}}{2014}]{heasoft}
{Nasa High Energy Astrophysics Science Archive Research Center (Heasarc)} 2014,
  {HEAsoft: Unified Release of FTOOLS and XANADU} (\mn@eprint {ascl}
  {1408.004})

\bibitem[\protect\citeauthoryear{{Negueruela}, {Gonz{\'a}lez-Fern{\'a}ndez},
  {Dorda}, {Marco}  \& {Clark}}{{Negueruela} et~al.}{2013}]{Negueruela13}
{Negueruela} I.,  {Gonz{\'a}lez-Fern{\'a}ndez} C.,  {Dorda} R.,  {Marco} A.,
  {Clark} J.~S.,  2013, in {Kervella} P.,  {Le Bertre} T.,   {Perrin} G.,  eds,
   EAS Publications Series Vol. 60, EAS Publications Series. pp 279--285
  (\mn@eprint {arXiv} {1303.1837}), \mn@doi{10.1051/eas/1360032}

\bibitem[\protect\citeauthoryear{{Neugent}}{{Neugent}}{2021}]{Neugent20}
{Neugent} K.~F.,  2021, \mn@doi [\apj] {10.3847/1538-4357/abd47b}, \href
  {https://ui.adsabs.harvard.edu/abs/2021ApJ...908...87N} {908, 87}

\bibitem[\protect\citeauthoryear{{Nicholl}}{{Nicholl}}{2018}]{Nicholl18}
{Nicholl} M.,  2018, \mn@doi [Research Notes of the American Astronomical
  Society] {10.3847/2515-5172/aaf799}, \href
  {https://ui.adsabs.harvard.edu/abs/2018RNAAS...2..230N} {2, 230}

\bibitem[\protect\citeauthoryear{{Nicholl} et~al.,}{{Nicholl}
  et~al.}{2016}]{Nicholl16}
{Nicholl} M.,  et~al., 2016, \mn@doi [\apj] {10.3847/0004-637X/826/1/39}, \href
  {https://ui.adsabs.harvard.edu/abs/2016ApJ...826...39N} {826, 39}

\bibitem[\protect\citeauthoryear{{Nordin} et~al.,}{{Nordin}
  et~al.}{2019}]{Nordin19}
{Nordin} J.,  et~al., 2019, \mn@doi [\aap] {10.1051/0004-6361/201935634}, \href
  {https://ui.adsabs.harvard.edu/abs/2019A&A...631A.147N} {631, A147}

\bibitem[\protect\citeauthoryear{{Nordin}, {Brinnel}, {Giomi}, {Santen},
  {Gal-Yam}, {Yaron}  \& {Schulze}}{{Nordin}
  et~al.}{2020}]{2020TNSTR1248....1N}
{Nordin} J.,  {Brinnel} V.,  {Giomi} M.,  {Santen} J.~V.,  {Gal-Yam} A.,
  {Yaron} O.,   {Schulze} S.,  2020, Transient Name Server Discovery Report,
  \href {https://ui.adsabs.harvard.edu/abs/2020TNSTR1248....1N} {2020-1248, 1}

\bibitem[\protect\citeauthoryear{{Nugent} et~al.,}{{Nugent}
  et~al.}{2006}]{Nugent+06}
{Nugent} P.,  et~al., 2006, \mn@doi [\apj] {10.1086/504413}, \href
  {https://ui.adsabs.harvard.edu/abs/2006ApJ...645..841N} {645, 841}

\bibitem[\protect\citeauthoryear{{O'Neill} et~al.,}{{O'Neill}
  et~al.}{2019}]{ONeill19}
{O'Neill} D.,  et~al., 2019, \mn@doi [\aap] {10.1051/0004-6361/201834566},
  \href {https://ui.adsabs.harvard.edu/abs/2019A&A...622L...1O} {622, L1}

\bibitem[\protect\citeauthoryear{{Pastorini} et~al.,}{{Pastorini}
  et~al.}{2007}]{Pastorini2007}
{Pastorini} G.,  et~al., 2007, \mn@doi [\aap] {10.1051/0004-6361:20066784},
  \href {https://ui.adsabs.harvard.edu/abs/2007A&A...469..405P} {469, 405}

\bibitem[\protect\citeauthoryear{{Perley}, {Barbarino}, {Sollerman},
  {Schweyer}, {Schulze}  \& {Yang}}{{Perley}
  et~al.}{2020}]{2020TNSCR1259....1P}
{Perley} D.,  {Barbarino} C.,  {Sollerman} J.,  {Schweyer} T.,  {Schulze} S.,
  {Yang} Y.,  2020, Transient Name Server Classification Report, \href
  {https://ui.adsabs.harvard.edu/abs/2020TNSCR1259....1P} {2020-1259, 1}

\bibitem[\protect\citeauthoryear{{Phillips} et~al.,}{{Phillips}
  et~al.}{2013}]{Phillips13}
{Phillips} M.~M.,  et~al., 2013, \mn@doi [\apj] {10.1088/0004-637X/779/1/38},
  \href {https://ui.adsabs.harvard.edu/abs/2013ApJ...779...38P} {779, 38}

\bibitem[\protect\citeauthoryear{{Pickles}}{{Pickles}}{1998}]{pickles+98}
{Pickles} A.~J.,  1998, \mn@doi [\pasp] {10.1086/316197}, \href
  {http://adsabs.harvard.edu/abs/1998PASP..110..863P} {110, 863}

\bibitem[\protect\citeauthoryear{{Podsiadlowski}, {Joss}  \&
  {Hsu}}{{Podsiadlowski} et~al.}{1992}]{Podsiadlowski1992}
{Podsiadlowski} P.,  {Joss} P.~C.,   {Hsu} J.~J.~L.,  1992, \mn@doi [\apj]
  {10.1086/171341}, \href
  {https://ui.adsabs.harvard.edu/abs/1992ApJ...391..246P} {391, 246}

\bibitem[\protect\citeauthoryear{{Poznanski}, {Prochaska}  \&
  {Bloom}}{{Poznanski} et~al.}{2012}]{Poznanski12}
{Poznanski} D.,  {Prochaska} J.~X.,   {Bloom} J.~S.,  2012, \mn@doi [\mnras]
  {10.1111/j.1365-2966.2012.21796.x}, \href
  {http://adsabs.harvard.edu/abs/2012MNRAS.426.1465P} {426, 1465}

\bibitem[\protect\citeauthoryear{{Rest} et~al.,}{{Rest} et~al.}{2005}]{rest+05}
{Rest} A.,  et~al., 2005, \mn@doi [\apj] {10.1086/497060}, \href
  {http://adsabs.harvard.edu/abs/2005ApJ...634.1103R} {634, 1103}

\bibitem[\protect\citeauthoryear{{Rest} et~al.,}{{Rest} et~al.}{2014}]{Rest14}
{Rest} A.,  et~al., 2014, \mn@doi [\apj] {10.1088/0004-637X/795/1/44}, \href
  {https://ui.adsabs.harvard.edu/abs/2014ApJ...795...44R} {795, 44}

\bibitem[\protect\citeauthoryear{{Rodr{\'\i}guez}, {Clocchiatti}  \&
  {Hamuy}}{{Rodr{\'\i}guez} et~al.}{2014}]{Rodriguez14}
{Rodr{\'\i}guez} {\'O}.,  {Clocchiatti} A.,   {Hamuy} M.,  2014, \mn@doi [\aj]
  {10.1088/0004-6256/148/6/107}, \href
  {https://ui.adsabs.harvard.edu/abs/2014AJ....148..107R} {148, 107}

\bibitem[\protect\citeauthoryear{{Roming} et~al.,}{{Roming}
  et~al.}{2005}]{Roming05}
{Roming} P. W.~A.,  et~al., 2005, \mn@doi [\ssr] {10.1007/s11214-005-5095-4},
  \href {https://ui.adsabs.harvard.edu/abs/2005SSRv..120...95R} {120, 95}

\bibitem[\protect\citeauthoryear{{Roy} et~al.,}{{Roy} et~al.}{2011}]{Roy11}
{Roy} R.,  et~al., 2011, \mn@doi [\apj] {10.1088/0004-637X/736/2/76}, \href
  {https://ui.adsabs.harvard.edu/abs/2011ApJ...736...76R} {736, 76}

\bibitem[\protect\citeauthoryear{{Ryder} et~al.,}{{Ryder}
  et~al.}{2018}]{Ryder18}
{Ryder} S.~D.,  et~al., 2018, \mn@doi [\apj] {10.3847/1538-4357/aaaf1e}, \href
  {https://ui.adsabs.harvard.edu/abs/2018ApJ...856...83R} {856, 83}

\bibitem[\protect\citeauthoryear{{Sana} et~al.,}{{Sana} et~al.}{2012}]{Sana12}
{Sana} H.,  et~al., 2012, \mn@doi [Science] {10.1126/science.1223344}, \href
  {https://ui.adsabs.harvard.edu/abs/2012Sci...337..444S} {337, 444}

\bibitem[\protect\citeauthoryear{{Sapir} \& {Waxman}}{{Sapir} \&
  {Waxman}}{2017}]{Sapir17}
{Sapir} N.,  {Waxman} E.,  2017, \mn@doi [\apj] {10.3847/1538-4357/aa64df},
  \href {https://ui.adsabs.harvard.edu/abs/2017ApJ...838..130S} {838, 130}

\bibitem[\protect\citeauthoryear{{Schechter}, {Mateo}  \& {Saha}}{{Schechter}
  et~al.}{1993}]{schechter+93}
{Schechter} P.~L.,  {Mateo} M.,   {Saha} A.,  1993, \mn@doi [\pasp]
  {10.1086/133316}, \href {http://adsabs.harvard.edu/abs/1993PASP..105.1342S}
  {105, 1342}

\bibitem[\protect\citeauthoryear{{Schlafly} \& {Finkbeiner}}{{Schlafly} \&
  {Finkbeiner}}{2011}]{Schlafly11}
{Schlafly} E.~F.,  {Finkbeiner} D.~P.,  2011, \mn@doi [\apj]
  {10.1088/0004-637X/737/2/103}, \href
  {http://adsabs.harvard.edu/abs/2011ApJ...737..103S} {737, 103}

\bibitem[\protect\citeauthoryear{{Schoeniger} \& {Sofue}}{{Schoeniger} \&
  {Sofue}}{1997}]{Schoeniger97}
{Schoeniger} F.,  {Sofue} Y.,  1997, \aap, \href
  {https://ui.adsabs.harvard.edu/abs/1997A&A...323...14S} {323, 14}

\bibitem[\protect\citeauthoryear{{Siebert}, {Dimitriadis}, {Polin}  \&
  {Foley}}{{Siebert} et~al.}{2020}]{Siebert20}
{Siebert} M.~R.,  {Dimitriadis} G.,  {Polin} A.,   {Foley} R.~J.,  2020,
  \mn@doi [\apjl] {10.3847/2041-8213/abae6e}, \href
  {https://ui.adsabs.harvard.edu/abs/2020ApJ...900L..27S} {900, L27}

\bibitem[\protect\citeauthoryear{{Silverman} et~al.,}{{Silverman}
  et~al.}{2017}]{silverman17}
{Silverman} J.~M.,  et~al., 2017, \mn@doi [\mnras] {10.1093/mnras/stx058},
  \href {https://ui.adsabs.harvard.edu/abs/2017MNRAS.467..369S} {467, 369}

\bibitem[\protect\citeauthoryear{{Smartt}}{{Smartt}}{2009}]{Smartt09}
{Smartt} S.~J.,  2009, \mn@doi [\araa] {10.1146/annurev-astro-082708-101737},
  \href {http://adsabs.harvard.edu/abs/2009ARA%26A..47...63S} {47, 63}

\bibitem[\protect\citeauthoryear{{Smartt}}{{Smartt}}{2015}]{smartt+15}
{Smartt} S.~J.,  2015, \mn@doi [\pasa] {10.1017/pasa.2015.17}, \href
  {http://adsabs.harvard.edu/abs/2015PASA...32...16S} {32, e016}

\bibitem[\protect\citeauthoryear{{Smartt}, {Eldridge}, {Crockett}  \&
  {Maund}}{{Smartt} et~al.}{2009}]{smartt+09a}
{Smartt} S.~J.,  {Eldridge} J.~J.,  {Crockett} R.~M.,   {Maund} J.~R.,  2009,
  \mn@doi [\mnras] {10.1111/j.1365-2966.2009.14506.x}, \href
  {http://adsabs.harvard.edu/abs/2009MNRAS.395.1409S} {395, 1409}

\bibitem[\protect\citeauthoryear{{Smartt} et~al.,}{{Smartt}
  et~al.}{2015}]{Smartt15}
{Smartt} S.~J.,  et~al., 2015, \mn@doi [\aap] {10.1051/0004-6361/201425237},
  \href {http://adsabs.harvard.edu/abs/2015A%26A...579A..40S} {579, A40}

\bibitem[\protect\citeauthoryear{{Smith}}{{Smith}}{2014}]{Smith14}
{Smith} N.,  2014, \mn@doi [\araa] {10.1146/annurev-astro-081913-040025}, \href
  {http://adsabs.harvard.edu/abs/2014ARA%26A..52..487S} {52, 487}

\bibitem[\protect\citeauthoryear{{Sollerman}, {Cumming}  \&
  {Lundqvist}}{{Sollerman} et~al.}{1998}]{Sollerman98}
{Sollerman} J.,  {Cumming} R.~J.,   {Lundqvist} P.,  1998, \mn@doi [\apj]
  {10.1086/305163}, \href
  {https://ui.adsabs.harvard.edu/abs/1998ApJ...493..933S} {493, 933}

\bibitem[\protect\citeauthoryear{{Sollerman} et~al.,}{{Sollerman}
  et~al.}{2021}]{Sollerman21}
{Sollerman} J.,  et~al., 2021, arXiv e-prints, \href
  {https://ui.adsabs.harvard.edu/abs/2021arXiv210714503S} {p. arXiv:2107.14503}

\bibitem[\protect\citeauthoryear{{Soraisam} et~al.,}{{Soraisam}
  et~al.}{2018}]{Soraisam18}
{Soraisam} M.~D.,  et~al., 2018, \mn@doi [\apj] {10.3847/1538-4357/aabc59},
  \href {https://ui.adsabs.harvard.edu/abs/2018ApJ...859...73S} {859, 73}

\bibitem[\protect\citeauthoryear{{Stritzinger} et~al.,}{{Stritzinger}
  et~al.}{2018}]{Stritzinger18}
{Stritzinger} M.~D.,  et~al., 2018, \mn@doi [\aap]
  {10.1051/0004-6361/201730842}, \href
  {https://ui.adsabs.harvard.edu/abs/2018A&A...609A.134S} {609, A134}

\bibitem[\protect\citeauthoryear{{Sukhbold}, {Ertl}, {Woosley}, {Brown}  \&
  {Janka}}{{Sukhbold} et~al.}{2016}]{Sukhbold16}
{Sukhbold} T.,  {Ertl} T.,  {Woosley} S.~E.,  {Brown} J.~M.,   {Janka} H.-T.,
  2016, \mn@doi [\apj] {10.3847/0004-637X/821/1/38}, \href
  {http://adsabs.harvard.edu/abs/2016ApJ...821...38S} {821, 38}

\bibitem[\protect\citeauthoryear{{Suntzeff} \& {Bouchet}}{{Suntzeff} \&
  {Bouchet}}{1990}]{Suntzeff90}
{Suntzeff} N.~B.,  {Bouchet} P.,  1990, \mn@doi [\aj] {10.1086/115358}, \href
  {https://ui.adsabs.harvard.edu/abs/1990AJ.....99..650S} {99, 650}

\bibitem[\protect\citeauthoryear{{Swift} et~al.,}{{Swift}
  et~al.}{2022}]{Swift22}
{Swift} J.~J.,  et~al., 2022, \mn@doi [\pasp] {10.1088/1538-3873/ac5aca}, \href
  {https://ui.adsabs.harvard.edu/abs/2022PASP..134c5005S} {134, 035005}

\bibitem[\protect\citeauthoryear{{Szalai}, {Zs{\'\i}ros}, {Fox}, {Pejcha}  \&
  {M{\"u}ller}}{{Szalai} et~al.}{2019}]{Szalai19}
{Szalai} T.,  {Zs{\'\i}ros} S.,  {Fox} O.~D.,  {Pejcha} O.,   {M{\"u}ller} T.,
  2019, \mn@doi [\apjs] {10.3847/1538-4365/ab10df}, \href
  {https://ui.adsabs.harvard.edu/abs/2019ApJS..241...38S} {241, 38}

\bibitem[\protect\citeauthoryear{{Szalai} et~al.,}{{Szalai}
  et~al.}{2021}]{Szalai21}
{Szalai} T.,  et~al., 2021, \mn@doi [\apj] {10.3847/1538-4357/ac0e2b}, \href
  {https://ui.adsabs.harvard.edu/abs/2021ApJ...919...17S} {919, 17}

\bibitem[\protect\citeauthoryear{{Tartaglia} et~al.,}{{Tartaglia}
  et~al.}{2021}]{Tartaglia21}
{Tartaglia} L.,  et~al., 2021, \mn@doi [\apj] {10.3847/1538-4357/abca8a}, \href
  {https://ui.adsabs.harvard.edu/abs/2021ApJ...907...52T} {907, 52}

\bibitem[\protect\citeauthoryear{{Teja}, {Singh}, {Sahu}, {Anupama}, {Kumar}
  \& {Nayana A.~J.}}{{Teja} et~al.}{2022}]{Teja22}
{Teja} R.~S.,  {Singh} A.,  {Sahu} D.~K.,  {Anupama} G.~C.,  {Kumar} B.,
  {Nayana A.~J.} 2022, arXiv e-prints, \href
  {https://ui.adsabs.harvard.edu/abs/2022arXiv220209412T} {p. arXiv:2202.09412}

\bibitem[\protect\citeauthoryear{{Teplitz}, {Capak}, {Brooke}, {Shenoy},
  {Brinkworth}, {Desai}, {Khan}  \& {Laher}}{{Teplitz} et~al.}{2010}]{SEIP}
{Teplitz} H.~I.,  {Capak} P.,  {Brooke} T.,  {Shenoy} S.,  {Brinkworth} C.,
  {Desai} V.,  {Khan} I.,   {Laher} R.,  2010, in {Mizumoto} Y.,  {Morita}
  K.~I.,   {Ohishi} M.,  eds,  Astronomical Society of the Pacific Conference
  Series Vol. 434, Astronomical Data Analysis Software and Systems XIX. p.~437

\bibitem[\protect\citeauthoryear{{Terreran} et~al.,}{{Terreran}
  et~al.}{2016}]{terreran2016}
{Terreran} G.,  et~al., 2016, \mn@doi [\mnras] {10.1093/mnras/stw1591}, \href
  {https://ui.adsabs.harvard.edu/abs/2016MNRAS.462..137T} {462, 137}

\bibitem[\protect\citeauthoryear{{Terreran} et~al.,}{{Terreran}
  et~al.}{2022}]{terreran+2022}
{Terreran} G.,  et~al., 2022, \mn@doi [\apj] {10.3847/1538-4357/ac3820}, \href
  {https://ui.adsabs.harvard.edu/abs/2022ApJ...926...20T} {926, 20}

\bibitem[\protect\citeauthoryear{{Tinyanont}, {Ridden-Harper}, {Foley},
  {Morozova}, {Kilpatrick}, {Dimitriadis}, {Rest}  \& {Wang}}{{Tinyanont}
  et~al.}{2021}]{Tinyanont21}
{Tinyanont} S.,  {Ridden-Harper} R.,  {Foley} R.~J.,  {Morozova} V.,
  {Kilpatrick} C.~D.,  {Dimitriadis} G.,  {Rest} A.,   {Wang} Q.,  2021, \mnras

\bibitem[\protect\citeauthoryear{{Tissera}, {Machado}, {Sanchez-Blazquez},
  {Pedrosa}, {S{\'a}nchez}, {Snaith}  \& {Vilchez}}{{Tissera}
  et~al.}{2016}]{Tissera16}
{Tissera} P.~B.,  {Machado} R. E.~G.,  {Sanchez-Blazquez} P.,  {Pedrosa} S.~E.,
   {S{\'a}nchez} S.~F.,  {Snaith} O.,   {Vilchez} J.,  2016, \mn@doi [\aap]
  {10.1051/0004-6361/201628188}, \href
  {https://ui.adsabs.harvard.edu/abs/2016A&A...592A..93T} {592, A93}

\bibitem[\protect\citeauthoryear{{Tomasella} et~al.,}{{Tomasella}
  et~al.}{2013}]{Tomasella13}
{Tomasella} L.,  et~al., 2013, \mn@doi [\mnras] {10.1093/mnras/stt1130}, \href
  {https://ui.adsabs.harvard.edu/abs/2013MNRAS.434.1636T} {434, 1636}

\bibitem[\protect\citeauthoryear{{Tonry}}{{Tonry}}{2011}]{Tonry11}
{Tonry} J.~L.,  2011, \mn@doi [\pasp] {10.1086/657997}, \href
  {https://ui.adsabs.harvard.edu/abs/2011PASP..123...58T} {123, 58}

\bibitem[\protect\citeauthoryear{{Valenti} et~al.,}{{Valenti}
  et~al.}{2016}]{Valenti16}
{Valenti} S.,  et~al., 2016, \mn@doi [\mnras] {10.1093/mnras/stw870}, \href
  {https://ui.adsabs.harvard.edu/abs/2016MNRAS.459.3939V} {459, 3939}

\bibitem[\protect\citeauthoryear{{Van Dyk} et~al.,}{{Van Dyk}
  et~al.}{2011}]{van-dyk+11}
{Van Dyk} S.~D.,  et~al., 2011, \mn@doi [\apjl] {10.1088/2041-8205/741/2/L28},
  \href {http://adsabs.harvard.edu/abs/2011ApJ...741L..28V} {741, L28}

\bibitem[\protect\citeauthoryear{{Van Dyk} et~al.,}{{Van Dyk}
  et~al.}{2014}]{van-dyk+14}
{Van Dyk} S.~D.,  et~al., 2014, \mn@doi [\aj] {10.1088/0004-6256/147/2/37},
  \href {http://adsabs.harvard.edu/abs/2014AJ....147...37V} {147, 37}

\bibitem[\protect\citeauthoryear{{Verhoelst}, {van der Zypen}, {Hony}, {Decin},
  {Cami}  \& {Eriksson}}{{Verhoelst} et~al.}{2009}]{verhoelst+09}
{Verhoelst} T.,  {van der Zypen} N.,  {Hony} S.,  {Decin} L.,  {Cami} J.,
  {Eriksson} K.,  2009, \mn@doi [\aap] {10.1051/0004-6361/20079063}, \href
  {http://adsabs.harvard.edu/abs/2009A%26A...498..127V} {498, 127}

\bibitem[\protect\citeauthoryear{{Walmswell} \& {Eldridge}}{{Walmswell} \&
  {Eldridge}}{2012}]{Walmswell12}
{Walmswell} J.~J.,  {Eldridge} J.~J.,  2012, \mn@doi [\mnras]
  {10.1111/j.1365-2966.2011.19860.x}, \href
  {https://ui.adsabs.harvard.edu/abs/2012MNRAS.419.2054W} {419, 2054}

\bibitem[\protect\citeauthoryear{{Weil}, {Fesen}, {Patnaude}  \&
  {Milisavljevic}}{{Weil} et~al.}{2020}]{Weil20}
{Weil} K.~E.,  {Fesen} R.~A.,  {Patnaude} D.~J.,   {Milisavljevic} D.,  2020,
  \mn@doi [\apj] {10.3847/1538-4357/aba4b1}, \href
  {https://ui.adsabs.harvard.edu/abs/2020ApJ...900...11W} {900, 11}

\bibitem[\protect\citeauthoryear{{Williams} et~al.,}{{Williams}
  et~al.}{2021}]{Williams2021}
{Williams} T.~G.,  et~al., 2021, \mn@doi [\aj] {10.3847/1538-3881/abe243},
  \href {https://ui.adsabs.harvard.edu/abs/2021AJ....161..185W} {161, 185}

\bibitem[\protect\citeauthoryear{{Wizinowich} et~al.,}{{Wizinowich}
  et~al.}{2006}]{LGSAO}
{Wizinowich} P.~L.,  et~al., 2006, \mn@doi [\pasp] {10.1086/499290}, \href
  {https://ui.adsabs.harvard.edu/abs/2006PASP..118..297W} {118, 297}

\bibitem[\protect\citeauthoryear{{Woosley} \& {Heger}}{{Woosley} \&
  {Heger}}{2007}]{woosley+07}
{Woosley} S.~E.,  {Heger} A.,  2007, \mn@doi [\physrep]
  {10.1016/j.physrep.2007.02.009}, \href
  {http://adsabs.harvard.edu/abs/2007PhR...442..269W} {442, 269}

\bibitem[\protect\citeauthoryear{{Woosley}, {Pinto}  \& {Ensman}}{{Woosley}
  et~al.}{1988}]{Woosley88}
{Woosley} S.~E.,  {Pinto} P.~A.,   {Ensman} L.,  1988, \mn@doi [\apj]
  {10.1086/165908}, \href
  {https://ui.adsabs.harvard.edu/abs/1988ApJ...324..466W} {324, 466}

\bibitem[\protect\citeauthoryear{{Woosley}, {Pinto}  \& {Hartmann}}{{Woosley}
  et~al.}{1989}]{Woosley89}
{Woosley} S.~E.,  {Pinto} P.~A.,   {Hartmann} D.,  1989, \mn@doi [\apj]
  {10.1086/168019}, \href
  {https://ui.adsabs.harvard.edu/abs/1989ApJ...346..395W} {346, 395}

\bibitem[\protect\citeauthoryear{{Yajima} et~al.,}{{Yajima}
  et~al.}{2019}]{Yajima2019}
{Yajima} Y.,  et~al., 2019, \mn@doi [\pasj] {10.1093/pasj/psz022}, \href
  {https://ui.adsabs.harvard.edu/abs/2019PASJ...71S..13Y} {71, S13}

\bibitem[\protect\citeauthoryear{{Yaron} et~al.,}{{Yaron}
  et~al.}{2017}]{Yaron17}
{Yaron} O.,  et~al., 2017, \mn@doi [Nature Physics] {10.1038/nphys4025}, \href
  {https://ui.adsabs.harvard.edu/abs/2017NatPh..13..510Y} {13, 510}

\bibitem[\protect\citeauthoryear{{Zapartas} et~al.,}{{Zapartas}
  et~al.}{2019}]{Zapartas19}
{Zapartas} E.,  et~al., 2019, \mn@doi [\aap] {10.1051/0004-6361/201935854},
  \href {https://ui.adsabs.harvard.edu/abs/2019A&A...631A...5Z} {631, A5}

\bibitem[\protect\citeauthoryear{{Zapartas}, {de Mink}, {Justham}, {Smith},
  {Renzo}  \& {de Koter}}{{Zapartas} et~al.}{2021}]{Zapartas21}
{Zapartas} E.,  {de Mink} S.~E.,  {Justham} S.,  {Smith} N.,  {Renzo} M.,   {de
  Koter} A.,  2021, \mn@doi [\aap] {10.1051/0004-6361/202037744}, \href
  {https://ui.adsabs.harvard.edu/abs/2021A&A...645A...6Z} {645, A6}

\bibitem[\protect\citeauthoryear{{de Jaeger} et~al.,}{{de Jaeger}
  et~al.}{2018}]{deJaeger18}
{de Jaeger} T.,  et~al., 2018, \mn@doi [\mnras] {10.1093/mnras/sty508}, \href
  {https://ui.adsabs.harvard.edu/abs/2018MNRAS.476.4592D} {476, 4592}

\bibitem[\protect\citeauthoryear{{de Jaeger} et~al.,}{{de Jaeger}
  et~al.}{2019}]{deJaeger19}
{de Jaeger} T.,  et~al., 2019, \mn@doi [\mnras] {10.1093/mnras/stz2714}, \href
  {https://ui.adsabs.harvard.edu/abs/2019MNRAS.490.2799D} {490, 2799}

\makeatother
\end{thebibliography}

\clearpage

\appendix

\onecolumn

\begin{longtable}{ccccc}
    \hline
    MJD  & Filter & Magnitude  & Uncertainty & Source \\
             &        & (AB mag)   & (mag)       &         \\
     \hline\hline
58976.095 & UVW1 & 14.855 & 0.059 & Swift \\
58976.097 & $U_S$ & 14.357 & 0.054 & Swift \\
58976.098 & $B$ & 14.955 & 0.045 & Swift \\
58976.099 & UVW2 & 15.025 & 0.067 & Swift \\
58976.102 & $V$ & 14.842 & 0.064 & Swift \\
58976.103 & UVM2 & 14.954 & 0.059 & Swift \\
58976.111 & $r^{\prime}$ & 15.077 & 0.015 & Las Cumbres \\
58976.156 & $g$ & 14.695 & 0.027 & Thacher \\
58976.161 & $r$ & 15.063 & 0.016 & Thacher \\
58976.164 & $i$ & 15.270 & 0.020 & Thacher \\
58976.170 & $z$ & 15.550 & 0.040 & Thacher \\
58976.174 & $V$ & 14.901 & 0.019 & Thacher \\
58976.300 & $g^{\prime}$ & 14.516 & 0.044 & Las Cumbres \\
58976.301 & $r^{\prime}$ & 15.043 & 0.076 & Las Cumbres \\
58976.302 & $i^{\prime}$ & 15.278 & 0.020 & Las Cumbres \\
58977.177 & $g$ & 14.542 & 0.013 & Thacher \\
58977.183 & $r$ & 14.773 & 0.011 & Thacher \\
58977.185 & $i$ & 14.939 & 0.011 & Thacher \\
58977.190 & $z$ & 15.143 & 0.029 & Thacher \\
58977.197 & $V$ & 14.619 & 0.013 & Thacher \\
58977.545 & $B$ & 14.695 & 0.023 & Lulin \\
58977.546 & $V$ & 14.470 & 0.017 & Lulin \\
58977.547 & $g$ & 14.555 & 0.020 & Lulin \\
58977.548 & $r$ & 14.759 & 0.013 & Lulin \\
58977.549 & $i$ & 14.884 & 0.014 & Lulin \\
58977.699 & UVW1 & 14.895 & 0.059 & Swift \\
58977.700 & $U_S$ & 14.246 & 0.054 & Swift \\
58977.700 & $B$ & 14.763 & 0.055 & Swift \\
58977.701 & UVW2 & 15.055 & 0.067 & Swift \\
58977.702 & $V$ & 14.432 & 0.063 & Swift \\
58977.702 & UVM2 & 14.904 & 0.059 & Swift \\
58977.730 & $u^{\prime}$ & 14.378 & 0.021 & Las Cumbres \\
58977.732 & $g^{\prime}$ & 14.422 & 0.012 & Las Cumbres \\
58977.733 & $r^{\prime}$ & 14.651 & 0.011 & Las Cumbres \\
58977.734 & $i^{\prime}$ & 14.849 & 0.010 & Las Cumbres \\
58978.176 & $g$ & 14.466 & 0.013 & Thacher \\
58978.182 & $r$ & 14.617 & 0.012 & Thacher \\
58978.185 & $i$ & 14.755 & 0.010 & Thacher \\
58978.193 & $z$ & 14.925 & 0.016 & Thacher \\
58978.196 & $V$ & 14.416 & 0.012 & Thacher \\
58978.228 & UVW1 & 14.966 & 0.059 & Swift \\
58978.230 & $U_S$ & 14.396 & 0.054 & Swift \\
58978.231 & $B$ & 14.763 & 0.055 & Swift \\
58978.232 & UVW2 & 15.397 & 0.068 & Swift \\
58978.235 & $V$ & 14.442 & 0.063 & Swift \\
58978.236 & UVM2 & 15.055 & 0.059 & Swift \\
58978.493 & $B$ & 14.712 & 0.019 & Lulin \\
58978.494 & $V$ & 14.454 & 0.012 & Lulin \\
58978.495 & $g$ & 14.492 & 0.014 & Lulin \\
58978.496 & $r$ & 14.605 & 0.009 & Lulin \\
58978.497 & $i$ & 14.769 & 0.010 & Lulin \\
58979.177 & $g$ & 14.449 & 0.011 & Thacher \\
58979.183 & $r$ & 14.542 & 0.009 & Thacher \\
58979.190 & $i$ & 14.672 & 0.009 & Thacher \\
58979.191 & $z$ & 14.805 & 0.014 & Thacher \\
58979.196 & $V$ & 14.486 & 0.011 & Thacher \\
58979.330 & $r$ & 14.681 & 0.003 & Pan-STARRS1 \\
58979.350 & UVW1 & 15.127 & 0.059 & Swift \\
58979.352 & $U_S$ & 14.506 & 0.054 & Swift \\
58979.353 & $B$ & 14.712 & 0.054 & Swift \\
58979.353 & UVW2 & 15.598 & 0.077 & Swift \\
58979.356 & $V$ & 14.450 & 0.062 & Swift \\
58979.357 & UVM2 & 15.286 & 0.059 & Swift \\
58979.365 & $i$ & 14.653 & 0.040 & Nickel \\
58979.410 & $u^{\prime}$ & 14.504 & 0.017 & Las Cumbres \\
58979.412 & $g^{\prime}$ & 14.435 & 0.009 & Las Cumbres \\
58979.413 & $r^{\prime}$ & 14.548 & 0.008 & Las Cumbres \\
58979.414 & $i^{\prime}$ & 14.672 & 0.008 & Las Cumbres \\
58979.417 & UVW1 & 15.308 & 0.059 & Swift \\
58979.418 & $U_S$ & 14.457 & 0.064 & Swift \\
58979.419 & $B$ & 14.714 & 0.064 & Swift \\
58979.419 & UVW2 & 15.709 & 0.077 & Swift \\
58979.421 & $V$ & 14.473 & 0.073 & Swift \\
58979.422 & UVM2 & 15.366 & 0.059 & Swift \\
58980.216 & UVW1 & 15.147 & 0.059 & Swift \\
58980.218 & $U_S$ & 14.506 & 0.054 & Swift \\
58980.219 & $B$ & 14.682 & 0.054 & Swift \\
58980.220 & UVW2 & 16.052 & 0.077 & Swift \\
58980.224 & $V$ & 14.430 & 0.053 & Swift \\
58980.225 & UVM2 & 15.608 & 0.059 & Swift \\
58980.511 & $B$ & 14.709 & 0.014 & Lulin \\
58980.512 & $V$ & 14.524 & 0.012 & Lulin \\
58980.513 & $g$ & 14.429 & 0.014 & Lulin \\
58980.514 & $r$ & 14.496 & 0.010 & Lulin \\
58980.515 & $i$ & 14.621 & 0.009 & Lulin \\
58980.720 & $u^{\prime}$ & 14.573 & 0.033 & Las Cumbres \\
58980.722 & $g^{\prime}$ & 14.384 & 0.015 & Las Cumbres \\
58980.723 & $r^{\prime}$ & 14.484 & 0.014 & Las Cumbres \\
58980.724 & $i^{\prime}$ & 14.624 & 0.015 & Las Cumbres \\
58981.180 & $g$ & 14.422 & 0.012 & Thacher \\
58981.183 & $r$ & 14.511 & 0.010 & Thacher \\
58981.190 & $i$ & 14.596 & 0.010 & Thacher \\
58981.196 & $z$ & 14.741 & 0.014 & Thacher \\
58981.616 & UVW1 & 15.630 & 0.059 & Swift \\
58981.617 & $U_S$ & 14.667 & 0.054 & Swift \\
58981.617 & $B$ & 14.734 & 0.055 & Swift \\
58981.617 & UVW2 & 16.529 & 0.078 & Swift \\
58981.619 & $V$ & 14.473 & 0.063 & Swift \\
58981.619 & UVM2 & 16.194 & 0.059 & Swift \\
58982.118 & $R$ & 14.500 & 0.031 & Auburn \\
58982.145 & $G$ & 14.434 & 0.032 & Auburn \\
58982.166 & $B$ & 14.858 & 0.041 & Auburn \\
58982.180 & $g$ & 14.436 & 0.011 & Thacher \\
58982.183 & $r$ & 14.514 & 0.009 & Thacher \\
58982.191 & $i$ & 14.659 & 0.009 & Thacher \\
58982.195 & $z$ & 14.763 & 0.014 & Thacher \\
58982.197 & $V$ & 14.449 & 0.012 & Thacher \\
58982.412 & UVW1 & 15.711 & 0.059 & Swift \\
58982.414 & $U_S$ & 14.727 & 0.054 & Swift \\
58982.415 & $B$ & 14.783 & 0.055 & Swift \\
58982.415 & UVW2 & 16.651 & 0.078 & Swift \\
58982.418 & $V$ & 14.452 & 0.063 & Swift \\
58982.419 & UVM2 & 16.336 & 0.068 & Swift \\
58982.705 & $u^{\prime}$ & 14.767 & 0.021 & Las Cumbres \\
58982.707 & $g^{\prime}$ & 14.417 & 0.300 & Las Cumbres \\
58983.207 & $g$ & 14.485 & 0.012 & Thacher \\
58983.212 & $r$ & 14.518 & 0.009 & Thacher \\
58983.218 & $i$ & 14.675 & 0.010 & Thacher \\
58983.221 & $z$ & 14.822 & 0.015 & Thacher \\
58983.229 & $V$ & 14.499 & 0.011 & Thacher \\
58983.280 & $g$ & 14.556 & 0.003 & Pan-STARRS1 \\
58983.290 & $z$ & 14.805 & 0.004 & Pan-STARRS1 \\
58984.116 & $R$ & 14.481 & 0.051 & Auburn \\
58984.162 & $B$ & 14.812 & 0.082 & Auburn \\
58984.181 & $g$ & 14.475 & 0.014 & Thacher \\
58984.189 & $i$ & 14.691 & 0.012 & Thacher \\
58984.196 & $z$ & 14.863 & 0.018 & Thacher \\
58984.200 & $V$ & 14.516 & 0.014 & Thacher \\
58984.666 & UVW1 & 16.187 & 0.068 & Swift \\
58984.667 & $U_S$ & 15.009 & 0.064 & Swift \\
58984.668 & $B$ & 14.803 & 0.055 & Swift \\
58984.668 & UVW2 & 17.071 & 0.078 & Swift \\
58984.671 & $V$ & 14.506 & 0.073 & Swift \\
58984.671 & UVM2 & 16.978 & 0.070 & Swift \\
58984.718 & $u^{\prime}$ & 14.973 & 0.019 & Las Cumbres \\
58984.720 & $g^{\prime}$ & 14.475 & 0.010 & Las Cumbres \\
58984.721 & $r^{\prime}$ & 14.513 & 0.009 & Las Cumbres \\
58984.722 & $i^{\prime}$ & 14.736 & 0.009 & Las Cumbres \\
58985.184 & $g$ & 14.515 & 0.013 & Thacher \\
58985.185 & $r$ & 14.527 & 0.011 & Thacher \\
58985.193 & $i$ & 14.733 & 0.010 & Thacher \\
58985.196 & $z$ & 14.839 & 0.015 & Thacher \\
58985.203 & $V$ & 14.553 & 0.012 & Thacher \\
58985.662 & UVW1 & 16.340 & 0.078 & Swift \\
58985.663 & $U_S$ & 15.099 & 0.064 & Swift \\
58985.663 & $B$ & 14.844 & 0.064 & Swift \\
58985.664 & UVW2 & 17.350 & 0.089 & Swift \\
58985.666 & $V$ & 14.554 & 0.083 & Swift \\
58985.666 & UVM2 & 17.277 & 0.080 & Swift \\
58986.181 & $g$ & 14.540 & 0.011 & Thacher \\
58986.188 & $r$ & 14.542 & 0.009 & Thacher \\
58986.194 & $i$ & 14.658 & 0.036 & Thacher \\
58986.197 & $z$ & 14.815 & 0.016 & Thacher \\
58986.200 & $V$ & 14.566 & 0.011 & Thacher \\
58986.270 & $g$ & 14.665 & 0.003 & Pan-STARRS1 \\
58986.476 & $u^{\prime}$ & 15.255 & 0.018 & Las Cumbres \\
58986.478 & $g^{\prime}$ & 14.575 & 0.013 & Las Cumbres \\
58986.740 & $u^{\prime}$ & 15.238 & 0.024 & Las Cumbres \\
58986.742 & $g^{\prime}$ & 14.538 & 0.013 & Las Cumbres \\
58986.743 & $r^{\prime}$ & 14.526 & 0.012 & Las Cumbres \\
58986.744 & $i^{\prime}$ & 14.699 & 0.013 & Las Cumbres \\
58988.185 & $g$ & 14.612 & 0.012 & Thacher \\
58988.190 & $r$ & 14.552 & 0.010 & Thacher \\
58988.194 & $i$ & 14.739 & 0.010 & Thacher \\
58988.199 & $z$ & 14.827 & 0.015 & Thacher \\
58988.205 & $V$ & 14.507 & 0.011 & Thacher \\
58988.538 & $u^{\prime}$ & 15.528 & 0.023 & Las Cumbres \\
58988.540 & $g^{\prime}$ & 14.626 & 0.014 & Las Cumbres \\
58988.541 & $r^{\prime}$ & 14.559 & 0.014 & Las Cumbres \\
58988.542 & $i^{\prime}$ & 14.770 & 0.014 & Las Cumbres \\
58989.218 & $g$ & 14.604 & 0.013 & Thacher \\
58989.225 & $r$ & 14.538 & 0.010 & Thacher \\
58989.228 & $i$ & 14.711 & 0.010 & Thacher \\
58989.233 & $z$ & 14.815 & 0.015 & Thacher \\
58989.241 & $V$ & 14.569 & 0.012 & Thacher \\
58989.864 & $B$ & 15.086 & 0.015 & Las Cumbres \\
58989.865 & $V$ & 14.546 & 0.011 & Las Cumbres \\
58989.867 & $r^{\prime}$ & 14.485 & 0.009 & Las Cumbres \\
58989.888 & $u^{\prime}$ & 15.693 & 0.019 & Las Cumbres \\
58989.890 & $g^{\prime}$ & 14.587 & 0.010 & Las Cumbres \\
58989.891 & $r^{\prime}$ & 14.509 & 0.008 & Las Cumbres \\
58989.892 & $i^{\prime}$ & 14.729 & 0.008 & Las Cumbres \\
58990.185 & $g$ & 14.595 & 0.013 & Thacher \\
58990.193 & $r$ & 14.540 & 0.010 & Thacher \\
58990.194 & $i$ & 14.736 & 0.010 & Thacher \\
58990.201 & $z$ & 14.786 & 0.014 & Thacher \\
58990.204 & $V$ & 14.582 & 0.012 & Thacher \\
58990.270 & $g$ & 14.668 & 0.003 & Pan-STARRS1 \\
58991.186 & $g$ & 14.622 & 0.010 & Thacher \\
58991.189 & $r$ & 14.503 & 0.009 & Thacher \\
58991.194 & $i$ & 14.694 & 0.009 & Thacher \\
58991.198 & $z$ & 14.794 & 0.013 & Thacher \\
58991.203 & $V$ & 14.562 & 0.011 & Thacher \\
58991.698 & $u^{\prime}$ & 15.881 & 0.021 & Las Cumbres \\
58991.700 & $g^{\prime}$ & 14.605 & 0.011 & Las Cumbres \\
58991.701 & $r^{\prime}$ & 14.494 & 0.009 & Las Cumbres \\
58991.702 & $i^{\prime}$ & 14.713 & 0.009 & Las Cumbres \\
58992.185 & $g$ & 14.671 & 0.012 & Thacher \\
58992.191 & $r$ & 14.536 & 0.010 & Thacher \\
58992.194 & $i$ & 14.692 & 0.010 & Thacher \\
58992.199 & $z$ & 14.808 & 0.015 & Thacher \\
58992.204 & $V$ & 14.572 & 0.011 & Thacher \\
58993.422 & $u^{\prime}$ & 16.083 & 0.019 & Las Cumbres \\
58993.424 & $g^{\prime}$ & 14.738 & 0.009 & Las Cumbres \\
58993.425 & $r^{\prime}$ & 14.556 & 0.008 & Las Cumbres \\
58993.426 & $i^{\prime}$ & 14.753 & 0.008 & Las Cumbres \\
58993.905 & UVW1 & 18.452 & 0.130 & Swift \\
58993.906 & $U_S$ & 16.135 & 0.075 & Swift \\
58993.907 & $B$ & 15.220 & 0.064 & Swift \\
58993.908 & UVW2 & $>$19.394 & -- & Swift \\
58993.910 & $V$ & 14.624 & 0.063 & Swift \\
58993.911 & UVM2 & $>$20.177 & -- & Swift \\
58994.189 & $g$ & 14.716 & 0.011 & Thacher \\
58994.194 & $r$ & 14.559 & 0.009 & Thacher \\
58994.196 & $i$ & 14.692 & 0.010 & Thacher \\
58994.203 & $z$ & 14.806 & 0.015 & Thacher \\
58994.207 & $V$ & 14.611 & 0.012 & Thacher \\
58995.188 & $g$ & 14.770 & 0.011 & Thacher \\
58995.194 & $r$ & 14.548 & 0.009 & Thacher \\
58995.196 & $i$ & 14.700 & 0.010 & Thacher \\
58995.202 & $z$ & 14.809 & 0.014 & Thacher \\
58995.205 & $V$ & 14.710 & 0.011 & Thacher \\
58995.552 & UVW1 & 18.727 & 0.147 & Swift \\
58995.553 & $U_S$ & 16.267 & 0.086 & Swift \\
58995.554 & $B$ & 15.301 & 0.065 & Swift \\
58995.555 & UVW2 & 19.915 & 0.235 & Swift \\
58995.557 & $V$ & 14.565 & 0.073 & Swift \\
58995.558 & UVM2 & $>$20.378 & -- & Swift \\
58996.268 & $g$ & 14.824 & 0.011 & Thacher \\
58996.272 & $r$ & 14.550 & 0.009 & Thacher \\
58996.279 & $i$ & 14.710 & 0.009 & Thacher \\
58996.282 & $z$ & 14.812 & 0.013 & Thacher \\
58996.290 & $V$ & 14.629 & 0.011 & Thacher \\
58996.528 & $u^{\prime}$ & 16.408 & 0.039 & Las Cumbres \\
58996.530 & $g^{\prime}$ & 14.848 & 0.017 & Las Cumbres \\
58996.531 & $r^{\prime}$ & 14.663 & 0.030 & Las Cumbres \\
58999.723 & $u^{\prime}$ & 16.711 & 0.021 & Las Cumbres \\
58999.725 & $g^{\prime}$ & 14.882 & 0.012 & Las Cumbres \\
58999.726 & $r^{\prime}$ & 14.561 & 0.010 & Las Cumbres \\
58999.727 & $i^{\prime}$ & 14.694 & 0.009 & Las Cumbres \\
59000.269 & $g$ & 14.914 & 0.017 & Thacher \\
59000.275 & $r$ & 14.585 & 0.015 & Thacher \\
59000.278 & $i$ & 14.686 & 0.017 & Thacher \\
59000.283 & $z$ & 14.766 & 0.028 & Thacher \\
59000.291 & $V$ & 14.687 & 0.019 & Thacher \\
59001.273 & $g$ & 14.918 & 0.025 & Thacher \\
59001.277 & $r$ & 14.564 & 0.020 & Thacher \\
59001.280 & $i$ & 14.668 & 0.027 & Thacher \\
59001.286 & $z$ & 14.742 & 0.059 & Thacher \\
59001.289 & $V$ & 14.610 & 0.029 & Thacher \\
59002.246 & $g$ & 14.953 & 0.015 & Thacher \\
59002.254 & $r$ & 14.621 & 0.013 & Thacher \\
59002.256 & $i$ & 14.709 & 0.013 & Thacher \\
59002.262 & $z$ & 14.759 & 0.022 & Thacher \\
59002.267 & $V$ & 14.717 & 0.016 & Thacher \\
59002.522 & UVW1 & 19.404 & 0.157 & Swift \\
59002.524 & $U_S$ & 16.992 & 0.080 & Swift \\
59002.524 & $B$ & 15.680 & 0.056 & Swift \\
59002.525 & UVW2 & $>$20.977 & -- & Swift \\
59002.528 & $V$ & 14.636 & 0.053 & Swift \\
59002.529 & UVM2 & $>$20.575 & -- & Swift \\
59002.792 & $u^{\prime}$ & 17.061 & 0.039 & Las Cumbres \\
59002.794 & $g^{\prime}$ & 14.962 & 0.015 & Las Cumbres \\
59002.795 & $r^{\prime}$ & 14.615 & 0.013 & Las Cumbres \\
59002.796 & $i^{\prime}$ & 14.732 & 0.014 & Las Cumbres \\
59003.264 & $B$ & 15.781 & 0.086 & Nickel \\
59003.267 & $V$ & 14.637 & 0.034 & Nickel \\
59003.269 & $r$ & 14.656 & 0.030 & Nickel \\
59003.271 & $i$ & 14.720 & 0.030 & Nickel \\
59003.337 & $u^{\prime}$ & 16.981 & 0.041 & Las Cumbres \\
59003.339 & $g^{\prime}$ & 15.031 & 0.013 & Las Cumbres \\
59003.340 & $r^{\prime}$ & 14.658 & 0.012 & Las Cumbres \\
59003.341 & $i^{\prime}$ & 14.772 & 0.011 & Las Cumbres \\
59004.248 & $g$ & 15.063 & 0.024 & Thacher \\
59004.254 & $r$ & 14.631 & 0.016 & Thacher \\
59004.257 & $i$ & 14.723 & 0.020 & Thacher \\
59004.262 & $z$ & 14.800 & 0.026 & Thacher \\
59004.268 & $V$ & 14.613 & 0.114 & Thacher \\
59004.310 & $r$ & 14.625 & 0.003 & Pan-STARRS1 \\
59004.806 & $u^{\prime}$ & 17.163 & 0.032 & Las Cumbres \\
59004.808 & $g^{\prime}$ & 14.980 & 0.013 & Las Cumbres \\
59004.809 & $r^{\prime}$ & 14.591 & 0.011 & Las Cumbres \\
59004.810 & $i^{\prime}$ & 14.657 & 0.010 & Las Cumbres \\
59006.473 & $u^{\prime}$ & 17.219 & 0.033 & Las Cumbres \\
59006.475 & $g^{\prime}$ & 15.061 & 0.012 & Las Cumbres \\
59006.476 & $r^{\prime}$ & 14.627 & 0.009 & Las Cumbres \\
59006.477 & $i^{\prime}$ & 14.720 & 0.008 & Las Cumbres \\
59007.036 & UVW1 & 19.605 & 0.168 & Swift \\
59007.038 & $U_S$ & 17.319 & 0.094 & Swift \\
59007.038 & $B$ & 15.803 & 0.056 & Swift \\
59007.039 & UVW2 & $>$20.995 & -- & Swift \\
59007.042 & $V$ & 14.698 & 0.053 & Swift \\
59007.042 & UVM2 & $>$20.587 & -- & Swift \\
59007.248 & $g$ & 15.050 & 0.017 & Thacher \\
59007.254 & $r$ & 14.646 & 0.014 & Thacher \\
59007.258 & $i$ & 14.710 & 0.013 & Thacher \\
59007.262 & $z$ & 14.800 & 0.017 & Thacher \\
59007.266 & $V$ & 14.764 & 0.016 & Thacher \\
59007.827 & $u^{\prime}$ & 17.347 & 0.033 & Las Cumbres \\
59007.829 & $g^{\prime}$ & 15.049 & 0.012 & Las Cumbres \\
59007.830 & $r^{\prime}$ & 14.604 & 0.010 & Las Cumbres \\
59007.831 & $i^{\prime}$ & 14.698 & 0.009 & Las Cumbres \\
59008.249 & $g$ & 15.049 & 0.018 & Thacher \\
59008.254 & $r$ & 14.650 & 0.013 & Thacher \\
59008.259 & $i$ & 14.773 & 0.044 & Thacher \\
59008.262 & $z$ & 14.740 & 0.019 & Thacher \\
59008.268 & $V$ & 14.765 & 0.017 & Thacher \\
59008.290 & $r$ & 14.673 & 0.003 & Pan-STARRS1 \\
59008.300 & $z$ & 14.766 & 0.004 & Pan-STARRS1 \\
59009.747 & $u^{\prime}$ & 17.494 & 0.031 & Las Cumbres \\
59009.749 & $g^{\prime}$ & 15.103 & 0.010 & Las Cumbres \\
59009.750 & $r^{\prime}$ & 14.646 & 0.008 & Las Cumbres \\
59009.751 & $i^{\prime}$ & 14.733 & 0.009 & Las Cumbres \\
59010.249 & $g$ & 15.087 & 0.014 & Thacher \\
59010.256 & $r$ & 14.745 & 0.015 & Thacher \\
59010.260 & $i$ & 14.691 & 0.011 & Thacher \\
59010.262 & $z$ & 14.792 & 0.015 & Thacher \\
59010.268 & $V$ & 14.803 & 0.012 & Thacher \\
59011.252 & $g$ & 15.111 & 0.012 & Thacher \\
59011.254 & $r$ & 14.680 & 0.011 & Thacher \\
59011.262 & $i$ & 14.718 & 0.010 & Thacher \\
59011.266 & $z$ & 14.769 & 0.015 & Thacher \\
59011.268 & $V$ & 14.799 & 0.013 & Thacher \\
59011.290 & $r$ & 14.684 & 0.003 & Pan-STARRS1 \\
59012.216 & UVW1 & 20.219 & 0.275 & Swift \\
59012.217 & $U_S$ & 17.695 & 0.122 & Swift \\
59012.218 & $B$ & 15.917 & 0.056 & Swift \\
59012.219 & UVW2 & $>$21.045 & -- & Swift \\
59012.222 & $V$ & 14.749 & 0.054 & Swift \\
59012.222 & UVM2 & $>$20.589 & -- & Swift \\
59012.251 & $g$ & 15.139 & 0.010 & Thacher \\
59012.254 & $r$ & 14.690 & 0.009 & Thacher \\
59012.259 & $i$ & 14.728 & 0.009 & Thacher \\
59012.267 & $z$ & 14.777 & 0.014 & Thacher \\
59012.268 & $V$ & 14.796 & 0.010 & Thacher \\
59012.361 & $u^{\prime}$ & 17.710 & 0.026 & Las Cumbres \\
59012.363 & $g^{\prime}$ & 15.155 & 0.009 & Las Cumbres \\
59012.364 & $r^{\prime}$ & 14.649 & 0.008 & Las Cumbres \\
59012.365 & $i^{\prime}$ & 14.715 & 0.009 & Las Cumbres \\
59013.251 & $g$ & 15.163 & 0.011 & Thacher \\
59013.256 & $r$ & 14.701 & 0.009 & Thacher \\
59013.262 & $i$ & 14.739 & 0.008 & Thacher \\
59013.264 & $z$ & 14.774 & 0.013 & Thacher \\
59013.268 & $V$ & 14.828 & 0.010 & Thacher \\
59013.834 & $u^{\prime}$ & 17.737 & 0.048 & Las Cumbres \\
59013.836 & $g^{\prime}$ & 15.183 & 0.014 & Las Cumbres \\
59014.253 & $g$ & 15.158 & 0.013 & Thacher \\
59014.254 & $r$ & 14.679 & 0.011 & Thacher \\
59014.262 & $i$ & 14.731 & 0.012 & Thacher \\
59014.267 & $z$ & 14.789 & 0.017 & Thacher \\
59014.269 & $V$ & 14.833 & 0.014 & Thacher \\
59014.788 & $u^{\prime}$ & 17.780 & 0.025 & Las Cumbres \\
59014.790 & $g^{\prime}$ & 15.166 & 0.009 & Las Cumbres \\
59014.791 & $r^{\prime}$ & 14.669 & 0.008 & Las Cumbres \\
59014.792 & $i^{\prime}$ & 14.713 & 0.008 & Las Cumbres \\
59015.240 & $V$ & 14.796 & 0.044 & Nickel \\
59015.242 & $r$ & 14.754 & 0.036 & Nickel \\
59015.244 & $i$ & 14.804 & 0.032 & Nickel \\
59015.252 & $g$ & 15.199 & 0.014 & Thacher \\
59015.258 & $r$ & 14.710 & 0.010 & Thacher \\
59015.263 & $i$ & 14.751 & 0.009 & Thacher \\
59015.266 & $z$ & 14.791 & 0.013 & Thacher \\
59015.269 & $V$ & 14.837 & 0.011 & Thacher \\
59015.280 & $g$ & 15.262 & 0.004 & Pan-STARRS1 \\
59016.252 & $g$ & 15.231 & 0.012 & Thacher \\
59016.256 & $r$ & 14.734 & 0.010 & Thacher \\
59016.261 & $i$ & 14.761 & 0.009 & Thacher \\
59016.266 & $z$ & 14.798 & 0.016 & Thacher \\
59016.269 & $V$ & 14.871 & 0.013 & Thacher \\
59017.123 & $R$ & 14.761 & 0.051 & Auburn \\
59017.128 & $G$ & 15.012 & 0.124 & Auburn \\
59017.131 & $B$ & 16.102 & 0.041 & Auburn \\
59017.992 & UVW1 & $>$19.831 & -- & Swift \\
59017.993 & $U_S$ & 18.003 & 0.265 & Swift \\
59017.994 & $B$ & 16.083 & 0.086 & Swift \\
59017.994 & UVW2 & $>$19.463 & -- & Swift \\
59017.997 & $V$ & 14.918 & 0.083 & Swift \\
59017.997 & UVM2 & $>$20.282 & -- & Swift \\
59018.250 & $g$ & 15.261 & 0.012 & Thacher \\
59018.258 & $r$ & 14.738 & 0.009 & Thacher \\
59018.263 & $i$ & 14.775 & 0.010 & Thacher \\
59018.268 & $z$ & 14.835 & 0.014 & Thacher \\
59018.269 & $V$ & 14.903 & 0.012 & Thacher \\
59018.310 & $i$ & 14.723 & 0.004 & Pan-STARRS1 \\
59018.320 & $g$ & 15.295 & 0.004 & Pan-STARRS1 \\
59020.164 & $R$ & 14.741 & 0.201 & Auburn \\
59022.218 & $g$ & 15.337 & 0.011 & Thacher \\
59022.223 & $r$ & 14.791 & 0.008 & Thacher \\
59022.225 & $i$ & 14.824 & 0.009 & Thacher \\
59022.230 & $z$ & 14.841 & 0.012 & Thacher \\
59022.237 & $V$ & 14.953 & 0.011 & Thacher \\
59023.219 & $r$ & 14.785 & 0.036 & Nickel \\
59023.222 & $i$ & 14.826 & 0.031 & Nickel \\
59023.285 & $g$ & 15.360 & 0.012 & Thacher \\
59023.288 & $r$ & 14.803 & 0.010 & Thacher \\
59023.296 & $i$ & 14.846 & 0.012 & Thacher \\
59023.298 & $z$ & 14.844 & 0.020 & Thacher \\
59023.303 & $V$ & 14.992 & 0.015 & Thacher \\
59023.775 & $B$ & 16.076 & 0.009 & Las Cumbres \\
59023.777 & $V$ & 14.936 & 0.007 & Las Cumbres \\
59023.779 & $r^{\prime}$ & 14.758 & 0.006 & Las Cumbres \\
59025.716 & $u^{\prime}$ & 18.692 & 0.101 & Las Cumbres \\
59025.718 & $g^{\prime}$ & 15.542 & 0.008 & Las Cumbres \\
59025.719 & $r^{\prime}$ & 14.822 & 0.006 & Las Cumbres \\
59025.720 & $i^{\prime}$ & 14.911 & 0.006 & Las Cumbres \\
59027.258 & $g$ & 15.447 & 0.017 & Thacher \\
59027.262 & $r$ & 14.885 & 0.013 & Thacher \\
59027.268 & $i$ & 14.847 & 0.014 & Thacher \\
59027.274 & $z$ & 14.878 & 0.023 & Thacher \\
59027.277 & $V$ & 15.068 & 0.018 & Thacher \\
59028.257 & $g$ & 15.546 & 0.028 & Thacher \\
59028.262 & $r$ & 14.877 & 0.020 & Thacher \\
59028.266 & $i$ & 14.840 & 0.022 & Thacher \\
59028.275 & $z$ & 14.901 & 0.038 & Thacher \\
59028.278 & $V$ & 15.116 & 0.027 & Thacher \\
59029.166 & $R$ & 14.921 & 0.101 & Auburn \\
59030.258 & $g$ & 15.607 & 0.020 & Thacher \\
59030.261 & $r$ & 14.946 & 0.015 & Thacher \\
59030.269 & $i$ & 14.931 & 0.017 & Thacher \\
59030.272 & $z$ & 14.950 & 0.031 & Thacher \\
59030.278 & $V$ & 15.187 & 0.025 & Thacher \\
59031.256 & $g$ & 15.604 & 0.021 & Thacher \\
59031.264 & $r$ & 14.986 & 0.014 & Thacher \\
59031.269 & $i$ & 14.975 & 0.015 & Thacher \\
59031.271 & $z$ & 15.008 & 0.024 & Thacher \\
59031.278 & $V$ & 15.189 & 0.023 & Thacher \\
59031.695 & $u^{\prime}$ & 19.114 & 0.167 & Las Cumbres \\
59031.697 & $g^{\prime}$ & 15.678 & 0.009 & Las Cumbres \\
59031.698 & $r^{\prime}$ & 14.946 & 0.007 & Las Cumbres \\
59031.699 & $i^{\prime}$ & 14.980 & 0.007 & Las Cumbres \\
59032.257 & $g$ & 15.665 & 0.021 & Thacher \\
59032.262 & $r$ & 15.004 & 0.013 & Thacher \\
59032.265 & $i$ & 14.978 & 0.013 & Thacher \\
59032.273 & $z$ & 14.992 & 0.020 & Thacher \\
59032.276 & $V$ & 15.299 & 0.021 & Thacher \\
59032.280 & $r$ & 14.999 & 0.003 & Pan-STARRS1 \\
59033.258 & $g$ & 15.731 & 0.021 & Thacher \\
59033.265 & $r$ & 14.963 & 0.014 & Thacher \\
59033.266 & $i$ & 15.048 & 0.014 & Thacher \\
59033.271 & $z$ & 15.020 & 0.020 & Thacher \\
59033.276 & $V$ & 15.321 & 0.024 & Thacher \\
59034.259 & $g$ & 15.838 & 0.025 & Thacher \\
59034.262 & $r$ & 15.082 & 0.014 & Thacher \\
59034.266 & $i$ & 15.094 & 0.014 & Thacher \\
59034.273 & $z$ & 15.127 & 0.022 & Thacher \\
59034.278 & $V$ & 15.341 & 0.026 & Thacher \\
59034.300 & $r$ & 15.128 & 0.004 & Pan-STARRS1 \\
59035.259 & $g$ & 16.013 & 0.036 & Thacher \\
59035.264 & $r$ & 15.155 & 0.017 & Thacher \\
59035.267 & $i$ & 15.149 & 0.020 & Thacher \\
59035.270 & $z$ & 15.144 & 0.026 & Thacher \\
59035.278 & $V$ & 15.428 & 0.034 & Thacher \\
59036.222 & $V$ & 15.471 & 0.038 & Nickel \\
59036.224 & $r$ & 15.239 & 0.037 & Nickel \\
59036.227 & $i$ & 15.221 & 0.035 & Nickel \\
59036.256 & $g$ & 16.018 & 0.054 & Thacher \\
59036.263 & $r$ & 15.268 & 0.022 & Thacher \\
59036.267 & $i$ & 15.178 & 0.020 & Thacher \\
59036.272 & $z$ & 15.150 & 0.051 & Thacher \\
59036.277 & $V$ & 15.677 & 0.060 & Thacher \\
59037.259 & $g$ & 16.148 & 0.033 & Thacher \\
59037.260 & $r$ & 15.300 & 0.018 & Thacher \\
59037.268 & $i$ & 15.282 & 0.020 & Thacher \\
59037.273 & $z$ & 15.316 & 0.030 & Thacher \\
59037.275 & $V$ & 15.851 & 0.056 & Thacher \\
59038.771 & $u^{\prime}$ & 19.624 & 0.224 & Las Cumbres \\
59038.773 & $g^{\prime}$ & 16.373 & 0.008 & Las Cumbres \\
59038.774 & $r^{\prime}$ & 15.447 & 0.006 & Las Cumbres \\
59038.775 & $i^{\prime}$ & 15.473 & 0.006 & Las Cumbres \\
59039.300 & $r$ & 15.584 & 0.005 & Pan-STARRS1 \\
59041.280 & $r$ & 15.918 & 0.004 & Pan-STARRS1 \\
59042.122 & $R$ & 15.901 & 0.121 & Auburn \\
59042.144 & $G$ & 16.814 & 0.218 & Auburn \\
59044.270 & $g$ & 17.515 & 0.017 & Pan-STARRS1 \\
59048.007 & UVW1 & $>$20.882 & -- & Swift \\
59048.009 & $U_S$ & $>$20.432 & -- & Swift \\
59048.009 & $B$ & 16.849 & 0.192 & Swift \\
59048.010 & UVW2 & $>$21.275 & -- & Swift \\
59048.014 & $V$ & 17.195 & 0.214 & Swift \\
59048.015 & UVM2 & $>$20.661 & -- & Swift \\
59048.280 & $g$ & 17.684 & 0.016 & Pan-STARRS1 \\
59050.745 & $g^{\prime}$ & 17.627 & 0.024 & Las Cumbres \\
59050.746 & $r^{\prime}$ & 16.466 & 0.013 & Las Cumbres \\
59050.747 & $i^{\prime}$ & 16.628 & 0.016 & Las Cumbres \\
59059.021 & UVW1 & $>$20.863 & -- & Swift \\
59059.024 & $U_S$ & $>$20.380 & -- & Swift \\
59059.025 & $B$ & 17.134 & 0.204 & Swift \\
59059.026 & UVW2 & $>$21.316 & -- & Swift \\
59059.030 & $V$ & 17.478 & 0.193 & Swift \\
59059.031 & UVM2 & $>$20.671 & -- & Swift \\
59060.367 & $g^{\prime}$ & 17.669 & 0.033 & Las Cumbres \\
59060.368 & $r^{\prime}$ & 16.602 & 0.018 & Las Cumbres \\
59060.369 & $i^{\prime}$ & 16.802 & 0.025 & Las Cumbres \\
59064.006 & UVW1 & $>$20.870 & -- & Swift \\
59064.009 & $U_S$ & $>$20.373 & -- & Swift \\
59064.010 & $B$ & 17.246 & 0.236 & Swift \\
59064.011 & UVW2 & $>$21.350 & -- & Swift \\
59064.016 & $V$ & 17.501 & 0.236 & Swift \\
59064.017 & UVM2 & $>$20.657 & -- & Swift \\
59069.120 & UVW1 & $>$20.935 & -- & Swift \\
59069.123 & $U_S$ & $>$20.375 & -- & Swift \\
59069.123 & $B$ & 17.461 & 0.199 & Swift \\
59069.124 & UVW2 & $>$21.344 & -- & Swift \\
59069.128 & $V$ & 17.569 & 0.267 & Swift \\
59069.719 & UVM2 & $>$20.564 & -- & Swift \\
59163.487 & $r^{\prime}$ & 17.745 & 0.106 & Las Cumbres \\
59167.503 & UVW1 & $>$20.737 & -- & Swift \\
59167.506 & $U_S$ & $>$20.184 & -- & Swift \\
59167.507 & $B$ & $>$19.199 & -- & Swift \\
59167.508 & UVW2 & $>$21.000 & -- & Swift \\
59167.513 & $V$ & $>$16.535 & -- & Swift \\
59167.514 & UVM2 & $>$20.525 & -- & Swift \\
59169.475 & $g^{\prime}$ & 18.801 & 0.036 & Las Cumbres \\
59169.477 & $r^{\prime}$ & 17.753 & 0.014 & Las Cumbres \\
59169.479 & $i^{\prime}$ & 18.161 & 0.023 & Las Cumbres \\
59180.506 & $g^{\prime}$ & 18.819 & 0.070 & Las Cumbres \\
59180.507 & $r^{\prime}$ & 17.980 & 0.041 & Las Cumbres \\
59180.508 & $i^{\prime}$ & 18.251 & 0.079 & Las Cumbres \\
59189.733 & $i^{\prime}$ & 18.352 & 0.129 & Las Cumbres \\
59194.922 & UVW1 & $>$20.789 & -- & Swift \\
59194.925 & $U_S$ & $>$20.244 & -- & Swift \\
59194.926 & $B$ & $>$19.406 & -- & Swift \\
59194.927 & UVW2 & $>$21.196 & -- & Swift \\
59194.932 & $V$ & $>$17.060 & -- & Swift \\
59194.933 & UVM2 & $>$20.620 & -- & Swift \\
59195.580 & $g$ & 19.085 & 0.041 & Pan-STARRS1 \\
59195.590 & $i$ & 18.566 & 0.028 & Pan-STARRS1 \\
59196.577 & $r$ & 18.167 & 0.102 & Thacher \\
59197.576 & $r$ & 18.202 & 0.064 & Thacher \\
59199.541 & $r$ & 18.229 & 0.052 & Thacher \\
59201.546 & $r$ & 18.478 & 0.064 & Thacher \\
59203.533 & $g^{\prime}$ & 18.974 & 0.094 & Las Cumbres \\
59203.534 & $r^{\prime}$ & 18.492 & 0.056 & Las Cumbres \\
59207.272 & UVW1 & $>$20.994 & -- & Swift \\
59207.273 & $U_S$ & $>$20.552 & -- & Swift \\
59207.274 & $B$ & $>$19.689 & -- & Swift \\
59207.275 & UVW2 & $>$21.407 & -- & Swift \\
59207.278 & $V$ & $>$18.613 & -- & Swift \\
59207.278 & UVM2 & $>$20.677 & -- & Swift \\
59209.293 & $g^{\prime}$ & 19.071 & 0.029 & Las Cumbres \\
59209.296 & $r^{\prime}$ & 18.377 & 0.021 & Las Cumbres \\
59209.299 & $i^{\prime}$ & 18.628 & 0.025 & Las Cumbres \\
59211.660 & $r$ & 18.531 & 0.037 & Pan-STARRS1 \\
59215.271 & $g^{\prime}$ & 19.049 & 0.142 & Las Cumbres \\
59215.272 & $r^{\prime}$ & 18.557 & 0.092 & Las Cumbres \\
59215.273 & $i^{\prime}$ & 18.714 & 0.089 & Las Cumbres \\
59216.650 & $r$ & 18.594 & 0.048 & Pan-STARRS1 \\
59217.026 & UVW1 & $>$20.936 & -- & Swift \\
59217.027 & $U_S$ & $>$20.628 & -- & Swift \\
59217.028 & $B$ & $>$19.768 & -- & Swift \\
59217.028 & UVW2 & $>$21.474 & -- & Swift \\
59217.031 & $V$ & $>$18.644 & -- & Swift \\
59217.031 & UVM2 & $>$20.686 & -- & Swift \\
59221.262 & $g^{\prime}$ & 19.043 & 0.032 & Las Cumbres \\
59221.265 & $r^{\prime}$ & 18.611 & 0.024 & Las Cumbres \\
59221.269 & $i^{\prime}$ & 18.796 & 0.032 & Las Cumbres \\
59227.121 & UVW1 & $>$20.985 & -- & Swift \\
59227.124 & $U_S$ & $>$20.608 & -- & Swift \\
59227.125 & $B$ & $>$19.749 & -- & Swift \\
59227.126 & UVW2 & $>$21.469 & -- & Swift \\
59227.131 & $V$ & $>$17.873 & -- & Swift \\
59227.132 & UVM2 & $>$20.688 & -- & Swift \\
59227.249 & $g^{\prime}$ & 19.391 & 0.061 & Las Cumbres \\
59227.250 & $r^{\prime}$ & 18.676 & 0.048 & Las Cumbres \\
59227.251 & $i^{\prime}$ & 18.865 & 0.063 & Las Cumbres \\
59229.519 & $g^{\prime}$ & 19.369 & 0.206 & Las Cumbres \\
59229.520 & $r^{\prime}$ & 18.695 & 0.051 & Las Cumbres \\
59229.521 & $i^{\prime}$ & 18.831 & 0.065 & Las Cumbres \\
59231.305 & $g^{\prime}$ & 19.271 & 0.063 & Las Cumbres \\
59231.600 & $r$ & 18.923 & 0.038 & Pan-STARRS1 \\
59231.610 & $g$ & 19.531 & 0.062 & Pan-STARRS1 \\
59237.313 & $g^{\prime}$ & 19.342 & 0.096 & Las Cumbres \\
59237.314 & $r^{\prime}$ & 18.878 & 0.055 & Las Cumbres \\
59237.315 & $i^{\prime}$ & 18.972 & 0.059 & Las Cumbres \\
59242.990 & UVW1 & $>$19.113 & -- & Swift \\
59242.992 & $U_S$ & $>$18.565 & -- & Swift \\
59242.992 & $B$ & $>$19.716 & -- & Swift \\
59242.993 & UVW2 & $>$20.754 & -- & Swift \\
59242.996 & $V$ & $>$16.857 & -- & Swift \\
59242.996 & UVM2 & $>$19.557 & -- & Swift \\
59243.570 & $r$ & 19.109 & 0.082 & Pan-STARRS1 \\
59243.580 & $z$ & 19.309 & 0.103 & Pan-STARRS1 \\
59243.962 & $g^{\prime}$ & 19.315 & 0.145 & Las Cumbres \\
59244.124 & UVW1 & $>$20.976 & -- & Swift \\
59244.126 & $U_S$ & $>$20.532 & -- & Swift \\
59244.127 & $B$ & $>$19.708 & -- & Swift \\
59244.128 & UVW2 & $>$21.424 & -- & Swift \\
59244.131 & $V$ & $>$17.786 & -- & Swift \\
59244.131 & UVM2 & $>$20.670 & -- & Swift \\
59245.430 & $g^{\prime}$ & 19.343 & 0.089 & Las Cumbres \\
59245.433 & $r^{\prime}$ & 19.260 & 0.068 & Las Cumbres \\
59245.435 & $i^{\prime}$ & 18.941 & 0.068 & Las Cumbres \\
59250.270 & $g^{\prime}$ & 19.308 & 0.078 & Las Cumbres \\
59250.271 & $r^{\prime}$ & 19.297 & 0.075 & Las Cumbres \\
59250.272 & $i^{\prime}$ & 19.107 & 0.063 & Las Cumbres \\
59255.039 & $g^{\prime}$ & 19.547 & 0.177 & Las Cumbres \\
59255.040 & $r^{\prime}$ & 19.345 & 0.081 & Las Cumbres \\
59255.041 & $i^{\prime}$ & 19.014 & 0.078 & Las Cumbres \\
59261.379 & $g^{\prime}$ & 19.695 & 0.153 & Las Cumbres \\
59261.380 & $r^{\prime}$ & 19.544 & 0.053 & Las Cumbres \\
59261.381 & $i^{\prime}$ & 19.076 & 0.063 & Las Cumbres \\
59262.570 & $g$ & 19.821 & 0.064 & Pan-STARRS1 \\
59266.198 & $g^{\prime}$ & 19.688 & 0.249 & Las Cumbres \\
59266.199 & $r^{\prime}$ & 19.508 & 0.055 & Las Cumbres \\
59266.200 & $i^{\prime}$ & 19.176 & 0.069 & Las Cumbres \\
59267.570 & $r$ & 19.672 & 0.062 & Pan-STARRS1 \\
59269.070 & UVW1 & $>$20.889 & -- & Swift \\
59269.073 & $U_S$ & $>$20.320 & -- & Swift \\
59269.074 & $B$ & $>$19.406 & -- & Swift \\
59269.075 & UVW2 & $>$21.133 & -- & Swift \\
59269.080 & $V$ & $>$17.442 & -- & Swift \\
59269.082 & UVM2 & $>$20.619 & -- & Swift \\
59269.520 & $z$ & 19.780 & 0.162 & Pan-STARRS1 \\
59269.530 & $r$ & 19.824 & 0.142 & Pan-STARRS1 \\
59270.410 & UVW1 & $>$20.583 & -- & Swift \\
59270.412 & $U_S$ & $>$20.067 & -- & Swift \\
59270.412 & $B$ & $>$19.182 & -- & Swift \\
59270.413 & UVW2 & $>$21.074 & -- & Swift \\
59270.415 & $V$ & $>$18.283 & -- & Swift \\
59270.416 & UVM2 & $>$20.585 & -- & Swift \\
59271.164 & $i^{\prime}$ & 19.214 & 0.286 & Las Cumbres \\
59277.540 & $r$ & 19.781 & 0.074 & Pan-STARRS1 \\
59283.145 & $r^{\prime}$ & 19.753 & 0.104 & Las Cumbres \\
59283.146 & $i^{\prime}$ & 19.343 & 0.144 & Las Cumbres \\
59285.222 & $V$ & 19.859 & 0.047 & Las Cumbres \\
59285.242 & $r^{\prime}$ & 19.755 & 0.041 & Las Cumbres \\
59288.072 & $r^{\prime}$ & 19.897 & 0.244 & Las Cumbres \\
59288.075 & $i^{\prime}$ & 19.503 & 0.286 & Las Cumbres \\
59295.420 & $g$ & 20.907 & 0.247 & Pan-STARRS1 \\
59305.480 & $r$ & $>$20.673 & -- & Pan-STARRS1 \\
59316.430 & $g$ & 20.533 & 0.224 & Pan-STARRS1 \\
59316.440 & $r$ & 20.502 & 0.173 & Pan-STARRS1 \\
59323.360 & $g$ & 20.462 & 0.145 & Pan-STARRS1 \\
59332.290 & $r$ & 20.608 & 0.206 & Pan-STARRS1 \\
59333.310 & $r$ & 20.841 & 0.206 & Pan-STARRS1 \\
59335.400 & $r$ & 21.002 & 0.246 & Pan-STARRS1 \\
59338.280 & $g$ & $>$21.360 & -- & Pan-STARRS1 \\
59338.290 & $i$ & 20.767 & 0.255 & Pan-STARRS1 \\
59346.280 & $g$ & 21.116 & 0.225 & Pan-STARRS1 \\
59346.290 & $r$ & 21.351 & 0.235 & Pan-STARRS1 \\
59349.310 & $g$ & $>$21.302 & -- & Pan-STARRS1 \\
59353.370 & $g$ & $>$21.167 & -- & Pan-STARRS1 \\
59360.330 & $r$ & $>$20.834 & -- & Pan-STARRS1 \\
59363.330 & $r$ & 21.520 & 0.272 & Pan-STARRS1 \\
59365.340 & $r$ & 21.338 & 0.229 & Pan-STARRS1 \\
59367.330 & $g$ & 21.639 & 0.297 & Pan-STARRS1 \\
59371.320 & $r$ & $>$20.948 & -- & Pan-STARRS1 \\
59371.330 & $g$ & $>$21.771 & -- & Pan-STARRS1 \\
59374.310 & $i$ & 21.050 & 0.255 & Pan-STARRS1 \\
59374.320 & $g$ & $>$21.713 & -- & Pan-STARRS1 \\
59377.290 & $g$ & $>$21.854 & -- & Pan-STARRS1 \\
59379.330 & $g$ & 21.284 & 0.273 & Pan-STARRS1 \\
59379.340 & $i$ & $>$21.369 & -- & Pan-STARRS1 \\
59387.280 & $r$ & $>$20.714 & -- & Pan-STARRS1 \\
59388.290 & $r$ & $>$21.784 & -- & Pan-STARRS1 \\
59389.280 & $i$ & $>$21.136 & -- & Pan-STARRS1 \\
59389.290 & $r$ & $>$21.540 & -- & Pan-STARRS1 \\
59391.280 & $r$ & $>$21.597 & -- & Pan-STARRS1 \\
59393.280 & $r$ & $>$21.805 & -- & Pan-STARRS1 \\
59394.290 & $r$ & $>$22.026 & -- & Pan-STARRS1 \\
59396.300 & $g$ & $>$22.875 & -- & Pan-STARRS1 \\
59396.310 & $z$ & $>$22.707 & -- & Pan-STARRS1 \\
59399.300 & $g$ & $>$21.655 & -- & Pan-STARRS1 \\
59400.290 & $i$ & $>$21.771 & -- & Pan-STARRS1 \\
59401.280 & $g$ & $>$22.301 & -- & Pan-STARRS1 \\
59401.290 & $r$ & $>$21.543 & -- & Pan-STARRS1 \\
59402.290 & $g$ & $>$22.502 & -- & Pan-STARRS1 \\
59403.280 & $g$ & $>$22.247 & -- & Pan-STARRS1 \\
59404.280 & $g$ & $>$21.669 & -- & Pan-STARRS1 \\
59405.290 & $g$ & $>$22.801 & -- & Pan-STARRS1 \\
59405.300 & $r$ & $>$21.069 & -- & Pan-STARRS1 \\
59423.489 & $F814W$ & 21.880 & 0.017 & HST \\
59423.495 & $F555W$ & 22.332 & 0.015 & HST \\ \hline
    \caption{\addtocounter{table}{4} Our ultraviolet and optical photometry and public observations of SN~2020jfo presented in this paper, including all available detections and upper limits from Las Cumbres, Nickel, Pan-STARRS1, {\it Swift}, Thacher, and Auburn telescopes.}
    \label{tab:photometry}
\end{longtable}

\end{document}